\documentclass{JHEP3}

\usepackage{amsmath}
\usepackage{bm}
\usepackage{bbm}
\usepackage{graphicx}
\usepackage{mathrsfs}
\usepackage{slashed}

\def\slashchar#1{\setbox0=\hbox{$#1$}     		% set a box for #1
   \dimen0=\wd0                                 	% and get its size
   \setbox1=\hbox{/} \dimen1=\wd1               	% get size of /
   \ifdim\dimen0>\dimen1                        	% #1 is bigger
      \rlap{\hbox to \dimen0{\hfil/\hfil}}      	% so center / in box
      #1                                        	% and print #1
   \else                                        	% / is bigger
      \rlap{\hbox to \dimen1{\hfil$#1$\hfil}}   	% so center #1
      /                                         	% and print /
   \fi}
\title{Particle-hole instability in the $AdS_4$ holography}

\author{E. Gubankova~\thanks{Also at ITEP, Moscow, Russia}\\
Institute for Theoretical Physics, J. W. Goethe-University,
D-60438 Frankfurt am Main, Germany\\
E-mail: \email{gubankova@th.physik.uni-frankfurt.de}
}

\abstract{
We show that particle-hole pairing is realized in the 
background of a charged black hole
in magnetic field.
The pairing instability occurs for sufficiently large fermion charges,
which correspond to the Fermi liquid regime. The critical 
temperature for Fermi liquids is proportional to the magnetic field
and vanishes as we approach
the non-Fermi liquid state.
The pairing order parameter leads to a relative shift 
of the Fermi surfaces
corresponding to the bulk fermions with spin up and down. 
The value of the shift
in Fermi momentum $k_F$ and the critical temperature $T_c$
are proportional to the effective density of states at the Fermi surface.
Our one-loop calculations provide a dual description
of the magnetic catalysis for the lowest Landau level in graphene. 
This analyses may be relevant 
for the antiferromagnetic behavior in the cuprate superconductors
and for the chiral spirals in the chiral magnetic effect.

We also discuss thermodynamic and transport properties of a system at the boundary
at zero magnetic field.
The scaling behavior of the specific heat is $c\sim T$ for Fermi liquid
and $c\sim T^{2\nu}$ for non-Fermi liquid, while the behavior of the DC conductivity
is the same $\sigma\sim T^{-2\nu}$ in both cases. 
While it can be difficult to extract transport and hydrodynamic from the lattice,
the $AdS/CFT$ approach provides a robust frame for nonperturbative calculation of these
properties.
}
\preprint{}

\begin{document}

%%%%%%%%%%%%%%%%%%%%%%%%%%%%%%%%%%%%%%%%%%%%%%%%%%%%%%%%%%%%%%%%%%%%%%%%%%%%%%%%
\section{Introduction}

Particle-hole pairing appears in different contexts in condensed matter physics.
We consider here magnetic catalysis, i.e., generation of the T-odd mass parameter in the presence of a magnetic field.
It is a well established phenomenon in $(2+1)$-dimensions and it is believed to explain the anomalous quantum Hall effect
in graphene, i.e., the appearance of the additional plateaus in the Hall conductivity $\sigma_{xy}$ for the lowest
Landau level \cite{Shovkovy:2d} (see \cite{Shovkovy:3d} for the magnetic catalysis in $(3+1)$-dimensions).
Electron-hole pairing is responsible for the spin density order parameter
and the antiferromagnetic nature of the cuprate superconductors at half filling. Spin-density wave in the form
of spin-orbit ordering can trigger the superconducting pairing, while both superconducting electron-electron and 
spin density wave electron-hole orders are essential to describe physics of Mott insulating and pseudogap phases.
Recently, there was an interest to the particle-hole pairing in the form of chiral spirals in the context
of the chiral magnetic effect and the quarkyonic matter \cite{Kharzeev:2010}. 

These phenomena involve strongly coupled physics. We therefore use the $AdS/CFT$ correspondence
which is a powerful tool in understanding strongly coupled quantum field theories. 
It is formulated as a duality between classical gravitational theory in the anti-de Sitter ($AdS$) space
and a strongly coupled conformal field theory ($CFT$) in the limit of large $N$ and large 't Hooft coupling $\lambda$
defined on the boundary of the $AdS$ space. Recently the $AdS/CFT$ correspondence was applied
to different phenomena which arise in the context of condensed matter systems \cite{ads-cond-mat1:2010,
ads-cond-mat2:2010,ads-cond-mat3:2010,ads-cond-mat4:2010,ads-cond-mat5:2010,ads-cond-mat6:2010}.
Many of the above studies were initiated by the original papers on a holographic superconductor \cite{Gubser:2008,Hartnoll:2008},
the non-Fermi liquid behavior \cite{Lee:2008}, and quantum phase transitions \cite{Cubrovic:2009}.
In particular, there have been significant developments in understanding the superconducting instability 
near a charged black hole. It was shown that charged black holes are unstable to forming hair,
which means that a (free) charged (or neutral) scalar field develops a vacuum expectation value
and breaks spontaneously the corresponding symmetry when put 
in the charged black hole background with asymptotic $AdS$ geometry \cite{Gubser:2008,Hartnoll:2008}.
This was linked to the Breitenlochner-Freedman instability 
that provides a gravitational mechanism for superconductivity: if the charge of the boson
is sufficiently large compared to its mass it will condense.
This mechanism does not give microscopic details behind the superconductivity like the BCS pairing does.
It provides the evidence for the bosonic condensate and suggests a holographic mechanism 
for the superconductivity. 
Using the Cooper pair
picture the critical temperature has been calculated in Ref.\cite{Hartman:2010}. 

In describing the particle-hole pairing we follow the same route as used to address superconductivity \cite{Hartnoll:2009}.
Both graphene and the cuprate superconductors are systems at finite charge density in $(2+1)$-dimensions.
Therefore the gravity dual description is given by a charged black hole in $(3+1)$-dimensional
anti-de Sitter space-time $AdS_4$. Strong coupling and large $N$ limit of the boundary theory translates into
a gravity theory at small curvature and low energy, which reduces to a universal sector of classical
Einstein gravity plus matter fields. The global $U(1)$ symmetry of the conformal field theory (CFT) with current $J_{\mu}$
is mapped to a $U(1)$ local gauge symmetry with a gauge field $A_M$ in the $AdS_4$. In the $AdS_4$,
$A_M$ is an actual (not background) $U(1)$ field, which is dynamic.             

In this paper we add a four-Fermi contact interaction between the charged fermions. We choose
the channel favoring the magnetic catalysis and look for the particle-hole instability that shows up
when the one-loop effective action has negative modes. There is important difference between showing
the instability for the bosonic field $<\Phi>\neq 0$ and for fermions $<\bar{\psi}\psi> \neq 0$.
Calculation for bosons is classical in the black hole background, 
whereas for fermions it involves one loop computation. We use variational approach where we utilize
the formula for the one loop fermion determinant expressed 
through a sum over quasinormal modes of the black hole \cite{Denef1:2009,Denef2:2009}.
As discussed in \cite{Denef1:2009,Denef2:2009},
the quasinormal modes are given
by the poles of the retarded fermion Green function. The structure of the poles for the retarded
Green function has been obtained in \cite{Faulkner1:2009} for various relative relations between 
the charge and the mass of the fermion. 
We also do a one-loop
calculation in the bulk to obtain a non-local in the radial direction Ginsburg-Landau action.
The latter calculation involves bulk fermion propagators and a radial profile for the pairing order
parameter. The idea of calculation follows the Ginsburg-Landau approach.

In this paper we consider application of particle-hole pairing to the magnetic catalysis.
Magnetic catalysis has been shown in $(2+1)$ \cite{Shovkovy:2d} and $(3+1)$ \cite{Shovkovy:3d} dimensional
relativistic models. The general result is that a constant magnetic field 
leads to the generation of a fermion dynamical mass even at the weakest attractive interaction
between fermions. The essence of the effect is that in the magnetic field the dimension
of the system effectively reduces $d\rightarrow d-2$,
i.e., to $(0+1)$ and $(1+1)$ dimensional systems, that favors the dynamics of
the particle-hole pairing (therefore the name of magnetic catalysis) \cite{Shovkovy:2d}. 
%(The physical reason for this reduction is the fact that the motion of charged particles 
%is restricted in directions perpendicular to the magnetic field.) 
%There is however no controversy with
%Mermin-Wagner-Coleman theorem which forbids spontaneous breaking of continuous symmetry in $0+1$ and $1+1$.
%This theorem is not applicable to this case \cite{shovkovy}. 
We can choose different forms of the four-Fermi interaction, that will 
generate a mass term for the fermions with needed symmetry properties. For simplicity, we choose
a contact interaction $G_{int}(\bar{\psi}i\Gamma^2\Gamma^5\psi)^2$ with the strength $G_{int}$
written through the mass scale of the interaction $G_{int}=1/M_{int}^2$. We show that
this interaction triggers the generation of the mass term
$\Delta\bar{\psi}i\Gamma^2\Gamma^5\psi$ which is odd both under time-reversal 
and parity transformations (see the representation of $\Gamma$ matrices). Contrary to the Dirac mass term
$m\bar{\psi}\psi$, the generation of the $T$-odd mass $\Delta\bar{\psi}i\Gamma^2\Gamma^5\psi$ does not break
any symmetry, e.g. $U(1)_L\times U(1)_R$ in the NJL model or spin (flavor) symmetry $SU(2)$ ($U(2)_{+}\times U(2)_{-}$)
in the case of graphene \cite{Shovkovy:2d}. Because no symmetry is spontaneously broken by the T-odd mass,
no gap opens in the spectrum. In this sense, 
the condensate $\Delta\bar{\psi}i\Gamma^2\Gamma^5\psi$ is similar to the spin density wave 
$<\psi^{\dagger}\vec{\sigma}\psi>$, where the former couples to the mass and the latter one to the chemical potential.    
%The chiral structure of the resulting theory is rather rich \cite{shovkovy}.         

The paper is organized in the following way. In section \ref{section:2} we introduce the black hole geometry
and consider the near horizon limit which is dual to the $IR\;CFT$. In section \ref{section:3}
we perform the variational calculation for the particle-hole pairing order parameter. 
In section \ref{section:4} we perform Ginsburg-Landau calculations in the $AdS_4$ bulk geometry,
and calculate the critical temperature. In section \ref{section:5} we consider thermodynamic and transport properties
of a system on the boundary at zero magnetic field.
Appendices contain solution of the Dirac equation
in magnetic field and calculation of the Landau levels in the $AdS_4$ holography (Appendix \ref{appendix:a}), calculation
of the conformal dimension in the $IR\;CFT_3$ (Appendix \ref{appendix:a'}),
solution of the Dirac equation in the $AdS_2$ and obtaining the $IR\;CFT_1$ conformal dimension (Appendix \ref{appendix:b}), 
derivation of the $AdS_2$ Green function (Appendix \ref{appendix:b'}), 
one-loop calculation in $(2+1)$-dimensional field theory (Appendix \ref{appendix:c}), calculation
of the critical temperature in the $AdS_4$ (Appendix \ref{appendix:d}).

\section{Dyonic black hole and infrared CFT}\label{section:2}

We consider $3$-dimensional conformal field theory (CFT) with global $U(1)$ symmetry that has a gravity dual.
At finite charge density and in the presence of magnetic field, the system can be described by a dyonic black hole
in 4-dimensional anti-de Sitter space-time, $AdS_4$, with the current $J_{\mu}$ in the CFT mapped to a $U(1)$ gauge field
$A_M$ in $AdS$. 

The action for a vector field $A_M$ coupled to $AdS_4$ gravity can be written as
\begin{equation}
S=\frac{1}{2\kappa^2}\int d^4x \sqrt{-g}\left( {\mathcal R} +\frac{6}{R^2} -\frac{R^2}{g_F^2}F_{MN}F^{MN}\right), 
\label{action1}
\end{equation}
where $g_F^2$ is an effective dimensionless gauge coupling and $R$ is the curvature radius of $AdS_4$.
The equations of motion following from eq.(\ref{action1})  are solved by the geometry of the dyonic black hole,
i.e., with both electric and magnetic charges,
\begin{equation}
ds^2=g_{MN}dx^Mdx^N = \frac{r^2}{R^2}(-fdt^2+d\vec{x}^2)+\frac{R^2}{r^2}\frac{dr^2}{f},
\label{ads4-metric1}
\end{equation}
where the redshift factor, $f$, and the vector field $A_M$ reflect the fact that the system
is at finite charge density and in the magnetic field,
\begin{eqnarray}
 f &=& 1+\frac{Q^2+H^2}{r^4}-\frac{M}{r^3},\nonumber\\
A_t &=& \mu\left(1-\frac{r_0}{r}\right),\;\; A_x = -{\mathcal H}y,
\label{ads4-metric2}
\end{eqnarray}
where we chose the Landau gauge; 
the chemical potential $\mu$ and the magnetic field ${\mathcal H}$ are given by 
\begin{equation}
 \mu=\frac{g_FQ}{R^2r_0},\;\; {\mathcal H} = \frac{g_F H}{R^4}.
\end{equation}
Here $r_0$ is the horizon radius determined by the largest positive root of the redshift factor,
$f(r_0)=0$,
\begin{eqnarray}
M=r_0^3 + \frac{Q^2+H^2}{r_0} 
\end{eqnarray}
and the CFT is defined at the boundary $r\rightarrow\infty$.
The geometry eqs.(\ref{ads4-metric1}),(\ref{ads4-metric2}) describes the boundary theory at a finite density,
i.e., a system in the medium at chemical potential $\mu$, with the charge, energy, and 
entropy densities given, respectively, by
\begin{eqnarray}
\rho = 2\frac{Q}{\kappa^2R^2g_F},\;\; 
\epsilon =\frac{M}{\kappa^2R^4},\;\;
s=\frac{2\pi}{\kappa^2}\frac{r_0^2}{R^2}.
\end{eqnarray}
The temperature of the system is identified with the Hawking temperature of the black hole,
$T_H\sim |f^{\prime}(r_0)|/4\pi$,
\begin{equation}
T=\frac{3r_0}{4\pi R^2}\left(1-\frac{Q^2+H^2}{3r_0^4}\right). 
\end{equation}
Since $Q$ and $H$ have dimensions of $[L]^2$, it is convenient to parametrize them as
\begin{equation}
 Q^2 = 3r_{*}^4, \;\;
Q^2+H^2 = 3r_{**}^4.
\label{q-and-h} 
\end{equation}
In terms of $r_0$, $r_{*}$ and $r_{**}$ the expressions are
\begin{eqnarray}
f &=& 1+\frac{3r_{**}^4}{r^4}-\frac{r_0^3+3r_{**}^4/r_0}{r^3},\nonumber\\
A_t &=& \mu\left(1-\frac{r_0}{r}\right),\;\;
A_x=-{\mathcal H},
\label{factor}
\end{eqnarray}
with 
\begin{equation}
\mu=\sqrt{3}g_F\frac{r_{*}^2}{R^2r_0},\;\;
{\mathcal H}=\sqrt{3}g_F\frac{\sqrt{r_{**}^4-r_{*}^4}}{R^4}. 
\end{equation}
The expressions for the charge, energy and entropy densities, and for the temperature
are simplified as
\begin{eqnarray}
\rho&=&\frac{2\sqrt{3}}{\kappa^2g_F}\frac{r_{*}^2}{R^2},\;\;
\epsilon=\frac{1}{\kappa^2}\frac{r_0^3+3r_{**}^4/r_0}{R^4},\;\;
s=\frac{2\pi}{\kappa^2}\frac{r_0^2}{R^2},\nonumber\\
T&=& \frac{3}{4\pi}\frac{r_0}{R^2}\left(1-\frac{r_{**}^4}{r_0^4}\right). 
\end{eqnarray}

In the first part of the paper we consider the zero temperature limit, i.e., extremal black hole, 
\begin{equation}
 T=0 \;\; \rightarrow  \;\; r_0=r_{**},
\label{zeroT}
\end{equation}
which in original variables is $Q^2+H^2=3r_0^4$. 
In the zero temperature limit, eq.(\ref{zeroT}), the redshift factor $f$, eq.(\ref{factor}),
develops a double zero at the horizon
\begin{equation}
 f=6\frac{(r-r_{**})^2}{r_{**}^2}+\cdots \,.
\end{equation}
As a result, near the horizon the $AdS_4$ metric reduces to $AdS_2 \times R^2$
with the curvature radius of $AdS_2$ given by
\begin{equation}
 R_2=\frac{1}{\sqrt{6}}R.
\end{equation}
This is a very important property of the metric, which simplifies calculations.  This metric reduction can be seen explicitly
by considering the scaling limit 
\begin{eqnarray}
&& r-r_{**}=\lambda\frac{R_2^2}{\zeta},\;\;
t= \frac{\tau}{\lambda},\nonumber\\
&& \lambda\rightarrow 0\; {\rm with}\; \zeta,\;\tau\; {\rm finite},
\label{scaling} 
\end{eqnarray}
then the metric eq.(\ref{ads4-metric1}) describes a black hole in $AdS_2\times R^2$
\begin{equation}
ds^2=\frac{R_2^2}{\zeta^2}(-d\tau^2+d\zeta^2)+\frac{r_{**}^2}{R^2}d\vec{x}^2,
\label{ads2-metric}
\end{equation} 
with 
\begin{equation}
A_{\tau}=\frac{g_F}{\sqrt{12}}\frac{r_{*}^2}{r_{**}^2}\frac{1}{\zeta},\;\;
A_x=-{\mathcal H}y. 
\end{equation}
Physically, the scaling limit eq.(\ref{scaling}) with finite $\tau$ corresponds to the long time limit 
of the original time coordinate $t$, which translates to the low frequency limit
of the boundary theory 
\begin{equation}
\frac{\omega}{\mu}\rightarrow 0, 
\end{equation}
where $\omega$ is the frequency conjugate to $t$. (One can think of $\lambda$ as being a frequency $\omega$).
Near the $AdS_4$ horizon, 
we expect that gravity of the $AdS_2$ region of an extremal dyonic black hole is described
by a $CFT_1$ dual. We refer to \cite{Faulkner1:2009} for an account of the $AdS_2/CFT_1$ duality. In what follows we use
the horizon of $AdS_2$ region at $\zeta\rightarrow \infty$ (coefficient in front of $d\tau$ vanishes at the horizon)
and the infrared $CFT$ ($IR\;CFT$) defined at the $AdS_2$ boundary, 
$\zeta=0$. The scaling picture eqs.(\ref{scaling}),(\ref{ads2-metric}) suggests that in the low frequency limit,
the $2$-dimensional boundary theory is described by this $IR\;CFT$ (which is a $CFT_1$). The Green function 
for operator ${\mathcal O}$ in the boundary theory is obtained as a small frequency expansion and by a matching
procedure of different regions along radial direction,
and is expressed through the Green function of the $IR\;CFT$ \cite{Faulkner1:2009}.

\section{Variational calculations of the paring gap}\label{section:3}

In this section we perform variational calculations of the particle-hole pairing gap
in the bulk. The logic of calculations is the same as in a field theory, except for arising
radial dependence of bulk quantities, e.g. for the gap parameter $\Delta(r)$ as opposed
to the BCS with a constant gap. The radial profile is important to keep, since it insures convergence
of radial integrals and for different bulk behavior characterizes different systems on the boundary.
Our variational calculations in the bulk are possible due to the one-loop formula for an effective action
obtained in \cite{Denef1:2009,Denef2:2009} and expressions for the poles of the fermion Green function
obtained in \cite{Faulkner1:2009}.

\subsection{Effective action for interacting fermions in a magnetic field}\label{section:3.1}
 
We consider a spinor field $\psi$ in the $AdS_4$ of charge $q$ and mass $m$,
which is dual to an operator ${\mathcal O}$ in the boundary
$CFT_3$ of charge $q$ and dimension  
\begin{equation}
\Delta_{\psi} = \frac{3}{2} + mR, 
\label{conformal-dimension}
\end{equation}
with $mR\geq \frac{1}{2}$ and corresponding to the ``stable'' $CFT$.  
In the black hole geometry, eq.(\ref{ads4-metric1}),
the quadratic action for $\psi$ is written as   
\begin{equation}
S_{0} = i\int d^4x \sqrt{-g}\left(\bar{\psi}\Gamma^{M}{\mathcal D}_{M}\psi-m\bar{\psi}\psi\right),
\label{free-action}
\end{equation}
where $\bar{\psi}=\psi^{\dagger}\Gamma^{\underline t}$, and
\begin{equation}
{\mathcal D}_M=\partial_M +\frac{1}{4}\omega_{abM}\Gamma^{ab}-iqA_M, 
\end{equation}
with $\omega_{abM}$ the spin connection, and $\Gamma^{ab}=\frac{1}{2}[\Gamma^a,\Gamma^b]$; here $M$
and $a,b$ denote the bulk space-time and tangent space indices respectively, and $\mu,\nu$
denote indices along the boundary directions, i.e. $M=(r,\mu)$.

As discussed in the introduction, we can add to $S_0$ the contact interacting part
\begin{eqnarray}
S_{int}=- \int d^4x \sqrt{-g} G_{int}(\bar{\psi}i\Gamma^{\hat{2}}\Gamma^{\hat{5}}\psi)
(\bar{\psi}i\Gamma^{\hat{2}}\Gamma^{\hat{5}}\psi), 
\label{interaction}
\end{eqnarray}
where $G_{int}=1/M_{int}^2$, $M_{int}$ is a mass scale of the interaction.
The representation for $\Gamma$ matrices is given by eq.(\ref{matrices}), and hat
indices on $\Gamma$ matrices always refer to tangent space indices.
In this representation of $\Gamma$ matrices, 
\begin{equation}
i\Gamma^{\hat{2}}\Gamma^{\hat{5}}= 
\left(\begin{array}{cc}
1 & 0 \\
0 & -1
\end{array}
\right).
\end{equation}
We also have $i\Gamma^{\hat{2}}\Gamma^{\hat{5}}= -\Gamma^{\hat{r}}\Gamma^{\hat{t}}\Gamma^{\hat{1}}$.
The form of interaction eq.(\ref{interaction}) is motivated by the form of the projectors eq.(\ref{projection})
which decouple $\psi$ into two components.

We add the magnetic Zeeman splitting of the spin degeneracy
\begin{equation}
S_{B}= \int d^4x \sqrt{-g}\; q{\mathcal H}\bar{\psi}\Gamma^{\hat{t}}\sigma^3\psi, 
\end{equation}
where $\sigma^3$ acts on spin indices. The resulting action is the following sum 
$S=S_0+S_{int}+S_{B}$.

We solve the four-Fermi interaction, $S_{int}$, in the mean-field approximation by performing standard Hubbard-Stratonovich
transformation.
We introduce a composite order parameter 
\begin{equation}
\Delta= 2G_{int}<\bar{\psi}i\Gamma^{\hat{2}}\Gamma^{\hat{5}}\psi>
\label{gap}
\end{equation}
and decouple the interaction into a quadratic form
\begin{equation}
S_{int}= \int d^4x \sqrt{-g} \left(\frac{\Delta^2}{4G_{int}}
-(\Delta\bar{\psi}i\Gamma^{\hat{2}}\Gamma^{\hat{5}}\psi+h.c.)\right). 
\end{equation}
The order parameter $\Delta$ is T-odd, i.e. 
$T(\Gamma^{\hat{0}}i\Gamma^{\hat{2}}\Gamma^{\hat{5}})T^{\dagger}=-(\Gamma^{\hat{0}}i\Gamma^{\hat{2}}\Gamma^{\hat{5}})$, with
\begin{equation}
T= C\Gamma^{5}=
i\left(\begin{array}{cc}
0 & -1 \\
1 & 0
\end{array}
\right),\;\;
C= \left(\begin{array}{cc}
-\sigma^2 & 0 \\
0 & -\sigma^2
\end{array}
\right)
\end{equation}
and charge conjugation is fixed by $C\Gamma^{\hat{t}}=\Gamma^{\hat{r}}$. The exact form of the four-Fermi interaction
is not important. With any interaction $G_{int}(\bar{\psi}\Gamma\psi)(\bar{\psi}\Gamma\psi)$ respecting the symmetries,
where $\Gamma$ stands for a collective combination of Gamma matrices, the gap given by eq.(\ref{}) can be generated.
Its effective action  is given by $S_{eff}=-i{\rm Tr}\left(\ln G^{-1}+\frac{1}{2}(G_0^{-1}G-1)\right)$ which satisfies
the stationarity condition (or gap equation) $\delta S_{eff}/\delta G=0$ \cite{Shovkovy:2d}. 

To get an effective action for $\Delta$, we intetgrate out the fermion fields
with the result
\begin{equation}
S_{eff}=\int d^4x \sqrt{-g} \left(\frac{|\Delta|^2}{4G_{int}}
-\frac{1}{2}{\rm Tr}\ln G^{-1}\right),
\label{one-loop}
\end{equation}
where the full fermion propagator $G(x,x')=<\psi(x)\bar{\psi}(x')>$ and its inverse is given by
\begin{equation}
G^{-1}(x,x')= \Gamma^{M}{\mathcal D}_{M}-m - \Delta i\Gamma^{\hat{2}}\Gamma^{\hat{5}} \pm q{\mathcal H}\Gamma^{\hat{t}}.
\end{equation}
In the one-loop effective action eq.(\ref{one-loop}), the trace and the logarithm are taken in the functional sense.
%${\rm Tr}$ implies trace over the spinor indices and integration 
%in the ${\rm AdS_4}$, $\int d^4 x$. 
The coordinate $x=\{r,t,\vec{x}\}$ includes the radial $r$ and $(2+1)$ boundary directions $\{t,\vec{x}\}$
in $AdS_4$. If we assume translational invariance along boundary directions, then the Fourier transform
is given by
\begin{equation}
 G(x,x')=T\sum_n\int\frac{d^2k}{(2\pi)^2}G(i\omega_n,k,r,r'){\rm e}^{-i\omega_n(t-t')+i\vec{k}(\vec{x}-\vec{x}')},
\end{equation}
where fermionic Matsubara frequency is $\omega_n=(2n+1)\pi T$. 
We furthermore assume that the condensate is a function of only the radial coordinate
in the $AdS_4$, $\Delta(r,\omega,k)=\Delta(r)$. The effective action is given by
\begin{equation}
S_{eff}=\frac{V_2}{T}\int dr \sqrt{-g} \left(\frac{|\Delta(r)|^2}{4G_{int}}
-\frac{T}{2}\sum_n\int\frac{d^2k}{(2\pi)^2}dr'{\rm Tr}\ln G^{-1}(i\omega_n,k,r,r')\right).
\label{one-loop}
\end{equation} 
In general, it is a difficult task to calculate one loop effective action in the bulk. Here we will use
the method suggested in \cite{Denef1:2009,Denef2:2009} to calculate the free energy (determinants) in a black hole background
as a sum over the quasinormal modes of the black hole.
We will also use the recent analytical results \cite{Faulkner1:2009} for the fermion quasinormal modes.

\subsection{Variational calculations of magnetic catalysis in a charged black hole geometry $AdS_4$}\label{section:3.2}

To calculate one-loop fermion action eq.(\ref{one-loop}) we need to know eigenvalues
of the Dirac equation eq.(\ref{dir2}), which can be written symbolically for each mode as \cite{Denef1:2009,Denef2:2009}
\begin{equation}
M(z,l)\Phi=\lambda(z,l)\Phi 
\end{equation}
with $z=i\omega_n$. The zero modes $\lambda(z_{*}(l),l)=0$ define as solutions
the quasinormal frequencies $z_{*}(l)$.
As was shown in \cite{Denef1:2009,Denef2:2009}, the quasinormal frequencies $z_{*}$ of a wave equation in a black hole space-time
are actually poles in the corresponding retarded Green function in the black hole background, where
\begin{equation}
M(i\omega_n,l)G(i\omega_n,l,r,r')=r^4\delta(r,r'). 
\end{equation}
Indeed representing the Green function as a sum over eigenfunctions we have
\begin{equation}
{\rm Tr} \frac{1}{M(i\omega_n,l)}=\int_0^{r_+}dr G(i\omega_n,l,r,r), 
\end{equation}
which is the usual representation of $G$. This equation is shown to be true for general complex $z=i\omega_n$
and $G$ satisfyes the ingoing boundary condition at $\omega\neq 0$ 
and regularity at $\omega=0$ at the horizon \cite{Denef1:2009,Denef2:2009}.
Therefore, as was shown in \cite{Denef1:2009,Denef2:2009}, the fermion determinant given 
by a sum over the quasinormal frequencies of the black hole,
i.e. when $M(z,l)$ has a zero eigenvalue, is equivalent to the sum over the poles of the analytically 
continued to complex frequencies fermion
Green function with ingoing boundary conditions at the horizon. This method is also used
in color superconductivity \cite{Alford:2006}.   

Analytic results have been obtained for the Green function in the $AdS_4$ \cite{Faulkner1:2009}.
(Numerically it has been obtained in \cite{Vegh:2009}.)
A general form for the retarded Green function is given by \cite{Faulkner1:2009,Faulkner2:2009,Faulkner:2010}
\begin{equation}
G_R(\omega,k)=\frac{B_{+}+B_{-}G^{IR}(\omega)}{A_{+}+A_{-}G^{IR}(\omega,k)}, 
\end{equation}
where the ratio of  numerator to denominator comes from expansion of the solution
of Dirac equation at the boundary $r\rightarrow\infty$, therefore the ratio $B/A$ is 
the $AdS_4$ retarded Green function, $G_R$.
The coefficients $A_{\pm}, B_{\pm}$ are expansions (rows) in small frequency $\omega$. 
The low-frequency limit is equivalent to the expansion near the horizon at small $\omega$,
where metric reduces to $AdS_4\rightarrow AdS_2\times R^2$. Therefore
$A_{+}\;(B_{+})$ and $A_{-}\;(B_{-})$ components arise from expansion (and matching procedure)
near the $AdS_2$ boundary, that relates ratio $A_{+}/A_{-}\;(B_{+}/B_{-})$ via 
the $IR$ Green function $G^{IR}$ obtained in the $AdS_2\times R^2$ calculation. 
Non-analytical frequency behavior of $G$ is controlled by the $IR\;CFT$, $G^{IR}$, while 
coefficients $A_{\pm}$ and $B_{\pm}$ carry the UV information. In section \ref{section:4}, we find 
the coefficients $A_{\pm}$, $B_{\pm}$ to the leading order in frequency, that requires solution
of the Dirac equation in the $AdS_4$.

It was found in \cite{Vegh:2009} and \cite{Faulkner1:2009}, that the fermion Green function develops a sharp pick indicating 
the existence of the Fermi surface and quasiparticle poles. Expansion of the Green function
near the Fermi surface is given by \cite{Faulkner1:2009}
\begin{eqnarray}
G_R(\omega,k) &=& \frac{(-h_1v_F)}{\omega-v_Fk_{\perp}+\Sigma(\omega,k_F)},
\nonumber\\
\Sigma(\omega,k_F) &=& hv_FG^{IR}(\omega,k_F)=hv_Fc(k_F)\omega^{2\nu_{k_F}}, 
\label{green-function}
\end{eqnarray}
where we keep notations for constants introduced in \cite{Faulkner1:2009}.
Here $k_{\perp}=k-k_F$, $h_1,h,v_F$ are governed by the UV physics, and were obtained numerically in \cite{Faulkner1:2009}
and here in section \ref{section:4}. Eq.(\ref{green-function}) gives the rough structure for the boundary Green function.
Further, in section \ref{section:4}, we use a more detailed description for $G_R$.  
As $T\rightarrow 0$, $\Sigma(\omega,k_F)\rightarrow \omega^{2\nu_{k_F}}$, therefore at zero temperature
there is no dependence on $\Delta$ coming from $G^{IR}$.
We are interested
in the poles of $G_R$ at zero temperature. In this section UV constants will not be important.
Theres are three characteristic regimes depending on the $\nu_{k_F}$,
the low energy ($\omega\ll \mu$) scaling dimension of the dual fermionic operator.
The poles of the Green function are located in the lower half complex 
plane at 
\begin{equation}
\omega_c(k)=\omega_{*}(k)-i\Gamma(k).
\label{pole}
\end{equation} 
For the three regimes we have the following \cite{Faulkner1:2009}, \cite{Faulkner:2010}

$\bullet$ For (quasi-) Fermi liquid, $\nu_{k_F}>\frac{1}{2}$,
\begin{equation}
\omega_{*}(k) = v_F(k-k_F)+\ldots,\;\;
\frac{\Gamma(k)}{\omega_{*}(k)}\sim (k-k_F)^{2\nu_{k_F}-1}\rightarrow 0, 
\label{landau1}
\end{equation}
and the residue of the pole is $Z=h_1v_F$. The pole represents a stable quasiparticle
as one approaches the Fermi surface, with linear dispersion relation and $v_F$ being the Fermi velocity,
vanishing decay width and a non-vanishing spectral weight $Z$ at the Fermi surface.

$\bullet$ For non-Fermi liquid, $\nu_{k_F}<\frac{1}{2}$,
\begin{equation}
\omega_{*}(k) = (k-k_F)^{\frac{1}{2\nu_{k_F}}},\;\;
\frac{\Gamma(k)}{\omega_{*}(k)} =  const, 
\label{landau2}
\end{equation}
and the residue of the pole is $Z\sim (k-k_F)^{\frac{1-2\nu_{k_F}}{2\nu_{k_F}}}\rightarrow 0$.
The pole represents an unstable quasiparticle
as one approaches the Fermi surface, with exponent in dispersion relation greater than one,
the imaginary part is comparable to the real part of the pole, and a vanishing
spectral weight $Z$ at the Fermi surface. Non-Fermi liquid is example of a Fermi surface
without sharp quasiparticle picks.

$\bullet$ For marginal Fermi liquid, $\nu_{k_F}=\frac{1}{2}$,
\begin{equation}
\Sigma(\omega) \approx \tilde{c}_1\omega\log\omega +id_1\omega,\;\;
\frac{d_1}{\tilde{c}_1} = -\frac{\pi}{1+{\rm e}^{-\frac{2\pi q}{\sqrt{12}}}}, 
\label{landau3}
\end{equation}
where $\tilde{c}_1<0$ and $d_1$ are real constants.  
The single-particle scattering rate is linear in $\omega$, while it is still suppressed
compared to the real part as the Fermi surface is approached, but the suppression is only logarithmic.
The quasiparticle residue also vanishes logarithmically at the Fermi surface.
We use this summary on quasiparticle poles below.

Following \cite{Denef1:2009,Denef2:2009}, we represent the fermion determinant in an effective action eq.(\ref{one-loop})
as a sum over poles of the retarded Green function in the black hole background.
We obtain an analog to eq.(\ref{action2}) of one-loop action
\begin{eqnarray}
S_{eff}= \frac{V_2}{T}\left(\int dr \sqrt{-g}
\frac{|\Delta(r)|^2}{4G_{int}}+\frac{T|q{\mathcal H}|}{2\pi}
\sum_{z_{*}[\Delta(r)]}\ln\left(\frac{1}{2\pi}
|\Gamma(\frac{iz_{*}[\Delta(r)]}{2\pi T}+\frac{1}{2})|^2\right)\right),
\label{action3}
\end{eqnarray} 
where $z_{*}[\Delta(r)]$ is a functional of the order parameter $\Delta(r)$,
$V_2$ is the boundary spatial volume.
In order to make a connection with the field theory eq.(\ref{action2}),
the following equations has been used for the complex frequency $z_{*}$ 
\begin{equation}
|\Gamma(\frac{1}{2}+iz)|^2=\frac{\pi}{\cosh(\pi z)}. 
\end{equation}
It was shown in \cite{Denef2:2009}, that it holds
for a complex $z$ by matching poles and zeros of the two meromorphic functions.
Here $|\Gamma(\frac{1}{2}+iz)|^2=\Gamma(\frac{1}{2}+iz)\Gamma(\frac{1}{2}-i\bar{z})$. 
We will consider the zero temperature limit, therefore as explained in \cite{Denef1:2009},
the sum in eq.(\ref{action3}) is saturated by one pole eq.(\ref{pole}),
\begin{equation}
z_{*}[\Delta(r)] = \omega_{*}[\Delta(r)]-i\Gamma[\Delta(r)],
\label{pole2}
\end{equation}
where the real and imaginary parts of the dispersion are functionals of the order parameter $\Delta(r)$.
As was shown in \cite{Denef2:2009},
equation (\ref{action3}) for the fermion determinant captures only the singular contributions
incorporated by the closest to $\omega=0$ pole eq.(\ref{pole}),
and smooth analytic terms are not improtant. 

We take the functional derivative,
\begin{equation}
\delta S_{eff}=\frac{V_2}{T}\int dr\sqrt{-g}\left(
\frac{2\Delta(r)}{4G_{int}}+\frac{T|q{\mathcal H}|}{2\pi}\frac{\delta}{\delta\Delta(r)}
\ln\left(\frac{1}{2\pi}
|\Gamma(\frac{iz_{*}[\Delta(r)]}{2\pi T}+\frac{1}{2})|^2\right)
\right)\delta\Delta(r), 
\end{equation}
with $\frac{\delta\Delta(r')}{\delta\Delta(r)}=\delta(r-r')$ and therefore
the dimension of the functional derivative $\frac{\delta}{\delta\Delta(r)}$
is $\frac{1}{[r\Delta]}$.
For the gap equation, $\frac{\delta S_{eff}}{\delta\Delta(r)}=0$, we have
\begin{equation}
\Delta(r)= \frac{G_{int}|q{\mathcal H}|}{\pi}
\frac{1}{\pi}\left(\frac{\delta\omega_{*}[\Delta(r)]}{\delta\Delta(r)}
{\rm Im}  \Psi(\frac{iz_{*}[\Delta(r)]}{2\pi T}+\frac{1}{2})
-\frac{\delta\Gamma[\Delta(r)]}{\delta\Delta(r)}
{\rm Re} \Psi(\frac{iz_{*}[\Delta(r)]}{2\pi T}+\frac{1}{2})\right), 
\label{gap-equation}
\end{equation}
where $\Psi$ is the digamma function, $\Psi(x)=\frac{d\ln\Gamma(x)}{dx}$.

In the zero temperature limit, $T\sim 0$, we have
\begin{eqnarray}
&& \hspace{-4cm} \Delta(r) = \frac{G_{int}|q{\mathcal H}|}{\pi}\frac{1}{\pi}
\left(\frac{\delta\omega_{*}[\Delta(r)]}{\delta\Delta(r)}
(\frac{\pi}{2}-\arctan\frac{\Gamma[\Delta(r)]}{\omega_{*}[\Delta(r)]})\right.\nonumber\\
&+&\left.\frac{\delta\Gamma[\Delta(r)]}{\delta\Delta(r)}
\ln\frac{2\pi T}{\sqrt{\omega_{*}[\Delta(r)]^2+\Gamma[\Delta(r)]^2}}\right).
\end{eqnarray}
Generally, it is difficult to find the dependence for the pole $z_{*}[\Delta(r)]$ \cite{Faulkner2:2009}.
However, here we have simplifications. First, the order parameter enters the Dirac equation essentially
as a mass term on the diagonal, i.e. it does not mix $\psi^{\dagger}$ and $\psi$ as a superconducting gap does.
Therefore in the pole of the Green function eq.(\ref{green-function}), the frequency is not affected,
and there is only a shift in the Fermi momentum $k_F$. Second, we can consider the order parameter $\Delta$
to be small. The procedure of finding the Fermi momentum is reduced to finding the bound state of the Schrodinger
equation with zero energy \cite{Faulkner1:2009} or to finding a solution to the Dirac equation which is normalizable
at the boundary \cite{Hartman:2010}. Since the order parameter contributes to the potential term in the Dirac equation,
the shift in $k_F$ is given by the first order perturbative correction
\begin{eqnarray}
&& k_F\rightarrow k_F \pm \delta k_F[\Delta(r)],\nonumber\\
&& \delta k_F[\Delta(r)] =  \frac{b h_1}{v_FR^3}\int dr \sqrt{-g} \psi^{0}(r)^{\dagger}\sigma^1\psi^{0}(r)\Delta(r),
\label{shift} 
\end{eqnarray}
where the zero modes $\psi^{0}$ are solutions of the (free, without $\Delta$) Dirac equation with $\omega=0$ and $k=k_F$
introduced in the next section, the signs $\pm$ refer to components $F_{1}/F_{2}$ of the Dirac equation (\ref{dir2}),
$b$ is a dimensionless constant which has to be determined from the equation for the Fermi momentum $k_F$ in the presence
of the gap $\Delta$.
Note that $\Gamma^{\hat{t}}={\rm diag}(i\sigma^1,i\sigma^1)$.
The unperturbed Fermi momentum is $k_F=\sqrt{q^2\mu^2-m^2}$ and $R$ is the $AdS$ radius.
In eqs.(\ref{landau1})-(\ref{landau3}),
the following substitution should be made $k\rightarrow \sqrt{2|q{\mathcal H}|l}$ (this substitution in the pole
was shown to be true in \cite{Denef1:2009} based on scaling arguments, see also Appendix \ref{appendix:a}).
Introducing magnetic field lowers the Fermi energy. For large magnetic field the lowest Landau level $l=0$ dominates
and higher Landau levels are not important,
while for small magnetic field all Landau levels should be included in the sum to correctly reproduce the limit of
zero magnetic field.    
Following \cite{Denef1:2009}, we tune magnetic field to the point when $k\sim k_F$, and once the pole crosses the Fermi surface
it is counted in the fermion determinant of the effective action.      

For the Fermi liquid, $\nu_{k_F}>\frac{1}{2}$,
\begin{equation}
\omega_{*}[\Delta(r)]=v_F\delta k_F[\Delta(r)], 
\;\;
\Gamma[\Delta(r)] \sim (\delta k_F[\Delta(r)])^{2\nu_{k_F}}.
\end{equation}
Near the Fermi surface we have for small $\Delta\sim 0$
\begin{eqnarray}
&& \frac{\delta\omega_{*}[\Delta(r)]}{\delta\Delta(r)}\sim \frac{h_1v_F^3}{R^3}\psi^{0}(r)^{\dagger}\sigma^1\psi^{0}(r),
\;\;
\frac{\delta\Gamma[\Delta(r)]}{\delta\Delta(r)}\sim
\frac{\psi^{0}(r)^{\dagger}\sigma^1\psi^{0}(r)}{R^4}(\delta k_F[\Delta(r)])^{2\nu_{k_F}-1}\rightarrow 0, 
\nonumber\\
&& \frac{\Gamma[\Delta(r)]}{\omega_{*}[\Delta(r)]}\sim
(\delta k_F[\Delta(r)])^{2\nu_{k_F}-1} \rightarrow 0.
\end{eqnarray}
Therefore at $T=0$, the gap equation gives the following solution
\begin{equation}
\Delta(r)= \frac{G_{int}|q{\mathcal H}|bh_1v_F^3}{2\pi R^3}
\psi^{0}(r)^{\dagger}\sigma^1\psi^{0}(r),
\label{solution-gap2}
\end{equation}
where $G_{int}=\frac{1}{M_F^2}$, and $b$ is a dimensionless constant.
The difference in factor $2$ with the $(2+1)$-dimensional case eq.(\ref{solution-gap})
is due to taking one pole eq.(\ref{pole}) instead of two poles in the field theory, 
which does not affect our conclusions.
Eq.(\ref{solution-gap2}) contains the radial profile of the order parameter, 
$\psi^{0}(r)^{\dagger}\sigma^1\psi^{0}(r)$ shown in Fig.(\ref{plot-delta}).
The prefactor in eq.(\ref{solution-gap2}) contains information about 
the magnetic catalysis for the lowest Landau level at $T=0$. 
At zero temperature,
the only solution is a nonzero gap which is proportional to the magnetic field and a radius
of the four-Fermi interaction $\Delta\sim \frac{1}{M_F}|q{\mathcal H}|$.
For a finite $T$, from eq.(\ref{gap-equation}), there exist also a trivial solution
$\Delta=0$, and a critical temperature $T_c$ separates the phases with zero and nonzero gaps.
This is in complete analogy with a $(2+1)$ field theory case \cite{Shovkovy:2d}. We consider the regime
around $T_c$ and the corresponding phase transition in the next section. 

For the non-Fermi liquid, $\nu_{k_F}<\frac{1}{2}$,
\begin{equation}
\omega_{*}[\Delta(r)]\sim \Gamma[\Delta(r)]\sim
(\delta k_F[\Delta(r)])^{\frac{1}{2\nu_{k_F}}}.
\end{equation}
Near the Fermi surface we have for small $\Delta\sim 0$
\begin{equation}
\frac{\delta\omega_{*}[\Delta(r)]}{\delta\Delta(r)}\sim\frac{\delta\Gamma[\Delta(r)]}{\delta\Delta(r)}
\sim (\delta k_F[\Delta(r)])^{\frac{1}{2\nu_{k_F}}-1}
\rightarrow 0,
\end{equation}
zeros for both term in the gap equation mean that there is no insability.
Therefore in this case 
\begin{equation}
\Delta(r)=0, 
\label{solution-gap3}
\end{equation}
i.e. no gap is generated in a magnetic field. Thus, for Fermi liquids, particle-hole pairing
$\bar{\psi}\Gamma\psi$, is favorable in the magnetic field, while non-Fermi liquids
do not support the pairing. The same conclusion has been reached for the case of the superconducting
pairing $<\psi\Gamma\psi>$ where $\Gamma$ contains $\Gamma^{\hat{5}}$ in \cite{Hartman:2010}. 
There, it was suggested that
taking the long-range four-Fermi interaction may generate the instability for non-Fermi liquids.
On a technical ground, the momentum/frequency dependent four-Fermi interaction will change
a simple shift in the Fermi momentum eq.(\ref{shift}), so that
the derivatives of the real and/or imaginary parts of the pole will not vanish.   
%Technically, the gap vanishes for non-Fermi liquids
%due to both vanishing derivatives of the real and imaginary parts of the pole. Therefore,
%taking the long-range four-Fermi interaction probably will not solve the problem.

\section{Pairing instability in the Ginsburg-Landau formalism} \label{section:4}

In this section we consider the pairing particle-hole instability using Ginsburg-Landau
approach in the bulk. Our calculation closely follows 
the leading order (one loop) Ginsburg-Landau procedure, with the only difference of using bulk
fermion propagators. The prescription to construct propagators and vertices on the gravity side
is given in \cite{Hartman:2010}, \cite{Faulkner:2010}.

\subsection{Microscopic calculations of magnetic catalysis. Ginsburg-Landau in a holographic approach}\label{section:4.1}

In the mean field approximation, the four-Fermi  (contact) interaction $\bar{\psi}(x)\Gamma\psi(x)\bar{\psi}(x)\Gamma\psi(x)$
gives the following bilinear terms
\begin{equation}
\bar{\psi}\Gamma\psi\Delta+\bar{\psi}\bar{\Gamma}\psi\Delta^{\dagger},
\end{equation}
where the second term is hermitian conjugate to the first one and 
$\bar{\Gamma}=\Gamma^{\hat{t}}\Gamma^{\dagger}\Gamma^{\hat{t}}$, and
the order parameter 
\begin{equation}
\Delta=G_{int} \langle \bar{\psi}\Gamma\psi \rangle 
\end{equation}
is a singlet (number) in spin space. Here
$\Gamma=i\Gamma^{\hat{2}}\Gamma^{\hat{5}}$. The one-loop (Euclidean) action is given to the second order in $\Delta$
\begin{eqnarray}
S^{(2)} = \int d^4x \sqrt{g}\frac{|\Delta|}{4G_{int}}-2\int d^4x d^4x' 
{\rm tr} {\mathcal G}(x',x) \Gamma \Delta(x) {\mathcal G}(x,x')\bar{\Gamma}\Delta(x')^{\dagger},  
\end{eqnarray}
where the Euclidean non-interacting Green function in the bulk is ${\mathcal G}(x,x')=-\langle\psi(x)\bar{\psi}(x')\rangle$.
Assuming translational invariance along the spacetime $\{\tau,\vec{x}\}$, we perform the Fourier transform 
\begin{equation}
{\mathcal G}(x,x') = T\sum_n \int\frac{d^2k}{(2\pi)^2}{\mathcal G}(r,r',i\omega_n,\vec{k})
{\rm e}^{-i\omega_n(\tau-\tau')+i\vec{k}(\vec{x}-\vec{x}')},
\end{equation}
where the radial coordinate is $r$ and the boundary directions are $\{\tau,\vec{x}\}$; the fermionic
Matsubara frequencies are $\omega_n=\pi T(2n+1)$. We assume that the gap depends only on the radial direction,
$\Delta=\Delta(u)$. The one-loop effective action is given by 
\begin{eqnarray}
S^{(2)} &=& \frac{V_2}{T}\int dr\sqrt{g(r)}\left(\frac{|\Delta|^2}{4G_{int}}
+\int dr'\sqrt{g(r')}\Delta(r)\Delta(r')^{*} F(r,r')\right)\nonumber\\ 
F(r,r') &=& -2T\sum_n\int\frac{d^2k}{(2\pi)^2}{\rm tr}
{\mathcal G}(r',r,i\omega_n,\vec{k})\Gamma {\mathcal G}(r,r',-i\omega_n,-\vec{k})\bar{\Gamma}. 
\label{action5}
\end{eqnarray}
We make analytic continuation of the Euclidean Green function into the lower (upper) half plane in imaginary frequency plane,
and use the following expressions relating Euclidean and retarded (advanced) Green functions \cite{Hartman:2010}  
\begin{eqnarray}
&& {\mathcal G}(z){\mathcal G}(-z)={\mathcal G}^R(z){\mathcal G}^A(-z) \nonumber\\
&&  {\mathcal G}^A(r,r')=-{\mathcal G}^R(r,r')^{*}, 
\end{eqnarray}
where $z=i\omega_n$, in order to rewrite the action in terms of retarded (advanced) Green functions 
\begin{equation}
F(r,r') = i\int\frac{d^2k}{(2\pi)^2}\int_{-\infty}^{\infty}\frac{d\Omega}{\pi}\tanh\frac{\Omega}{2T}
{\rm tr} {\mathcal G}^R(r',r,\Omega,\vec{k})^{*}\Gamma {\mathcal G}^R(r,r',-\Omega,-\vec{k})\bar{\Gamma}, 
\label{kernel}
\end{equation}
where we substituted the Matsubara sum by the contour integral, $i\omega_n\rightarrow z$, and
$\Omega$ is on the real axis of $z$.
In order to calculate this integral, we express the bulk Green function through the boundary one
as given in \cite{Hartman:2010}. The bulk Green function is a solution of the free Dirac equation, eq.(\ref{dirac5}),
\begin{equation}
D(\Omega,k){\mathcal G}^R(r,r',\Omega,k)=\frac{1}{\sqrt{-g}}i\delta(r,r'), 
\end{equation}
with the free radial Dirac operator $D(\Omega,k)$, which includes the mass term 
and the magnetic field but has zero gap, $\Delta=0$.
The bulk Green function is constructed through the modes $\psi(r)$, $\psi={\rm e}^{-i\Omega t+ikx}\psi(r)$,
which are solutions of the free Dirac equation eq.(\ref{dirac5}) 
\begin{equation}
D(\Omega,k)\psi_{radial}(r)=0. 
\end{equation}
Due to the choice of the Gamma matrices, eq.(\ref{matrices}), $\psi$ decouples into two-component spinors,
$\psi=(\psi_1,\psi_2)^{T}$, which are eigenfunctions with definite eigenvalue of 
$\Gamma^{\hat{r}}\Gamma^{\hat{t}}\Gamma^{\hat{1}}$.
Therefore the bulk retarded Green function 
has the block-diagonal form, eq.(\ref{block-diagonal}), where
the components ${\mathcal G}_\alpha$, $\alpha=1,2$, are constructed from the solutions of the Dirac equation as \cite{Hartman:2010}
\begin{equation}
{\mathcal G}_{\alpha}^R(r,r')=\frac{G_{\alpha}(\Omega,k)}{R^3}\times\left\{
\begin{array}{c}
-\psi_{\alpha}^{bdy}(r)\tilde{\psi}_{\alpha}^{in}(r')\;\; r>r'\\
-\psi_{\alpha}^{in}(r)\tilde{\psi}_{\alpha}^{bdy}(r')\;\; r<r' 
\end{array} \right.,
\label{bulk-boundary}
\end{equation}
with $\tilde{\psi}_{\alpha}=i\psi_{\alpha}^{T}\sigma^{1}$. Note that 
$\Gamma^{\hat{0}}\equiv \Gamma^{\hat{t}}={\rm diag}(i\sigma^1,i\sigma^1)$
and the dimension of $\psi$ is $[\psi]\sim L^{3/2}$; the minus sign comes from the definition
of the bulk Green fuction ${\mathcal G}=-<\psi\bar{\psi}>$.
Here the prefactor arises from the Wronskian \cite{Hartman:2010},
and includes the retarded Green function of the boundary field theory (later we refer to it as the boundary Green function).
Since the Wronskian is a constant related to the conserved charge current, it is simpler to calculate it at the conformal 
boundary \cite{Hartman:2010} with the result given by eq.(\ref{bulk-boundary}).

There is the following reasoning behind contruction of the bulk Green function in eq.(\ref{bulk-boundary}).
At the boundary, $r,r'\rightarrow \infty$, the behavior is given by two terms (we omit prefactors)
\begin{equation}
G\sim  r^{\Delta_{\psi}-d} + r^{-\Delta_{\psi}}, 
\end{equation}
and we put to zero the non-normalizable part $\sim r^{\Delta_{\psi}-d}$,
and leave the normalizable part $\sim r^{-\Delta_{\psi}}$.
At the horizon, $r,r'\rightarrow r_0$, there are two terms
\begin{equation}
G\sim {\rm e}^{-ikr_0} +{\rm e}^{ikr_0}, 
\end{equation}
where we throw away the outgoing solution ${\rm e}^{-ikr_0}$,
and leave the ingoing one ${\rm e}^{ikr_0}$. In this way we obtain the retarded Green function.
Note that
${\rm e}^{\pm ikr_0}\sim (r-r_0)^{\pm i\omega/T}$ with the Lorentzian signature.
In the Euclidean space, this means that when the behavoir of the Green function is fixed at the horizon,
$r,r'\rightarrow r_0$ as $G\sim (r-r_0)^{\pm \omega/T}$,
then the retarded Green function has asymptotics $G_R\sim (r-r_0)^{\omega/T}$, with $\omega>0$
(and corresponds to the ingoing solution), and the advanced Green function 
has asymptotics $G_A\sim (r-r_0)^{-\omega/T}$, with $\omega<0$
(and corresponds to the outgoing solution). 

The boundary conditions are fixed as follows for the solutions of the Dirac equation.
The  (free) solution of the Dirac equation has the following behavior near the boundary, 
$r\rightarrow\infty$, eq.(\ref{boundary}) with $\Delta=0$,   
\begin{equation}
 \psi_{\alpha}\sim a_{\alpha}r^{mR}
\left(\begin{array}{c}
0\\
1 
\end{array}\right) 
+b_{\alpha}r^{-mR}
\left(\begin{array}{c}
1\\
0 
\end{array}\right).
\end{equation}
The two spinors $(1,0)$ and $(0,1)$ are eigenstates of $\Gamma^{\hat{r}}$ with opposite eigenvalues, implying that
$a_{\alpha}$ and $b_{\alpha}$ are canonically conjugate (in a radial Hamiltonian slicing) \cite{Iqbal:2009}. Therefore
a boundary condition must be imposed on one, with the other allowed to fluctuate. For $mR>\frac{1}{2}$, we choose
the fluctuating piece to be the normalizable mode at the boundary (with regular behavior) proportonal to $(1,0)$.
This gives us the solution $\psi^{bdy}$. More generally,
the quantization choice for $mR>\frac{1}{2}$ is to impose the boundary condition $a_{\alpha}=0$ on the fluctuating mode.
Another quantization choice for $mR>\frac{1}{2}$ is discussed in \cite{Iqbal:2009,Faulkner1:2009}. For the solution $\psi^{in}$
we impose ingoing boundary conditions at the horizon, $r=r_0$. Thus, there are two normalizable solutions 
with the following behavior
at the conformal boundary $r\rightarrow \infty$
\begin{eqnarray}
\psi_{\alpha}^{bdy}&=& r^{-mR}
\left(\begin{array}{c}
1\\
0 
\end{array}\right) 
\nonumber\\
\psi_{\alpha}^{in} &=& \frac{1}{G_{\alpha}} r^{mR}
\left(\begin{array}{c}
0\\
1 
\end{array}\right) 
+r^{-mR}
\left(\begin{array}{c}
1\\
0 
\end{array}\right),
\label{definition-solution}
\end{eqnarray}
where the boundary Green function $G_{\alpha}$ was introduced in $\psi_{\alpha}^{in}$ following its definition
being proportinal to $b_{\alpha}/a_{\alpha}$ in the ingoing solution. Using representation eq.(\ref{bulk-boundary}),
we can show that ${\mathcal G}^{R}(r,r',\Omega,\vec{k})^{*}=-{\mathcal G}^{A}(r,r',\Omega,\vec{k})$, since
$\psi^{bdy\;*}=\psi^{bdy}$ and $\psi^{in\;*}=\psi^{out}$. 
Also ${\mathcal G}^{A}(r,r',\Omega,\vec{k})=\Gamma^{\hat{t}}{\mathcal G}^{R}(r',r,\Omega,\vec{k})^{\dagger}\Gamma^{\hat{t}}$,
since $i\sigma^1{\mathcal G}^{R}_{\alpha}(r',r,\Omega,\vec{k})^{\dagger}i\sigma^1
=-{\mathcal G}^{R}_{\alpha}(r,r',\Omega,\vec{k})^{*}$. Therefore we can rewrite the kernel eq.(\ref{kernel}) in equivalent form
\begin{equation}
F(r,r') = -i\int\frac{d^2k}{(2\pi)^2}\int_{-\infty}^{\infty}\frac{d\Omega}{\pi}\tanh\frac{\Omega}{2T}
{\rm tr} \Gamma^{\hat{t}} {\mathcal G}^R(r,r',\Omega,\vec{k})^{\dagger}\Gamma^{\hat{t}}\Gamma 
{\mathcal G}^R(r,r',-\Omega,-\vec{k})\bar{\Gamma}. 
\label{kernel2}
\end{equation}
Using  the relation eq.(\ref{bulk-boundary}), we obtain
\begin{eqnarray}
F(r,r') &=& \frac{i}{R^6}\int\frac{d^2k}{(2\pi)^2}\int_{-\infty}^{\infty}\frac{d\Omega}{\pi}\tanh\frac{\Omega}{2T}
G_1(\Omega,\vec{k})^{*}G_1(-\Omega,-\vec{k})\times\nonumber\\
&&\left\{\begin{array}{c}
\psi_{2}^{in}(r',\Omega, \vec{k})^{\dagger}\sigma^1\psi_{2}^{in}(r',-\Omega,-\vec{k})
\psi_{1}^{bdy}(r,\Omega, \vec{k})^{\dagger}\sigma^1\psi_{1}^{bdy}(r,-\Omega,-\vec{k}),\;\; r>r'\\
\psi_{2}^{bdy}(r',\Omega, \vec{k})^{\dagger}\sigma^1\psi_{2}^{bdy}(r',-\Omega,-\vec{k})
\psi_{1}^{in}(r,\Omega, \vec{k})^{\dagger}\sigma^1\psi_{1}^{in}(r,-\Omega,-\vec{k}),\;\; r<r'
\end{array} \right.\nonumber\\
&&+(1 \leftrightarrow2),
\label{kernel3}
\end{eqnarray} 
where we used $\psi_{\alpha}^{in}(-\Omega)=\psi_{\alpha}^{out}$ and
$\psi_{\alpha}^{in}(\Omega)^{*}=\psi_{\alpha}^{out}(\Omega)$ which follows
from the definition of in(out)going solution, the solution $\psi^{bdy}$ is real, 
and $\psi_1(-\vec{k})=\psi_2(\vec{k})$ which follows 
from the symmetry of the Dirac equation \cite{Vegh:2009}; we omitted other arguments by $\psi$'s.   

All quantities in eq.(\ref{kernel3}) can be obtained numerical. However, at low temperatures $T\ll\mu$,
calculations can be done analytically, due to the fact that the main contribution comes
from the pole in a retarded Green function describing the Fermi surface, i.e.,
the closest to the origin pole $\omega_{*}\sim 0$ as $k\sim k_F$. The same argument that physics
occurs close to the Fermi surface in the bulk was used in \cite{Denef1:2009} describing phenomena in magnetic field
and for the BCS theory in a holographic approach by \cite{Hartman:2010}. This repeats the reasoning 
of the field theory, as in the BCS theory. Close to the Fermi surface and at low temperatures,
i.e. at $T\ll\Omega\ll\mu$, the boundary Green function, $G\equiv G_1$,
to the leading order is given by \cite{Faulkner1:2009}
\begin{equation}
G(\Omega,k)=\frac{(-h_1v_F)}{\Omega-v_Fk_{\perp}-h_2v_F{\rm e}^{i\theta-i\pi\nu}\Omega^{2\nu}}, 
\label{green-function}
\end{equation}
where $k_{\perp}=k-k_F$ is the perpendicular distance of the momentum from the Fermi surface,
$h_1$ and $v_f$ are real constants and are calculated below, $h_2$ is positive and the phase
$\theta$ is such that poles of the Green function are in the lower half complex frequncy plane,
$\nu$ is the zero temperature conformal dimension at the Fermi momentum, $\nu\equiv \nu_{k_F}$,
given by eq.(\ref{conformal2}) with $\Delta=0$.

We perform the momentum and frequency integrals in eq.(\ref{kernel3}). We make the same assumption
as in \cite{Denef1:2009} and as in our variational calculations.
We consider only contributions near the Fermi momentum $k\sim k_F$ (see discussion after eq.(\ref{shift})). 
Therefore, in the leading order, there is no
momentum dependence in the boundary Green functions, and the momentum
integral $d^2k$ is trivially performed, which gives in the magnetic field 
a factor $|q{\mathcal H}|$. In the frequency integral, for $T\ll\Omega$, we substitute
$\tanh\frac{\Omega}{2T}\rightarrow 1$, and we have
\begin{eqnarray}
F(r,r') &=& \frac{1}{R^6}{\rm Re}
\int_{0}^{\infty}\frac{d\Omega}{\pi^2}\;\frac{h_1}{-\Omega/v_F+h_2{\rm e}^{-i\theta+i\pi\nu}\Omega^{2\nu}}
\;\frac{h_1}{\Omega/v_F+h_2{\rm e}^{i\theta+i\pi\nu}\Omega^{2\nu}}\times\nonumber\\
&&\left\{\begin{array}{c}
\psi_{2}^{in}(r',\Omega)^{\dagger}\sigma^1\psi_{2}^{in}(r',-\Omega)
\psi_{1}^{bdy}(r,\Omega)^{\dagger}\sigma^1\psi_{1}^{bdy}(r,-\Omega)\;\; r>r'\\
\psi_{2}^{bdy}(r',\Omega)^{\dagger}\sigma^1\psi_{2}^{bdy}(r',-\Omega)
\psi_{1}^{in}(r,\Omega)^{\dagger}\sigma^1\psi_{1}^{in}(r,-\Omega)\;\; r<r'
\end{array} \right.,
\label{kernel4}
\end{eqnarray}
where the wave functions are evaluated at the Fermi momentum.
In the frequency integral,
depending on the critical exponent $\nu$ either the first or the second term in each denominator
dominates. To make this comparison, note that $h_2$ has a dimension, i.e. $h_2\sim \mu^{1-2\nu}$.
At small frequencies $\Omega\ll\mu$,
for $\nu>\frac{1}{2}$ the first term $\sim\Omega$ dominates, while for $\nu<\frac{1}{2}$ 
the second term $\sim\mu\left(\frac{\Omega}{\mu}\right)^{2\nu}$ dominates, as also shown in \cite{Faulkner1:2009}.
As discussed in \cite{Hartman:2010}, the range $T\ll\Omega\ll\mu$ implies that the fermion wavefunctions should be
evaluated at $\Omega=0$ and $k=k_F$ in the extremal $T=0$ black hole background. This means, they are exactly
zero modes at the Fermi surface, $\psi^{bdy}=\psi^{in}=\psi^{0}$, which will be calculated below.

For $\nu>\frac{1}{2}$, the leading behavior of eq.(\ref{kernel4}) at small temperatures $T\ll \mu$ is 
\begin{equation}
 F(r,r')=-\frac{|q{\mathcal H}|h_1^2v_F^3}{2\pi^2R^6}\frac{1}{T}\; \psi^{0}(r)^{\dagger}\sigma^1\psi^{0}(r)
\psi^{0}(r')^{\dagger}\sigma^1\psi^{0}(r'),
\end{equation}
where $\psi^{0}$ is the $T=0$ fermion zero mode at the fermi surface.

For $\nu<\frac{1}{2}$, the leading behavior of eq.(\ref{kernel4}) at small temperatures $T\ll\mu$ is
\begin{equation}
 F(r,r')=-\frac{|q{\mathcal H}|h_1^2v_F}{2\pi^2h_{2}^2 R^6}\cos(2\pi\nu)\frac{\mu^{1-4\nu}-T^{1-4\nu}}{1-4\nu}\;
\psi^{0}(r)^{\dagger}\sigma^1\psi^{0}(r)
\psi^{0}(r')^{\dagger}\sigma^1\psi^{0}(r'),
\end{equation}
that for $\nu<\frac{1}{4}$ becomes to the leading order in expansion $T/\mu$ temperature independent
\begin{equation}
 F(r,r')=-\frac{|q{\mathcal H}|h_1^2v_F}{2\pi^2h_{2}^2 R^6}\cos(2\pi\nu)\frac{\mu^{1-4\nu}}{1-4\nu}\;
\psi^{0}(r)^{\dagger}\sigma^1\psi^{0}(r)
\psi^{0}(r')^{\dagger}\sigma^1\psi^{0}(r'),
\end{equation}   
and for $\frac{1}{4}<\nu<\frac{1}{2}$ has a wrong sign for the kernel $F$ to give a nontrivial solution of the gap equation.
This means that there is no instability in particle-antiparticle pairing for $\nu<\frac{1}{2}$. We observed
this already before in variational calculations, where non-Fermi liquids did not support 
$\psi\bar{\psi}$-pairing while the Fermi liquids did, see eqs.(\ref{solution-gap2}) and (\ref{solution-gap3}).   

For the Fermi liquids, we obtain the equation for the critical temperature 
from the effective action eq.(\ref{action5})
\begin{equation}
\frac{\Delta(r)}{2G_{int}}+\int dr'\sqrt{-g(r')}\Delta(r')F(r',r)=0,
\end{equation}
which is an analog of the gap equation but with free propagators (Green functions).
As in \cite{Hartman:2010}, we use the factorisation of $F(r,r')$ to write an ansatz for the gap function
\begin{equation}
\Delta^{0}(r)\sim\psi^{0}(r)^{\dagger}\sigma^{1}\psi^{0}(r).
\label{gap} 
\end{equation}
Using this ansatz back in the equation, we get the following critical temperature for $\nu>\frac{1}{2}$ 
\begin{equation}
T_c=\frac{G_{int}|q{\mathcal H}|h_1^2v_F^3}{\pi^2R^6}\int dr\sqrt{-g(r)}(\psi^{0}(r)^{\dagger}\sigma^1\psi^{0}(r))^2.
\label{critical-temperature}
\end{equation}
The radial form of the gap function eq.(\ref{gap}) was obtained by variational calculations in eq.(\ref{solution-gap2}).
Eq.(\ref{solution-gap2}) can be considered as a one loop mass gap equation (self-energy correction)
with  a four-Fermi vertex $G_{int}$
and the bulk fermion propagator ${\mathcal G}^R(r,r')$ given by eq.(\ref{bulk-boundary}) and modified as
in eq.(\ref{shift}) to include the gap. The mass gap equation reads
$\Delta(r)\sim G_{int}\int\frac{d^2k}{(2\pi)^2}\int d\omega {\mathcal G}(\omega,\vec{k},r,r)$,
where integration gives $\sim |q{\mathcal H}|h_1v_F$ and the resulting gap as in eq.(\ref{solution-gap2}).  
Substituting eq.(\ref{solution-gap2}) in eq.(\ref{shift}) for $\delta k_F$,
we have
\begin{eqnarray}
T_c &\sim & v_F\delta k_F \sim \frac{G_{int}v_F^3}{R^3}N_{eff},\nonumber\\ 
N_{eff} &\sim & \frac{|q{\mathcal H}|h_1^2}{R^3}dr\sqrt{-g(r)}(\psi^{0}(r)^{\dagger}\sigma^1\psi^{0}(r))^2,
\label{density-of-states}
\end{eqnarray}
where $N_{eff}$ is the effective density of states at the Fermi surface.    

In the next section we calculate the constants $h_1$ and $v_F$ for the different fermion charges $q$, and find
the zero mode wave function $\psi^{0}(r)$. We follow procedure outlined in \cite{Hartman:2010}. We then find the critical
temperature $T_c$ as a function of the charge $q$ and the magnetic field.

\subsection{Solving for the zero modes and finding the critical temperature}\label{section:4.2}

In this section we follow the procedure of \cite{Hartman:2010}. We solve the Dirac equation for the zero mode in black hole
background, that amounts to finding a solution at zero frequencies $\omega=0$ (we use here $\omega\equiv \Omega$)
in the $T=0$ background. We will also set mass $m=0$. 
As was shown in \cite{Hartman:2010}, analytic solution can be found in this case.
The Dirac equation in the magnetic field 
\begin{equation}
 {\mathcal D}_M\psi=0, 
\end{equation}  
where ${\mathcal D}_M=\partial_M +\frac{1}{4}\omega_{abM}\Gamma^{ab}-iqA_M$ is defined in Appendix \ref{appendix:a}.
The Dirac equation can be written explicitly including the spin connection as 
\begin{eqnarray}
&& \hspace{-2cm} \left(-\frac{\sqrt{g_{ii}}}{\sqrt{g_{rr}}}\sigma^1\partial_r + \sqrt{g_{ii}}i\sigma^2m 
-\frac{\sqrt{g_{ii}}}{\sqrt{-g_{tt}}}\sigma^3(\omega+qA_t)
+\frac{\sqrt{g_{ii}}}{\sqrt{-g_{tt}}}\sigma^1
\frac{1}{2}\omega_{\hat{t}\hat{r}t}\right.\nonumber\\
&-&\left.\sigma^1\frac{1}{2}\omega_{\hat{x}\hat{r}x}
-\sigma^1\frac{1}{2}\omega_{\hat{y}\hat{r}y}-\lambda\right)\otimes 1
\left(\begin{array}{c}
\psi_1\\
\psi_2
\end{array}\right)=0,
\end{eqnarray}
where $\lambda\rightarrow \sqrt{2|q{\mathcal H}|l}$ is the eigenvalue for the Landau levels which takes into account
$\{x,y\}$-plane physics in the magnetic field (Appendix A), and $\psi=(\psi_1,\psi_2)^{T}$, and the $\Gamma$
matrices are given by eq.(\ref{matrices}). In the basis eq.(\ref{matrices}) the two components decouple,
therefore below we solve for the first component. 
Substituting the spin connection we have 
\begin{equation}
\left(-\frac{r^2\sqrt{f}}{R^2}\sigma^1\partial_r +\frac{r}{R}i\sigma^2m-\frac{1}{\sqrt{f}}\sigma^3(\omega+qA_t)
-\sigma^1\frac{r\sqrt{f}}{2R^2}(3+\frac{rf'}{2f})-\lambda\right)F=0,
\end{equation}
with $F=(y_1,y_2)$. As in the $AdS_2$, it is convenient to change the basis eq.(\ref{transformation}) 
\begin{equation} 
\left(\begin{array}{c}
\tilde{y_1}\\
\tilde{y_2}
\end{array}
\right) = \left(
\begin{array}{cc}
1 & -i \\
-i &  1
\end{array} \right)
\left(\begin{array}{c}
 y_1 \\
 y_2
\end{array}
\right),
\end{equation}
that simplifies the second order differential equation for one component.
The Dirac equation is given
\begin{equation}
\left(-\frac{r^2\sqrt{f}}{R^2}\sigma^1\partial_r +\frac{r}{R}i\sigma^3m+\frac{1}{\sqrt{f}}\sigma^2(\omega+qA_t)
-\sigma^1\frac{r\sqrt{f}}{2R^2}(3+\frac{rf'}{2f})-\lambda\right)\tilde{F}=0,
\end{equation}
with $\tilde{F}=(\tilde{y}_1,\tilde{y}_2)^{T}$. 

We introduce dimensionless variables with the goal to scale away the $AdS_4$ radius $R$ and
the horizon radius $r_0$
\begin{eqnarray}
&& r\rightarrow r_0 r,\;\; m\rightarrow \frac{m}{R},\;\;
r_{*}\rightarrow r_0 r_{*},\;\; r_{**}\rightarrow r_0 r_{**}
\nonumber\\
&& M\rightarrow r_0^3 M,\;\; Q\rightarrow r_0^2 Q,\;\;, H\rightarrow r_0^2H
\end{eqnarray}  
and 
\begin{eqnarray}
&& (t,\vec{x})\rightarrow \frac{R^2}{r_0}(t,\vec{x}),\;\;
A_{M}\rightarrow \frac{r_0}{R^2}A_M,\;\; \omega\rightarrow \frac{r_0}{R^2}\omega,\;\;
\nonumber\\
&& \lambda\rightarrow \frac{r_0}{R^2}\lambda,\;\; T\rightarrow \frac{r_0}{R^2}T,
\nonumber\\
&& ds^2\rightarrow R^2 ds^2.
\label{dimensionless1}
\end{eqnarray}
In the new variables we have 
\begin{eqnarray}
T&=&\frac{3}{4\pi}(1-r_{**}^4),\;\;
f=1+\frac{3r_{**}^4}{r^4}-\frac{1+3r_{**}^4}{r^3},
\nonumber\\
A_t&=&\mu(1-\frac{1}{r}),\;\;
\mu=\sqrt{3}g_Fr_{*}^2,
\label{dimensionless2}
\end{eqnarray} 
and the metric is given by
\begin{equation}
ds^2=r^2(-fdt^2+d\vec{x}^2)+\frac{1}{r^2}\frac{dr^2}{f},
\end{equation}
with the horizon at $r=1$, and the conformal boundary at $r\rightarrow \infty$; the red shift factor is
\begin{equation}
f=1+\frac{3r_{**}^4}{r^4}-\frac{1+3r_{**}^4}{r^3}.
\end{equation}
The Dirac equation is given by
\begin{equation}
\left(-r^2\sqrt{f}\sigma^1\partial_r + ri\sigma^3m+\frac{1}{\sqrt{f}}\sigma^2(\omega+qA_t)
-\sigma^1\frac{r\sqrt{f}}{2}(3+\frac{rf'}{2f})-\lambda\right)\tilde{F}=0.
\end{equation}
We set $m=0$. Then                        
we get the following second order equations for each component
\begin{eqnarray}
 && \hspace{-2cm}\left(r^4f\partial_r^2+(5r^3f+r^4f')\partial_r+\frac{15}{4}r^2f+2r^3f'+\frac{r^4f''}{4}\right.\nonumber\\
&+&\left.\frac{1}{f}((\omega+qA_t)\pm \frac{ir^2f'}{4})^2\mp r^2iqA'_t-\lambda^2
\right) \tilde{y}_{1;2}=0,
\end{eqnarray}
with $A_t=\mu(1-\frac{1}{r})$, 
and the upper/lower sign is for $\tilde{y}_1/\tilde{y}_2$.
At $T=0$, from eq.(\ref{dimensionless2}) we have $r_{**}=1$,
and the red shift factor develops the double zero near the horizon,
\begin{equation}
f= \frac{(r-1)^2(r^2+2r+3)}{r^4}.
\end{equation}
Due to this fact, the metric near horizon reduces to $AdS_2\times R^2$, and calculations in $AdS_4$
are possible to do analytically at small frequencies \cite{Faulkner1:2009},\cite{Hartman:2010}. We will utilize that below.

We introduce a new radial variable $z$,
\begin{equation}
r=\frac{1}{1-z}, 
\end{equation}
then the second order differential equation is given by
\begin{eqnarray}
&&\hspace{-3cm} \left(f\partial^2_z+(\frac{3f}{1-z}+f')\partial_z+\frac{15f}{4(1-z)^2}
+\frac{3f'}{2(1-z)}+\frac{f''}{4}\right.\nonumber\\
&+&\left.\frac{1}{f}((\omega+qA_t)\pm \frac{if'}{4})^2
\mp iqA'_t-\lambda^2
\right)\tilde{y}_{1;2}=0, 
\label{dirac-spinconnection}
\end{eqnarray}
with 
\begin{eqnarray}
f &=& 3z^2(z-z_0)(z-\bar{z}_0),\;\; z_0=\frac{1}{3}(4+i\sqrt{2}),
\nonumber\\
A_t &=& \mu z,\;\; \mu=\sqrt{3}g_F r_{*}^2, 
\label{parameters}
\end{eqnarray}
and horizon is at $z=0$, the conformal boundary is at $z=1$.
Due to the double zero of the red shift factor near the horizon, the second order differential equation
eq.(\ref{dirac-spinconnection}) can be solved analytically at $\omega=0$ \cite{Hartman:2010}.

We put for completness the second order equations for the components when spin connection has been eliminated
using transformation eq.(\ref{rescale})
\begin{equation}
\psi=(r^3\sqrt{f})^{-1/2}\Phi,
\label{rescale2}
\end{equation}
where equations for $\psi$ contain the spin connection, and for $\Phi$ do not. The second order equations
in dimensionless variables without spin connection read
\begin{equation}
\left(r^4f\partial_r^2+(2r^3f+\frac{r^4f'}{2})\partial_r+
\frac{r^4f'^2}{16f}+\frac{1}{f}(\omega+qA_t\pm\frac{ir^2f'}{4})^2
\mp r^2iqA'_t-\lambda^2\right)\tilde{y}_{1;2}=0,
\end{equation}
and we use the same notations for the components for $\Phi=(\tilde{y}_1,\tilde{y}_2)^{T}$ as for $\psi$;
again the upper/lower sign is for $\tilde{y}_1/\tilde{y}_2$.
Using the radial coordinate $z$, these equations are written as
\begin{equation}
\left(f\partial_z^2+\frac{f'}{2}\partial_z+\frac{f'^{2}}{16f}
+\frac{1}{f}(\omega+qA_t\pm\frac{if'}{4})^2\mp iqA'_t-\lambda^2
\right)\tilde{y}_{1;2}=0.  
\end{equation}
These equations look simpler than the corresponding equations with spin connection. However,
they are not readily recognized by MAPLE program, and we will not use them. Writing the transformation
which removes the spin connection explicitly as
\begin{equation}
\psi=\left(\frac{z}{(1-z)^3}\sqrt{3(z-z_0)(z-\bar{z}_0)}\right)^{-1/2}. 
\end{equation}
gives an idea about the prefactor that we should expect to get in the solution. As mentioned, these equations 
are not recognized my MAPLE. We therefore proceed with eq.(\ref{dirac-spinconnection}) which contains spin connection.

Near the horizon, $z=0$, we have $f=6z^2$ and
\begin{equation}
6z^2\tilde{y}''+12z\tilde{y}'+(\frac{3}{2}+\frac{(q\mu)^2}{6}-\lambda^2)\tilde{y}=0 
\end{equation}
and the same for $\tilde{z}$, giving the behavior near horizon
\begin{equation}
\tilde{y}_1\sim\tilde{y}_2\sim z^{-\frac{1}{2}\pm \nu},\;\; \nu=\frac{1}{6}\sqrt{6\lambda^2-(q\mu)^2}, 
\end{equation}
with $\mu=\sqrt{3}g_Fr_{*}^2$.
We will be interested in this scaling exponent given at the Fermi momentum, 
\begin{equation}
\nu\rightarrow \nu_{k_F}=\frac{1}{6}\sqrt{6k_F^2-3q^2g_F^2r_{*}^4}, 
\end{equation}
which is the conformal dimension found in eq.(\ref{conformal}). Note that $r_{**}=1$ at $T=0$.

Putting $\omega=0$ eqs.(\ref{dirac-spinconnection},\ref{parameters}) and using MAPLE,
we find the analytic solution for the zero mode \cite{Hartman:2010} (see also \cite{GradsteinRyzhik}). 
The first solution with regular behavior $z^{-\frac{1}{2}+\nu}$ at the horizon, $z\sim 0$, is given by
\begin{eqnarray}
\tilde{y}_{1;2}^{0} &=& N_{1;2}
(z-1)^{\frac{3}{2}}z^{-\frac{1}{2}+\nu_{\lambda}}(z-\bar{z}_0)^{-\frac{1}{2}-\nu_{\lambda}}
\left(\frac{z-z_0}{z-\bar{z}_0}\right)^{\frac{1}{4}(-1\mp \sqrt{2}q\mu/z_0)},\nonumber\\
&\times&  {}_2F_1\left(\frac{1}{2}+\nu_{\lambda}\mp\frac{\sqrt{2}}{3}q\mu,\nu_{\lambda}\pm i\frac{q\mu}{6},
1+2\nu_{\lambda},\frac{2i\sqrt{2}z}{3 z_0(z-\bar{z}_0)}\right),  
\label{solution5}
\end{eqnarray}
where ${}_2F_1$ is the hypergeometric function, $N_1, N_2$ are normalizations defined later,
and upper/lower sign is for $\tilde{y}_1/\tilde{y}_2$; 
the role of momentum is played by $\lambda\rightarrow \sqrt{2|q{\mathcal H}l|}$.
The second solution, with behavior $z^{-\frac{1}{2}-\nu}$ at the horizon, 
is obtained by replacing $\nu_{\lambda}\rightarrow -\nu_{\lambda}$ in eq.(\ref{solution5})
\begin{eqnarray}
\tilde{\eta}_{1;2}^{0}= \tilde{N}_{1;2}\left(\frac{\tilde{y}_{1;2}^0}{N_{1;2}}\;\; 
{\rm with}\;\; \nu_\lambda\rightarrow -\nu_{\lambda}\right),
\label{solution6}
\end{eqnarray}  
and it will be required to have a regular behavior at $z\sim 0$ for small frequencies.
Since normalization factors are constants, we find their relative weight by substituting solutions
back into first order differential equations eq.(\ref{dirac-spinconnection},\ref{parameters}) at $z\sim 0$,
\begin{equation}
\frac{N_1}{N_2}=-\frac{6i\nu_{\lambda}+q\mu}{\sqrt{6}\lambda}\left(\frac{z_0}{\bar{z}_0}\right)^{q\mu/\sqrt{2}z_0},\;\;
\frac{\tilde{N}_1}{\tilde{N}_2}=\frac{6i\nu_{\lambda}-q\mu}{\sqrt{6}\lambda}\left(\frac{z_0}{\bar{z}_0}\right)^{q\mu/\sqrt{2}z_0}. 
\label{normalization}
\end{equation}
The same relations are obtained when calculations are done for any $z$.
The zero mode equals to $\psi^0=\frac{1}{2}(\tilde{y}_1+i\tilde{y}_2,\tilde{y}_2+i\tilde{y}_1)$ with 
$\tilde{y}_{1;2}=\tilde{y}_{1;2}^{0}+\tilde{\eta}_{1;2}^{0}$.

\begin{figure}[h!]
\begin{center}
\includegraphics[width=10cm]{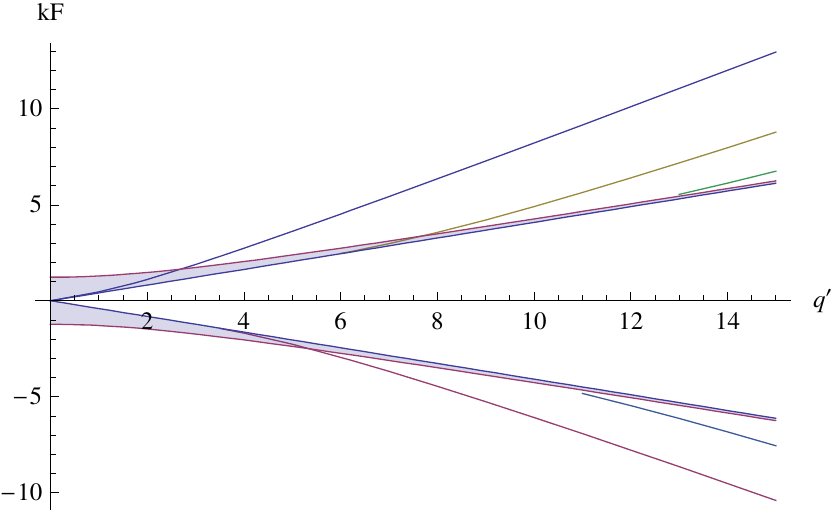}
\caption{Fermi momentum $k_F$ vs. charge of the fermion field $q'=\sqrt{3}q$. We choose $r_{*}=1$, $g_{F}=1$, therefore
$\mu=\sqrt{3}$. The inner (closer to x-axis) line of the filled/shaded
area is $\nu_{k_F}=0$ and the outer line is $\nu_{k_F}=\frac{1}{2}$, so that the shaded region corresponds to 
$0<\nu_{k_F}<\frac{1}{2}$. At a given $q$ there are multiple Fermi surfaces. From left to right are the first, second
etc. Fermi surfaces. They dissapear at $\nu_{k_F}=0$. Positive and negative $k_F$ correspond to Fermi surfaces
in $G_1$ and $G_2$ Green functions respectively. At $q=1$, which on the plot is $q'\sqrt{3}$ $k_F\approx 0.92$
in agreement with \cite{Vegh:2009}. The first Fermi surface hits the border-line between a Fermi and non-Fermi liquids
$\nu=\frac{1}{2}$ at $q'=2.71$.}
\label{plot-kf2}
\end{center}
\end{figure}

To obtain the Fermi momentum, we can follow \cite{Faulkner1:2009} and search for a normalizable solution of the Dirac equation
with the certain boundary conditions, which is equivalent of looking for a bound state in a zero-energy Schrodinger equation.
We use an alternative way suggested in \cite{Hartman:2010}, that also uses that the solution should be regular at the horizon and 
obey certain falloff conditions near the boundary of $AdS_4$. To construct $\psi^{bdy}$, we required the boundary condition
$a=0$ on the fluctuating mode, i.e. at the conformal boundary $z\rightarrow 1$
\begin{equation}
\psi^{0}=\frac{1}{2}\left(
\begin{array}{c}
\tilde{y}_1+i\tilde{y}_2\\
\tilde{y}_2+i\tilde{y}_1                     
\end{array}\right) \sim (1-z)^{3/2}
\left(
\begin{array}{c}
1\\
0                     
\end{array}\right) +\ldots.
\end{equation}
Therefore the equation for the Fermi momentum $k_F$ is
\begin{equation}
\lim_{z\rightarrow 1}(z-1)^{-3/2}(\tilde{y}_2+i\tilde{y}_1)=0. 
\label{kf}
\end{equation}
Using the zero mode solution eq.(\ref{solution5}) in the eq.(\ref{kf}), we have
\begin{equation}
\frac{{}_2F_1(1+\nu_{k_F}+\frac{iq\mu}{6}),\frac{1}{2}+\nu_{k_F}-\frac{\sqrt{2}q\mu}{3},1+2\nu_{k_F},\frac{2}{3}(1-i\sqrt{2})}
{{}_2F_1(\nu_{k_F}+\frac{iq\mu}{6},\frac{1}{2}+\nu_{k_F}-\frac{\sqrt{2}q\mu}{3},1+2\nu_{k_F},\frac{2}{3}(1-i\sqrt{2}))}=
\frac{6\nu_{k_F}-iq\mu}{k_F(-2i+\sqrt{2})},
\label{kf2} 
\end{equation}
with $\nu_{k_F}=\frac{1}{6}\sqrt{6k_{F}^2-(q\mu)^2}$.
We solve the equation for the Fermi surface numerically, using MATHEMATICA to evaluate the hypergeometric functions.
The solutions of eq.(\ref{kf2}) are depicted in Fig.(\ref{plot-kf2}). There are multiple Fermi surfaces for a given $q$.
Following \cite{Hartman:2010}, the largest $|k_F|$ is called the first Fermi surface, the next $|k_F|$ the second Fermi surface,
and so on. For all further plots, we choose $r_{*}=1$ and $g_F=1$; therefore $\mu=\sqrt{3}$. We recover a result
of the numerical solution of the Dirac equation \cite{Vegh:2009}: for $q=1$ which is $\mu q=\sqrt{3}$ we have
$k_F=0.9185$. In Fig.(\ref{plot-kf2}), positive and negative $k_F$ correspond to the Fermi surfaces in the Green functions
$G_1$ and $G_2$. The relation between two components when $m=0$ is $G_2(\omega,k)=-\frac{1}{G_1(\omega,k)}$ \cite{Vegh:2009}, 
therefore Fig.(\ref{plot-kf2}) is not symmetric with respect to $k_F=0$ axis.

\begin{figure}[h!]
\begin{center}
\includegraphics[width=10cm]{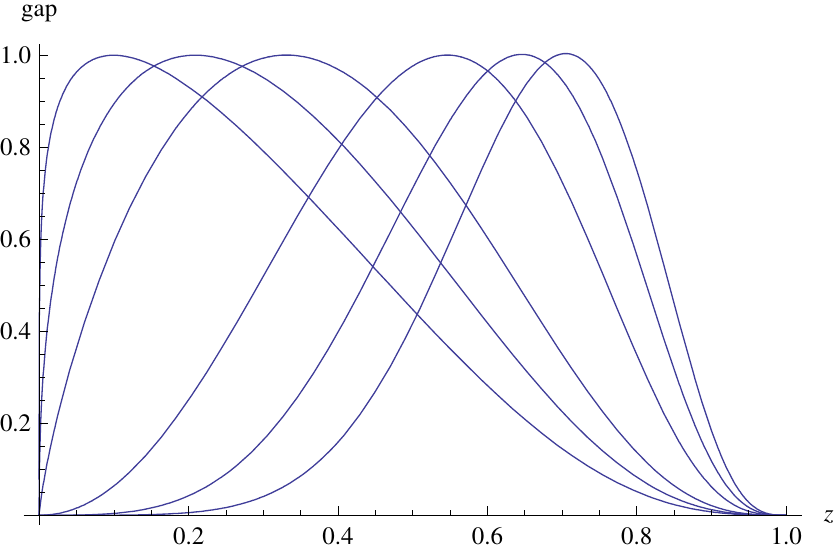}
\caption{Wavefunction of a pairing mode $\Delta^0=\psi^{0\dagger}\sigma^1\psi^{0}$ 
as a function of the radial coordinate $z$, 
with the horizon at $z=0$ and the boundary at $z=1$, for different
values of the charge $q'=\sqrt{3}q$ for the first Fermi surface. We set $r_{*}=1$, $g_F=1$. From left to right
the values of the charge are $q'=\{3,3.4,4,6,8,10\}$. The curves are normalized to have their maxima all the same.
At small charge, non-Fermi liquid, the wave function is supported near the horizon. At large charge, Fermi liquid,
the wave function is supported away from the horizon.}
\label{plot-delta}
\end{center}
\end{figure}

We substitute the Fermi momentum into the zero mode solution eq.(\ref{solution5}) and get the radial profile
for the pairing gap function $\Delta^{0}$ given by eq.(\ref{gap}). We plot $\Delta^{0}(z)$ for different charges,
Fig.(\ref{plot-delta}). The curves are normalized to have the same maxima. Charges are increased from left to right.
For large charge, when $\nu>\frac{1}{2}$, the zero modes are supported away from the horizon, while at smaller charge,
when $\nu\rightarrow \frac{1}{2}$, the zero mode functions are supported near the horizon. The same tendency was first
observed for the Cooper pairing \cite{Hartman:2010,Faulkner2:2009}. This means that for non-Fermi liquids, at $\nu<\frac{1}{2}$,
the physics of the Fermi surface is captured by the near horizon ${\rm AdS}_2\times {\rm R}^2$ region.

We express the boundary Green function through the zero mode solutions. To do that we should obtain the solution at small
but nonzero frequency and expand the Green function to the leading order in $\omega$.
We follow the matching procedure of \cite{Hartman:2010} between the solution in the ``near'' (to horizon) region 
(``inner'' in terminology of \cite{Faulkner1:2009}), $z\ll 1$,
and in the ``far'' (from horizon) region (``outer'' in terminology of \cite{Faulkner1:2009}), 
$z\gg \omega$. Matching the two solutions gives a solution on the full spacetime as long as the near and far regions overlap,
$\omega\ll 1$.

Near the horizon, $z\ll 1$ $f(z)=6z^2$, and
the second order wave equation,
eq.(\ref{dirac-spinconnection}), becomes
\begin{equation}
\left( z^2\partial^2_z+2z\partial_z+\frac{1}{36}\left((q\mu+\frac{\omega}{z})^2\pm 6i\frac{\omega}{z}+9-6\lambda^2\right)
\right) \tilde{y}_{1;2} = 0. 
\end{equation} 
Using MATHEMATICA, we obtain the following solutions
\begin{equation}
\tilde{y}_{1;2}^{near}=C_{1;2}
z^{-\frac{1}{2}-\nu_{\lambda}}{\rm e}^{-\frac{i\omega}{6z}}
{}_1F_1\left(\frac{1}{2}\mp \frac{1}{2}+\nu_{\lambda}+\frac{iq\mu}{6},1+2\nu_{\lambda},\frac{i\omega}{3z}
\right) +   D_{1;2} (\nu_{\lambda} \rightarrow -\nu_{\lambda}),
\end{equation}
with upper/lower sign is for $\tilde{y}_1/\tilde{y}_2$.
Requiring that the solution is ingoing at the horizon $z\sim 0$, $\sim {\rm e}^{+i\omega/6z}$, fixes
the ratio
\begin{equation}
\frac{C_1}{D_1}\sim \frac{C_2}{D_2}\sim
\frac{\Gamma(-2\nu_{\lambda})\Gamma(1+\nu_{\lambda}-\frac{iq\mu}{6})}
{\Gamma(2\nu_{\lambda})\Gamma(1-\nu_{\lambda}-\frac{iq\mu}{6})} (-i\omega)^{2\nu_{\lambda}}. 
\label{coefficients}
\end{equation}
The near horizon solution is in the matching region $z\gg \omega$,
\begin{equation}
\tilde{y}_{1;2}^{near}=A_{1;2}z^{-\frac{1}{2}-\nu_{\lambda}}+B_{1;2}z^{-\frac{1}{2}+\nu_{\lambda}}, 
\label{matching1}
\end{equation}
that corresponds to the $AdS_2$ boundary, given in eq.(\ref{ads2-solutions}). 
From eqs.(\ref{matching1},\ref{coefficients}) we have
\begin{equation}
G^{IR}\sim\frac{B_1}{A_1}\sim \frac{B_2}{A_2}\sim
\frac{\Gamma(-2\nu_{\lambda})\Gamma(1+\nu_{\lambda}-\frac{iq\mu}{6})}
{\Gamma(2\nu_{\lambda})\Gamma(1-\nu_{\lambda}-\frac{iq\mu}{6})} (-i\omega)^{2\nu_{\lambda}}, 
\label{irg}
\end{equation} 
since by definition the ratio $B_1/A_1$ gives the 
retarded Green function in the 
IR CFT living on the boundary of ${\rm AdS}_2$. 
We calculated it in the Appendix \ref{appendix:c}, eq.(\ref{green-result}).

\begin{figure}[h!]
\begin{center}
\includegraphics[width=10cm]{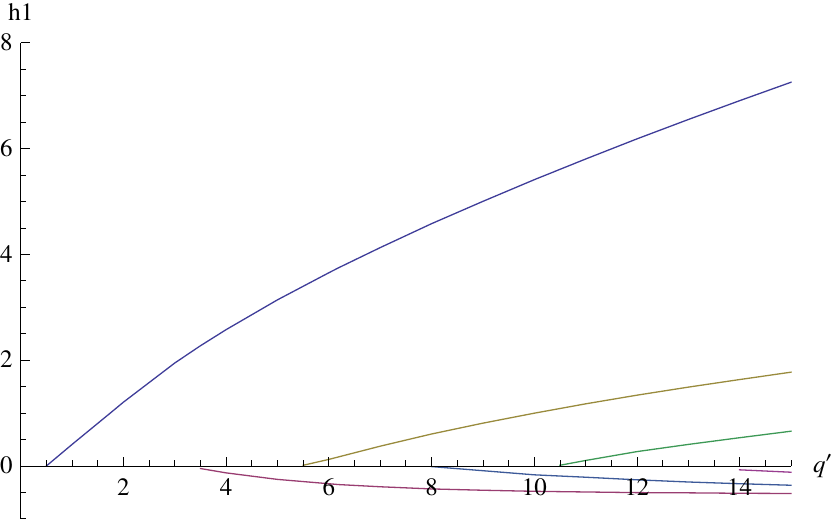}
\caption{Constant $h_1$, reflecting the UV physics of the $AdS_4$ bulk, vs. charge $q'=\sqrt{3}q$.
%eq.(\ref{constant4}).
It vanishes at $\nu_{k_F}=0$. The multiple lines are for various Fermi surfaces, in ascending order with the first fermi surface
on the left. Note, $h_1$ has the same sign as $k_F$. As above, positive and negative $k_F$ correspond to Fermi surfaces
in the Green functions $G_1$ and $G_2$ respectively.}
\label{plot-h1}
\end{center}
\end{figure}

In the asymptotic far region $z\gg\omega$, the second order wave equation is solved perturbatively in $\omega$,
\begin{equation}
\tilde{y}_{1;2}^{far}=\tilde{y}_{1;2}^{(0)}+\omega\tilde{y}_{1;2}^{(1)}, 
\end{equation}
where $\tilde{y}_{1;2}^{(0)}$ includes the zero modes found in before,
$\tilde{y}^{(0)}_{1;2}=\tilde{y}^{0}_{1;2}+\tilde{\eta}^{0}_{1;2}$. 
Expanding $\tilde{y}^{(0)}$ in the matching region, $z\ll 1$,
\begin{eqnarray}
\tilde{y}^{far}_{1;2}&=&\tilde{N}_{1,2}S_{1;2}(\nu)z^{-\frac{1}{2}-\nu}
+N_{1;2}S_{1;2}(-\nu)z^{-\frac{1}{2}+\nu} +O(\omega)\nonumber\\
S_{1;2}(\nu)&=& (-1)^{3/2}(-\bar{z}_0)^{-\frac{1}{2}+\nu_{\lambda}}
\left(\frac{z_0}{\bar{z}_0}\right)^{-\frac{1}{4}\mp\frac{\sqrt{2}q\mu}{4z_0}}. 
\label{matching2}
\end{eqnarray}
Comparing the near solution eq.(\ref{matching1}) and the far solution eq.(\ref{matching2})
in the matching region $\omega\ll z\ll 1$, we get
\begin{equation}
\frac{\tilde{N}_1}{N_1}=(-\bar{z}_0)^{-2\nu_{\lambda}}G^{IR}, 
\label{normalization2}
\end{equation}
that determines the relative contribution of $\tilde{y}^{0}$ and $\tilde{eta}^{0}$. We also have relations
for $N_2$ and $\tilde{N}_2$ given in terms of $N_1$ and $\tilde{N}_1$, eq.(\ref{normalization}).
Comparing eq.(\ref{irg}) and eq.(\ref{normalization2}), $\omega$ and $z$ scale with the same power in the solution
$\tilde{y}$ around the horizon. From eq.(\ref{normalization2}) follows that
\begin{equation}
\tilde{\eta}_{1;2}^{0}\sim G^{IR}\sim G^{IR}\sim\omega^{2\nu_{\lambda}}. 
\label{solution-eta}
\end{equation}
The first order correction $\tilde{y}_{1;2}^{(1)}$ satisfies an inhomogeneous second order wave equation
with $\tilde{y}_{1;2}^{(0)}$ as the source. To calculate the retarded boundary Green function, we need
only the leading asymptotic behavior near the boundary $z\rightarrow 1$. The asymptotic behavior can be found by integrating
the Dirac equation (\ref{dirac-spinconnection}) as in Appendix C of \cite{Faulkner1:2009} with the result \cite{Hartman:2010}
\begin{equation}
\tilde{y}_{2}^{(1)}+i\tilde{y}_{1}^{(1)}=2i(1-z)^3 \frac{\int_{0}^{1} dz \sqrt{g/g_{tt}}
(|\tilde{y}_1^{0}|^2+|\tilde{y}_2^0|^2)}{\tilde{y}_{1}^{0\;*}-i\tilde{y}_{2}^{0\;*}}.
\label{correction2} 
\end{equation}
Note, that from eq.(\ref{correction2}) and eq.(\ref{solution-eta}), only $\tilde{\eta}$ depends on the frequency,
while all other wavefunctions are independent of $\omega$.

\begin{figure}[h!]
\begin{center}
\includegraphics[width=10cm]{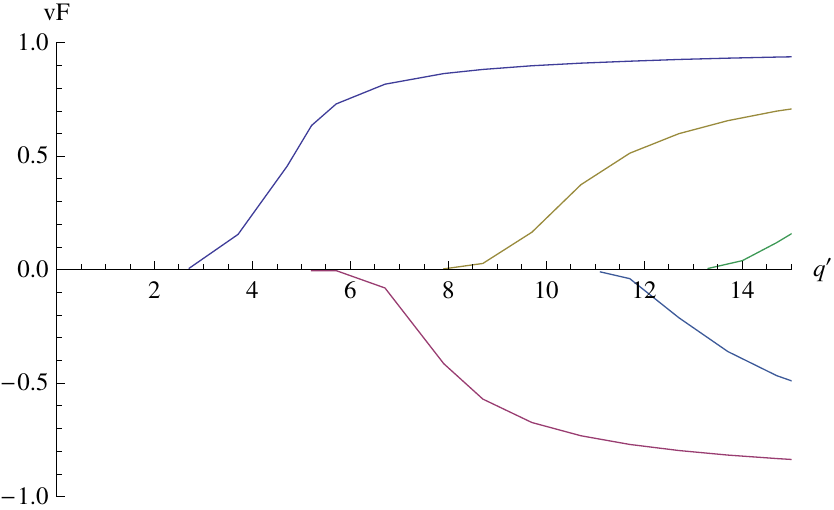}
\caption{Constant $v_F$, reflecting the UV physics of the $AdS_4$ bulk, vs. charge $q'=\sqrt{3}q$.
%eq.(\ref{constant5}).
It vanishes at $\nu_{k_F}=\frac{1}{2}$. For the first Fermi surface it happens ($\nu_{k_F}=\frac{1}{2}$) at $q'=2.71$.
%(compare with Fig.(\ref{plot-kf2})). 
The multiple lines are for various Fermi surfaces, in ascending order with the first fermi surface
on the left. Note, $v_F$ has the same sign as $k_F$. As above, positive and negative $k_F$ correspond to Fermi surfaces
in the Green functions $G_1$ and $G_2$, respectively.}
\label{plot-vf}
\end{center}
\end{figure}

The Green function of the dual field theory defined on the boundary of ${\rm AdS}_4$, as introduced
in eq.(\ref{definition-solution}), is
\begin{equation}
G=\lim_{z\rightarrow 1}\frac{\tilde{y}_1+i\tilde{y}_2}{\tilde{y}_2+i\tilde{y}_1}. 
\end{equation}
Expanding in small frequency $\omega$,
\begin{equation}
G=\lim_{z\rightarrow 1}\frac{\tilde{y}_{1}^{0}+i\tilde{y}_{2}^{0}+\tilde{\eta}_{1}^{0}+i\tilde{\eta}_{2}^{0}}
{\tilde{y}_{2}^{0}+i\tilde{y}_{1}^{0}+\tilde{\eta}_{2}^{0}+i\tilde{\eta}_{1}^{0}
+\omega(\tilde{y}_2^{(1)}+i\tilde{y}_1^{(1)})+O(\omega^2)}, 
\end{equation}
where the zero mode wavefunctions $\tilde{y}_{1;2}^{0}$, $\tilde{\eta}_{1;2}^{0}$ are defined in   
eq.(\ref{ads2-solutions}) with normalization given in eq.(\ref{normalization},\ref{normalization2})
and asymptotic behavoir of the last term in denominator given in eq.(\ref{correction2}). Near
the Fermi surface, $k_{\perp}=k-k_F$, and at $T=0$ the Green function is written \cite{Faulkner1:2009}
\begin{equation}
G=\frac{(-h_1v_F)}{\omega-v_Fk_{\perp}-h_2v_F{\rm e}^{i\theta-i\pi\nu}\omega^{2\nu_{k_F}}}, 
\end{equation}
which was used in our calculations for $T_c$, eq.(\ref{green-function}). Note that all quantities here   
 In the above formula, the last term in the denominator
comes from $\tilde{\eta}_{1;2}^{0}$ and includes $G^{IR}\sim \omega^{2\nu_{k_F}}$, i.e. it is determined
by the $IR\;AdS_2$ physics near the horizon. Other terms are determined by the UV phyics 
of the $AdS_4$ bulk. The constants $h_1$ and $v_F$ are 
\begin{eqnarray}
h_1 &=& \lim_{z\rightarrow 1}\frac{\tilde{y}_{1}^{0}+i\tilde{y}_{2}^{0}}
{\partial_{k}(\tilde{y_{2}^{0}}+i\tilde{y}_{1}^{0})},\label{constant4}\\
v_F &=& \frac{1}{h_1}\left(\int_{0}^{1} dz \sqrt{g/g_{tt}}\psi^{0\dagger}\psi^{0}\right)^{-1}
\lim_{z\rightarrow 1}\frac{|\tilde{y}_{1}^{0}+i\tilde{y}_{2}^{0}|^2}{(1-z)^3}, 
\label{constant5}
\end{eqnarray}
where all wavefunctions are evaluated at $k=k_F$; the zero mode has components 
$\psi^0=(\tilde{y}_{1}^{0}+i\tilde{y}_{2}^{0},\tilde{y}_{2}^{0}+i\tilde{y}_{1}^{0})$;
and the wavefunctions $\tilde{y}_{1;2}^{0}$ are given by analytic expressions, eq.(\ref{solution5}).
The constants $h_2, \theta$ can be also obtained from this expression. The constants $h_1$
and $v_F$ are dimensionless in eqs.(\ref{constant4}),(\ref{constant5}), i.e. the scaling is 
(from dimensional to dimensionless) $h_1\rightarrow \frac{1}{r_0}h_1$, 
$v_F\rightarrow \frac{R^3}{r_0^3}v_F$. Both constants are real. They are plotted as function
of $q\mu$ in Figs.(\ref{plot-h1}),(\ref{plot-vf}). 
The Fermi velocity vanishes at the horizontal line $v_F=0$ when $\nu_{k_F}=\frac{1}{2}$.
The wavefunction renormalization $h_1$ vanishes when $\nu_{k_F}=0$. The multiple lines in each plot are
for various Fermi surfaces, starting with the first FS at the most left.
Positive and negative $v_F$, $h_1$
correspond to the Fermi surfaces in the Grenn functions $G_1$ and $G_2$ respectively. Both $v_F$ and $h_1$
have the same sign as $k_F$.
The Fermi velocity decreases at small charges and tends to the speed of light at large charges. Geometrically
this means, that as the charge is lowered the zero mode wavefunction is supported near the black hole horizon,
Fig.(\ref{plot-delta}), where the gravitational redshift reduces the local speed of light as compared
to the boundary value. This was observed also in \cite{Hartman:2010,Faulkner1:2009}.

\begin{figure}[h!]
\begin{center}
\includegraphics[width=10cm]{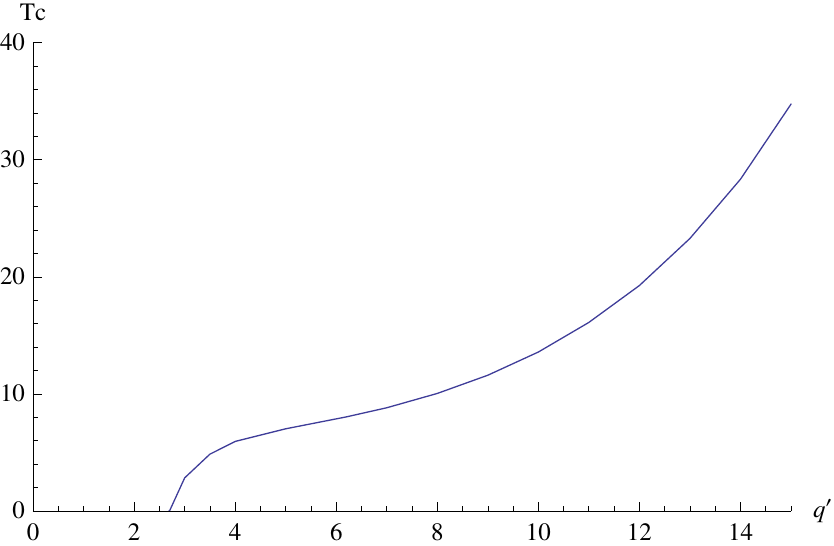}
\caption{Critical temperature $T_c$ vs. the charge $q'=\sqrt{3}q$ for the first Fermi surface.
% eq.(\ref{tc}).
Only parametric dependence is shown, and the $1/\pi^2$ factor is not included.
We set $r_{*}=1$, $g_F=1$. Note that $T_c$ vanishes around $q'=2.71$ which corresponds 
to $\nu_{k_F}=\frac{1}{2}$ for the first Fermi surface. This plot illustrates that pairing is supported
only for $\nu_{k_F}>\frac{1}{2}$ which is the region of Fermi liquids.}
\label{plot-tc}
\end{center}
\end{figure}

The dimensionless critial temperatute is given by
\begin{equation}
T_c=\frac{G_{int}|q{\mathcal H}|h_1^2v_F^3}{\pi^2}\int dz\sqrt{-g}(\psi^{0}(z)^{\dagger}\sigma^1\psi^{0}(z))^2,
\label{tc}
\end{equation}
with transformation to the dimensionless variables given by
\begin{eqnarray}
r\rightarrow r_0 r,\;\;
G_{int}\rightarrow \left(\frac{r_0^2}{R^3}\right)^{-2} G_{int},\;\; 
{\mathcal H}\rightarrow \frac{r_0^2}{R^4}{\mathcal H},\;\;
h_1\rightarrow \frac{r_0}{R^2}h_1,\;\;
\psi^{0}\rightarrow R^{3/2} \psi^{0}. 
\end{eqnarray}   
Using the analytic expression for the zero mode eq.(\ref{solution5}), and the results for $h_1,v_F,k_F$
calculated for a given charge $q$, we plot $T_c$ as function of charge for the first Fermi surface, Fig.(\ref{plot-tc}).
The next Fermi surfaces give smaller contributions and are not depicted on the plot. The critial temperature
vanishes exactly for $\nu=\frac{1}{2}$, which is for the first Fermi surface is 
at $q\mu=\sqrt{3}q=2.71$ and with $k_F=1.65$. Therefore there is no pairing for $\nu\leq \frac{1}{2}$.
This happens due to the fact that
to the leading order the density of states eq.(\ref{density-of-states}) vanishes for the non-Fermi liquids.
This conclusion agrees with our variational calculations, where pairing
occurs only for the Fermi liquids, while non-Fermi liquids do not support pairing. Note, that in principle
fermions from different Fermi surfaces can participate in pairing. 

\begin{figure}[h!]
\begin{center}
\includegraphics[width=10cm]{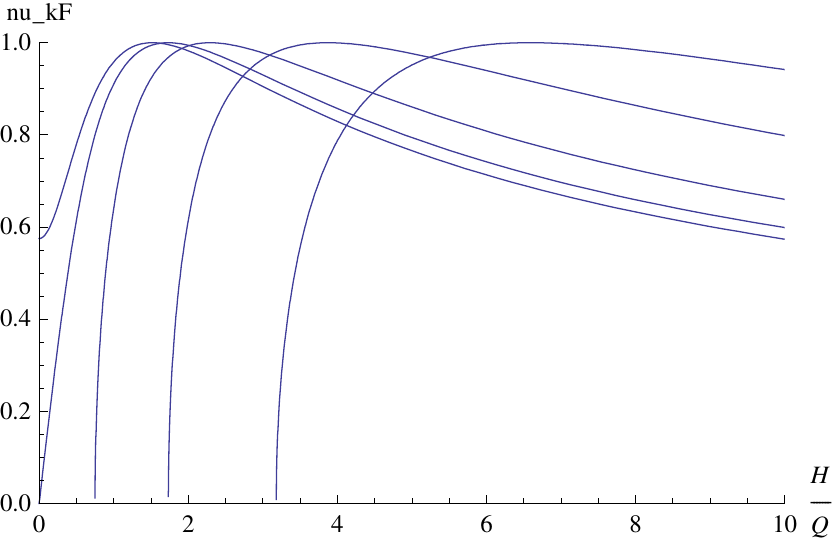}
\includegraphics[width=10cm]{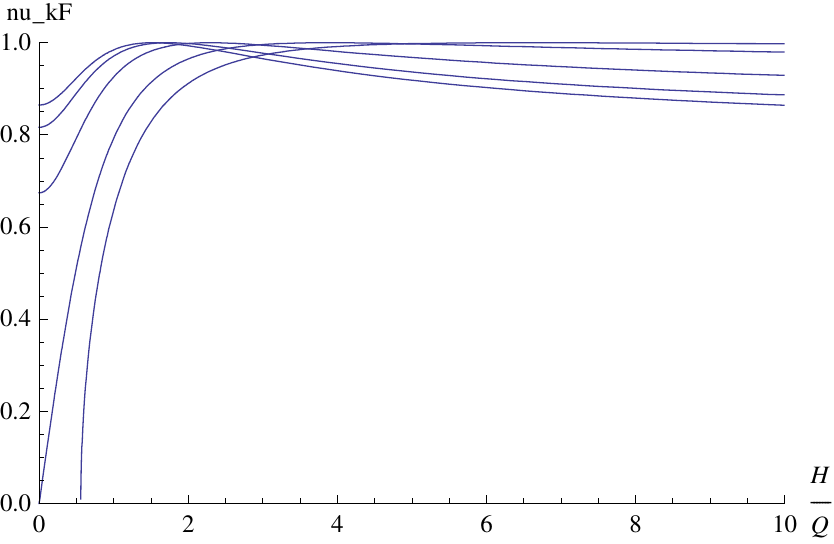}
\caption{Conformal dimension of the fermion operator in the $IR\;CFT$ $\nu_{k_F}$ as a function of the ratio $\frac{H}{Q}$ 
for different values for $q'$.
We used parametrization $\nu\sim\sqrt{a/\sqrt{1+x^2}+b-c/(1+x^2)}$ with $x=\frac{H}{Q}$.
The curves from right to left correspond to increasing charges $q'$, $a=\{0.3,0.5,0.8,1,1.1\}$, while
$b$ and $c=1$ are kept fixed. Top plot is for $b=0$, zero gap,
and bottom plot is for $b=0.5$, nonzero gap. At small and intermediate charges, the Fermi liquid regime
and hence the particle-hole pairing are realized at a threashold value of the magnetic field,
which is consistent with magnetic catalysis in graphene with impurities. 
}
\label{plot-nu}
\end{center}
\end{figure} 

Now we analyze behavior of the boundary field theory with magnetic field.
The conformal dimension of the fermionic operator ${\mathcal O}$ in the $IR\;CFT$ eq.(\ref{conformal2}) is
\begin{equation}
\nu_{k_F}=\sqrt{\frac{1}{6}\left(\frac{k_F^2}{r_{**}^2}+\Delta^2\right)-\frac{q^2g_F^2}{12}\frac{r_{*}^4}{r_{**}^4}}, 
\label{conformal}
\end{equation}
where it is taken in the chiral limit $m=0$ and at the Fermi surface $k_F=q\mu$, and
the dimensionless variables introduced before are used.
We rewrite this expression in terms of $Q$ and $H$, using eq.(\ref{q-and-h}), as
\begin{equation}
\nu_{k_F}=\sqrt{\frac{k_F^2}{\sqrt{12}Q\sqrt{1+(H/Q)^2}}+\frac{\Delta^2}{6}
-\frac{q^2g_F^2}{12(1+(H/Q)^2)}}. 
\end{equation}
Writing $\nu_{k_F}$ as a function of $x=H/Q$, 
$\nu\sim\sqrt{\frac{a}{\sqrt{1+x^2}}+b-\frac{c}{1+x^2}}$, we plot $\nu_{k_F}$ vs. $x$ for different parameters
$a,b,c$, Fig.(\ref{plot-nu}). Multiple curves correspond to increasing values for $a$ starting from the right to left curves,
while $b$ and $c$ are kept constant. This corresponds to increase of the fermion charge $q$.
At small charges the Fermi momentum $k_F$ is small (see Fig.(\ref{plot-kf2})), that corresponds to small $a$ (curves to the right).
All curves show the rapid growth at the beginning, then saturation at some maximum, and fall off which happens at large enough $x$
where we are not interested any more.    
For small charges, the rise to $\nu=\frac{1}{2}$ happens relatively quickly, and means there is a narrow window
of magnetic field for non-Fermi liquids. However, the region for $\frac{1}{2}<\nu<1$, the case of Fermi liquids, corresponds to
a much wider range of the magnetic field.
For small and intermediate charges, regime of the Fermi liquid requires a threashold magnetic field,
when $\nu=\frac{1}{2}$ is reached. This is consistent with the magnetic catalysis in a system with dissipations,
e.g., graphene with impurities,
where there is a nonzero width and the particle-hole gap is induced at a threashold value of the magnetic field \cite{Shovkovy:2d}. 
For large charges, corresponding to curves to the left, there is only a regime of Fermi liquids (no non-Fermi liquids are possible)
and pairing is supported for any magnetic field. Top and bottom pannels are plots for zero 
and nonzero gaps, respectively. 

In this paper we have imagined the existance of an ``experiemntal knob'' which can be used to adjust
the $UV$ scaling dimension of fermionic operator. As noted in \cite{Faulkner:2010}, the most useful knob will likely depend
on the $UV$ geometry into which this $AdS_2$ is embedded. In our case, an external magnetic field ${\mathcal H}$
will allow one to tune the $IR$ scaling dimension and to explore different sectors of the boundary field theory. 

In analogy to the superconducting instability $<\psi\psi>$ of the black hole, we find that the Breitenlohner-Freedman bound
can also be broken in case of $<\bar{\psi}\psi>$ condensate for large enough fermion charges $q$. 
Using the dimensionless variables, the conformal dimension of the bosonic operator in the $IR\;CFT$ 
in the presence of a magnetic field is
\begin{equation}
\nu=\sqrt{\frac{1}{6}\left(\frac{2|q{\mathcal H}|l}{r_{**}^2}+m^2\right)-\frac{q^2g_F^2}{12}\frac{r_{*}^4}{r_{**}^4}+\frac{1}{4}}, 
\end{equation}
which is obtained from eq.(\ref{conformal}) by $k_F^2\rightarrow 2|q{\mathcal H}|l$, the mass gap $\Delta\rightarrow m$
with $m$ being the mass of the bosonic field living in the bulk.
Here the last term $\frac{1}{4}$ distingushes the bosonic case from the fermionic one.
The conformal dimension of the bosonic operator in the $UV\;CFT$, $\Delta_{\phi}$, in the dimensionless variables is given by
\begin{equation}
\Delta_{\phi}=\frac{3}{2}+\sqrt{m^2+\left(\frac{3}{2}\right)^2}. 
\end{equation}
Breitenlochner-Freedman (BF) bound is broken when conformal dimension becomes imaginary.
As in the case with superconductor, there is a parameter range where the conformal dimension in the $IR$ 
(coming from the near $AdS_2$ horizon) is imaginary and the conformal dimension in the $UV$
(coming from the $AdS_4$ bulk) is real. Expressing the mass $m$ through $\Delta_{\phi}$, the condition
breaking the BF bound is
\begin{equation}
q^2g_F^2\frac{r_{*}^4}{r_{**}^4}\geq 2\Delta_{\phi}(\Delta_{\phi}-3)+3+\frac{4R^4|q{\mathcal H}|l}{r_{**}^2}.
\end{equation}
Here we restored the dimension.

\section{Equation of state and transport properties of the boundary field theory at zero magnetic field}\label{section:5}

In this section we consider thermodynamics of the boundary field theory, namely we obtain an equation of state
and find the scaling behavior of the specific heat with temperature. Then we consider transport properties
of the system on the boundary, specifically we calculate the DC conductivity and analyze its scaling behavior. 
We do not specify the boundary theory.
Instead we use the ``dressed'' by the gravity fermion propagators obtained from the $AdS_{3+1}/CFT_{2+1}$ analyzes 
in \cite{Faulkner1:2009}. 
As a result we obtain behavior of systems with properties ranging from Fermi and marginal
liquids to non-Fermi liquids. In particular we reproduce correct temperature scaling  
for the DC conductivity and specific heat in case of the Fermi and marginal liquids.
Since the two-point Green function of the boundary theory has been obtained using 
the $AdS/CFT$ correspondence, it is ``exact'' in terms of gauge coupling corrections. Therefore the lowest
order diagrams on the field theory side should suffice. Of course we lack the knowledge of
the ``dressed'' by the gravity gauge-fermion vertex. For the quantities considered below, the scaling behavior
does not change when vertex corrections are added.

\subsection{Equation of state and specific heat}

An effective potential in the CJT formalism is given by \cite{CJT}
\begin{eqnarray}
\Gamma_{eff} = \frac{1}{2}{\rm Tr}\ln S^{-1} +\frac{1}{2}{\rm Tr}(S_0^{-1}S-1)+\Gamma_2[S], 
\end{eqnarray}
where $S$ is a dressed fermion propagator, $\Gamma_2$ is the sum of all two-particle irreducible (2PI)
diagrams, and trace ${\rm Tr}$ involves integration $\int d^2x$.
The last two terms can be simplified
with the help of Dyson-Schwinger equation, to give
\begin{eqnarray}
\Gamma_{eff} =  \frac{1}{2}{\rm Tr}\ln S^{-1} -\frac{1}{4}{\rm Tr}(\Sigma S), 
\end{eqnarray}
where the self-energy is $\Sigma=S^{-1}-S_0^{-1}$.

We use ``dressed'' by the gravity retarded and advanced fermion propagators \cite{Faulkner1:2009}
\begin{eqnarray}
G_R(\omega,\vec{k}) &=& \frac{(-h_1v_F)}{\omega-v_F k_{\perp}+\Sigma(\omega,k_F)},\nonumber\\
G_A(\omega,\vec{k}) &=& -G_R(\omega,\vec{k})^{*} =-\frac{(-h_1v_F)}{\omega-v_F k_{\perp}+\Sigma^{*}(\omega,k_F)}, 
\label{fermion-propagator}
\end{eqnarray}
where the momentum is counted from the Fermi surface $k_{\perp}=k-k_F$, $h_1$ and $v_F$ are real constants
(we keep the same notations for the constants as in \cite{Faulkner1:2009}).
The self energy $\Sigma=hv_Fc(k_F)\omega^{2\nu_{k_F}}$ contains the real and imaginary parts,
$\Sigma=\Sigma_1+i\Sigma_2$, with imaginary part coming from scattering processes of a fermion in the bulk,
e.g. via pair creation, and scattering into the black hole. The spectral function defined as 
$A(\omega,\vec{k})=\frac{1}{\pi}{\rm Im}G_R(\omega,\vec{k})$ is
\begin{eqnarray}
A(\omega,\vec{k})&=& \frac{1}{\pi}\frac{h_1v_F \Sigma_2(\omega,k_F)}{(\omega-v_Fk_{\perp}+\Sigma_1(\omega,k_F))^2+
\Sigma_2(\omega,k_F)^2}.
\label{spectral}
\end{eqnarray}
Exactly due to inelastic/dissipative processes we are able to calculate transport coefficients, which will be infinite otherwise.
However, the imaginary self energy gives rise to a branch cut in the fermion propagator along ${\rm Im}\omega=0$
in a complex $\omega$ plane \cite{Denef1:2009,Basagoiti:2002,Tomoi}. Therefore in calculation of Matsubara sum
we should take into account contributions from poles and discontinuities along branch cuts \cite{Basagoiti:2002,Tomoi}
\begin{eqnarray}
T\sum_{odd\;m}F(i\omega_m) = \sum_{poles}n(z_i)Res(F,z=z_i)-\sum_{cuts}\int_{-\infty}^{\infty}
\frac{d\zeta}{2\pi i}n(\zeta)Disc\;F, 
\label{contour-integral}
\end{eqnarray}
with analytical continuation $i\omega_m\rightarrow z$, and $n(x)$ is the Fermi distribution function.  
One can use either $n(x)\equiv n(\frac{x}{T})$ or $\tanh(\frac{x}{2T})$ functions with prefactors $(-\frac{1}{2\pi i})$ and
$(-\frac{1}{4\pi i})$ respectively in the contour integral which give the same result for the observables. 
Calculation of Matsubara sums using perturbative expansion in the imaginary part of the self-energy 
has been developed in \cite{Armen}.  

For simplicity we take $(-h_1v_F)\rightarrow 1$ which will not change our results qualitatively.
Using the retarded fermion propagator eq.(\ref{fermion-propagator}), an effective potential is
\begin{eqnarray}
\Gamma_{eff} &\rightarrow & -\frac{1}{4\pi i} \frac{V_2}{T}\int\frac{d^2k}{(2\pi)^2} 
\int_C dz \tanh\frac{z}{2T} \times\nonumber\\
&&\hspace{1.5cm} T\left(\frac{1}{2}\ln\frac{z-v_F k_{\perp}+\Sigma(z,k_F)}{T}
-\frac{1}{4}\frac{\Sigma(z,k_F)}{z-v_F k_{\perp}+\Sigma(z,k_F)}\right),
\end{eqnarray}
where we substituted Matsubara sum by the contour integral. The original contour
$C_0$ going around the poles along imaginary $z$-axis was deformed in the contour $C$ going along the real $z$ axis
and $\Gamma$ being arcs at infinity with vanishing contribution \cite{Basagoiti:2002}.
In case of a real self-energy the result for the contour integration is (see Appendix \ref{appendix:c}
and \cite{Denef1:2009})  
\begin{eqnarray}
\Gamma_{eff} &\rightarrow & \frac{V_2}{T}\int\frac{d^2k}{(2\pi)^2}\sum_{z_{*}} 
\left(\frac{1}{2}T\ln\left( 2\cosh \frac{z_{*}}{2T}\right)
+\frac{1}{4} \Sigma(z_{*})\tanh\frac{z_{*}}{2T}\right),
\end{eqnarray}
where $z_{*}$ are the poles of the retarded propagator, and the sum over all possibloe poles is taken.
As was shown in \cite{Denef1:2009}, when a self-energy and hence poles include imaginary part, 
the following substitution of hyperbolic functions with $\Gamma$ functions should be made \cite{GradsteinRyzhik}
\begin{eqnarray}
 |\Gamma(\frac{1}{2}+iz)|^2 &=& \frac{\pi}{\cosh (\pi z)},\nonumber\\
 |\Gamma(iz)|^2 &=& \frac{\pi}{z \sinh(\pi z)}. 
\label{generalization}
\end{eqnarray}
Using the relation between the effective potential and the pressure, $p=\frac{T}{V_2}\Gamma_{eff}$,
we get an equation of state
\begin{eqnarray}
p = \int\frac{d^2k}{(2\pi)^2}\sum_{z_{*}}\left(-\frac{1}{2} 
T\ln\left(\frac{1}{2\pi}|\Gamma(\frac{iz_{*}}{2\pi T}+\frac{1}{2})|^2\right)
+\frac{1}{4} \frac{\Sigma(z_{*})|\Gamma(\frac{iz_{*}}{2\pi T}+\frac{1}{2})|^2}{\frac{|z_{*}|}
{2\pi T}|\Gamma(\frac{iz_{*}}{2\pi T})|^2}\right),
\label{eq-of-state}
\end{eqnarray}
where summation over complex poles $z_{*}$ is performed.
As in our previous calculations we take only the contribution of the nearest to $\omega=0$ pole 
eqs.(\ref{landau1}),(\ref{landau2}),(\ref{landau3}), and the self-energy $\Sigma(z)\sim z^{2\nu}$.
Near the Fermi surface, the one-loop contribution
dominates over the self-energy term for the Fermi liquids $\nu>\frac{1}{2}$, while
the sel-energy becomes important for the non-Fermi liquids $\nu<\frac{1}{2}$. 

Having calculated the pressure, we can obtain other thermodynamic quantities, e.g. the entropy, 
the specific heat, and the particle number density, respectively,
\begin{eqnarray}
s= \frac{\partial p}{\partial T},\;\; 
c=T\frac{\partial s}{\partial T},\;\;
n=\frac{\partial p}{\partial \mu}
\end{eqnarray}
with $\mu\equiv k_F$.

We find the temperature dependence for the specific heat. The first term in eq.(\ref{eq-of-state})
contributes to the specific heat
\begin{eqnarray}
&& \sim\frac{1}{T^2}\int\frac{d^2k}{(2\pi)^2}{\rm Re}\left(z_{*}^2\Psi^{\prime}(\frac{iz_{*}}{2\pi T}+\frac{1}{2})
+z_{*}^{*\,2}\Psi^{\prime}(-\frac{iz^{*}}{2\pi T}+\frac{1}{2})
\right),\nonumber\\
&& \frac{1}{T^2}\int\frac{d^2k}{(2\pi)^2}{\rm Re}\left(\sim z_{*}T\Psi(\frac{iz_{*}}{2\pi T}+\frac{1}{2});\;
\sim z_{*}^{*}T\Psi(\frac{-iz_{*}^{*}}{2\pi T}+\frac{1}{2})\right)
\end{eqnarray}
where $\Psi^{\prime}(x)=\frac{d\Psi}{dx}=\frac{d^2\ln\Gamma}{dx^2}$.
The second term in eq.(\ref{eq-of-state}) gives to the specific heat
the following contributions
\begin{eqnarray}
\frac{1}{T^2}\int\frac{d^2k}{(2\pi)^2}{\rm Re} \left(
\sim T\Sigma(z_{*})F[\Gamma];\;\sim z_{*}\Sigma(z_{*})F[\Gamma];\; \sim \frac{z_{*}^2\Sigma(z_{*})}{T} F[\Gamma]
\right), 
\end{eqnarray}
where $F[\Gamma]$ denotes a combination of $\Gamma$ functions and its first and second derivatives.
Here momentum integration is performed around the Fermi surface, $d^2 k\rightarrow k_F dk_{\perp}$
with $k_{\perp}=k-k_F$, the poles $z_{*}=\omega_c-i\Gamma$
are given by eqs.(\ref{landau1}),(\ref{landau2}),(\ref{landau3}) for the three cases of interest,
and $\Sigma(z)\sim z^{2\nu}$.

For the Fermi liquid $\nu>\frac{1}{2}$, $z_{\perp}\sim k_{\perp}$ (the real part is dominant).
The first term gives
$\frac{1}{T^2}\int dk_{\perp}z_{*}^2\rightarrow  T$ and the same behavior from the other combination.
In the second term we have $\Sigma\sim k_{\perp}^{2\nu}$. Therefore the second term gives 
$\frac{1}{T^2}\int dk_{\perp}\Sigma(z_{*}) z_{*} \rightarrow T^{2\nu}$
and the same behavior for the other two combinations. 
In eq.(\ref{eq-of-state}) for the pressure, the one-loop term dominates over the self-energy for $\nu>\frac{1}{2}$.
Therefore for Fermi liquid at low temperatures we have
\begin{eqnarray}
c\sim T. 
\end{eqnarray}
This result reproduces correctly the linear temperature behavior of the heat capacity 
known for the Fermi liquids. 

For the non-Fermi liquid $\nu<\frac{1}{2}$, $z_{\perp}\sim k_{\perp}^{\frac{1}{2\nu}}$ 
(for both real and imaginary parts). The first term gives 
$\frac{1}{T^2}\int dk_{\perp} k_{\perp}^{\frac{1}{\nu}}\rightarrow T^{\frac{1}{\nu}-1}$
and $\frac{1}{T^2}\int dk_{\perp} k_{\perp}^{\frac{1}{2\nu}}T\rightarrow T^{\frac{1}{2\nu}}$.
The second term gives 
$\frac{1}{T^2}\int dk_{\perp} \Sigma(z_{*}) T\rightarrow T^{2\nu}$ and subleading behavior for the other
two combinations. For $\nu<\frac{1}{2}$, the self-energy dominates over the one-loop in the pressure.
Therefore for non-Fermi liquid at low temperatures we have
\begin{eqnarray}
c\sim T^{2\nu}. 
\end{eqnarray}
This result for the heat capacity reflects the scaling behavior of the self-energy.

For $\nu=\frac{1}{2}$ all obtained above terms are $\sim T$. Therefore for the marginal liquids we have 
$c\sim T$.

We repeat derivation of equation of state using the spectral function eq.(\ref{spectral}).
Density of states can be written through a spectral function as follows
\begin{eqnarray}
n = T\sum_{m}\int\frac{d^2k}{(2\pi)^2}A(i\omega_m,\vec{k})
 \rightarrow  -\frac{1}{4\pi i} \int\frac{d^2k}{(2\pi)^2}\int_C dz A(z,\vec{k})f(z), 
\end{eqnarray}
where $f(z)=\tanh(\frac{z}{2T})$. One can use also the Fermi distribution function $f(z)=n(z)$
with a prefactor $(-\frac{1}{2\pi i})$, which gives the same result for observables.
The pressure is given by
\begin{eqnarray}
p=\int_{-\infty}^{\mu} d\mu' n, 
\end{eqnarray}
where in our case $\mu\equiv k_F$. 
For simplicity we again take $h_1v_F\rightarrow 1$. We expand the spectral function eq.(\ref{spectral})
with respect to the imaginary part of the self-energy,
which we consider to be small in this calculation \cite{Armen}
\begin{eqnarray}
&& A(z,\vec{k}) \approx 2\pi\delta(z-z_{*})-\Sigma_2(z,k_F){\cal P^{\prime}}\frac{1}{z-z_{*}}, \nonumber\\
&& {\cal P^{\prime}}\frac{1}{z-z_{*}} \equiv \frac{\partial}{\partial z}\left({\cal P}\frac{1}{z-z_{*}}\right),
\end{eqnarray}
where the pole of the propagator $z_{*}$ is a solution of the equation $z-v_Fk_{\perp}-\Sigma_1(z,k_F)=0$,
which does not contain imaginary part of the self energy $\Sigma_2$. 
Substituting this representation in the equation for the pressure, we have
\begin{eqnarray}
p = -\frac{1}{4\pi i} \frac{d^2k}{(2\pi)^2}\int_{-\infty}^{k_F} dk_{F}^{\prime}\int_{-\infty}^{\infty} dz
\left(2\pi \delta(z-z_{*})+\Sigma_2(z){\cal P^{\prime}}\frac{1}{z_{*}-z}\right)f(z). 
\end{eqnarray}
The frequency integral in the first term gives familiar expression for the number density
\begin{eqnarray}
n=\int\frac{d^2k}{(2\pi)^2} f(z_{*}), 
\end{eqnarray}
where usually $f$ is a Fermi distribution function, and the dispersion relation is given by $z_{*}$ (in standard notations
$z_{*} \rightarrow \varepsilon_k$).
Here we have $f(x)=\tanh(\frac{x}{2})$. Therefore integrating over $k_F$ gives 
$\int d k_{F}^{\prime} \tanh\frac{z_{*}}{2}\rightarrow \ln(2\cosh\frac{z_{*}}{2})$ where to the leading order $z_{*}\sim (k-k_F)$.
In the second term we interchange the order of integration in $z$ and $k_F$. Therefore,
$\int_{-\infty}^{k_F}dk_F^{\prime}{\cal P^{\prime}}\frac{1}{z_{*}(k_F^{\prime})-z}\rightarrow - \frac{1}{z_{*}(k_F)-z}$, 
to the leading order there is no $k_F$ dependence in $\Sigma_2(z)\sim z^{2\nu}$. The second integral is
$\frac{1}{2\pi i}\int_{-\infty}^{\infty}dz \Sigma_2(z,k_F)f(z)\frac{1}{z_{*}-z}\rightarrow \Sigma_2(z_{*})f(z_{*})$.
Combining all the terms together we have
\begin{eqnarray}
p & = & \int\frac{d^2k}{(2\pi)^2}\sum_{z_{*}} 
\left(\frac{1}{2}T\ln\left( 2\cosh \frac{z_{*}}{2T}\right)
+\frac{1}{4} \Sigma_2(z_{*})\tanh\frac{z_{*}}{2T}\right),
\end{eqnarray}
where $z_{*}$ is the pole of the fermion propagator without the imaginary part $\Sigma_2$,
and there is a summation over poles.
If we take $z_{*}$ to be the pole of the full propagator, $z_{*}$ becomes imaginary
and generalization of hyperbolic functions to the $\Gamma$ functions is necessary eq.(\ref{generalization}).
We arrive then at eq.(\ref{eq-of-state}) for the pressure of the system.

\subsection{DC conductivity}

We calculate the DC conductivity in the boundary theory using the ``dressed'' by gravity retarded/advanced
fermion propagators eq.(\ref{fermion-propagator}).
To make the calculations complete, we need the ``dressed'' vertex, to satisfy Ward identities.
As was argued in \cite{Faulkner:2010}, the boundary vertex which is obtained from the bulk one
can be approximated by a constant in the low temperature limit. Also, according to \cite{Basagoiti:2002},
the vertex has only singularities of the product of the Green functions. Therefore, dressing the vertex
will not change temperature dependence of the DC conductivity \cite{Basagoiti:2002}.

Using linear response theory, we have
\begin{equation}
 \sigma = -\frac{\partial}{\partial\omega}{\rm Im}\Pi_{AA}(\omega,\vec{k}=0)|_{\omega=0},
\label{conductivity}
\end{equation}
which is a Kubo formula for conductivity. 
Here the polarization operator is given by
\begin{eqnarray}
\Pi_{AA}(i\nu_n,0)=\int\frac{d^2k}{(2\pi)^2}T\sum_{\omega_m}G(i\omega_m+i\nu_n,\vec{k})
\Lambda_A(i\omega_m+i\nu_n,i\omega_m,\vec{k})G(i\omega_m,\vec{k})\Lambda_A^{(0)}(\vec{k}),\nonumber\\ 
\end{eqnarray}
where the fermion frequency is $\omega_m=(2m+1)\pi T$, and boson frequency is $\nu_n=2n\pi T$,
and in the low temperature limit $\Lambda_A(i\omega_m+i\nu_n,i\omega_m,\vec{k})=\Lambda_A^{(0)}(\vec{k})$. 
Usually the most difficult step is to take the Matsubara sum. Here we do it in two ways. First,
analytically continuing in the complex plane $i\omega_m\rightarrow z$ and 
replacing the Matsubara sum by the contour integral with the Fermi distribution function
$n(x)=\frac{1}{{\rm e}^x+1}$
whose poles are at the Matsubara frequencies along the imaginary axis. 
Second, using the spectral representation. In both cases we follow \cite{Basagoiti:2002},
where transport coefficients are calculated with propagators including imaginary parts.

In the first way, we have for the fermion Matsubara sum
\begin{eqnarray}
H(i\nu_n,\vec{k})=T\sum_{\omega_m}G(i\omega_m+i\nu_n,\vec{k})G(i\omega_m,\vec{k})\rightarrow
-\frac{1}{2\pi i} \int_C dz G(z+i\nu_n,\vec{k})G(z,\vec{k})n(z),\nonumber\\  
\label{contour-integral}
\end{eqnarray}
where the contour along the imaginary z-axis can be deformed to contour C which goes go along the two brunch cuts,
${\rm Im}Z=0$ and ${\rm Im}z=-\nu_n$, and the large arcs $\Gamma$ with vanishing contribution \cite{Basagoiti:2002}.
The fermion propagator has a branch cut along ${\rm Im}z=0$ \cite{Tomoi},\cite{Basagoiti:2002}. 
Therefore we can rewrite
\begin{eqnarray}
H(i\nu_n) &=& -\frac{1}{2\pi i}\int_{-\infty}^{\infty}d\zeta n(\zeta)G(i\nu_n +\zeta)(G_R(\zeta)-G_A(\zeta))
\nonumber\\
&-&\frac{1}{2\pi i}\int_{-\infty}^{\infty}d\zeta n(\zeta)G(-i\nu_n +\zeta)(G_R(\zeta)-G_A(\zeta)), 
\end{eqnarray}
where the difference of the retarded and advanced functions in the first bracket is due to the discontinuity
along ${\rm Im}z=0$ and in the second bracket due to the discontinuity along ${\rm Im}z=-\nu_n$.
This contribution corresponds to the second term in eq.(\ref{contour-integral}), 
and there are no pole contributions \cite{Basagoiti:2002}. 
We use the usual prescription for the retarded and advanced Green functions, $G_R=G(\omega+i0^{+})$ and 
$G_A=G(\omega-i0^{+})$. We suppress momentum indices.
Taking $i\nu_n\rightarrow \omega+i0^{+}$, we have
\begin{eqnarray}
H(\omega) &=& -\frac{1}{2\pi i}\int_{-\infty}^{\infty}d\zeta n(\zeta)G_R(\omega+\zeta)(G_R(\zeta)-G_A(\zeta))
\nonumber\\
&-& \frac{1}{2\pi i}\int_{-\infty}^{\infty}d\zeta n(\zeta+\omega)G_A(\omega+\zeta)(G_R(\zeta+\omega)-G_A(\zeta+\omega)),
\end{eqnarray}
where we changed the integration variable in the second integral $\zeta-\omega\rightarrow \zeta$.
In the limit $\omega\rightarrow 0$, the dominant contribution comes from the pair $G_RG_A$, and
it is inversely proportional to the distance between the poles
given by the imaginary part $\Sigma_2$. Combinations $G_RG_R$ and $G_AG_A$ with the poles on one side of real axis
make a much smaller contribution due to cancellation between the residues at the poles. Therefore,
as $\omega\sim 0$, we have
\begin{eqnarray}
H(\omega,\vec{k})\rightarrow -\frac{1}{2\pi i} \int_{-\infty}^{\infty} d\zeta
(n(\zeta+\omega)-n(\zeta))G_R(\zeta+\omega)G_A(\zeta), 
\end{eqnarray}
and
\begin{eqnarray}
{\rm Im}\Pi_{AA}(\omega,0) &=& \frac{1}{2\pi}\int\frac{d^2k}{(2\pi)^2}\Lambda_A^{(0)}(\vec{k})
\int_{-\infty}^{\infty}\frac{d\zeta}{2\pi}(n(\zeta+\omega)-n(\zeta))G_R(\zeta+\omega,\vec{k})\times\nonumber\\
&& \Lambda_A(\zeta+\omega+i0^{+},\zeta-i0^{-},\vec{k})G_A(\zeta,\vec{k}). 
\end{eqnarray}
In small $T$ limit the vertex is a constant. 
We integrate around the Fermi surface, therefore momentum integral is
$\int\frac{d^2k}{(2\pi)^2}\rightarrow \frac{k_Fdk_{\perp}}{(2\pi)^2}$ with $k_{\perp}=k-k_F$.
We exchange the order of integration and do first the momentum integral \cite{Hartman:2010},\cite{Faulkner:2010}.
For $\omega\sim 0$, we have
\begin{eqnarray}
&& \hspace{-2cm}\int_{-\infty}^{\infty} \frac{dk_{\perp}}{2\pi} 
\frac{1}{(\frac{\zeta}{v_F}-k_{\perp}+\Sigma(\zeta,k_F)+i0^{+})
(\frac{\zeta}{v_F}-k_{\perp}+\Sigma^{*}(\zeta,k_F)-i0^{+})}=\nonumber\\
&& \frac{i}{\Sigma(\zeta,k_F)-\Sigma^{*}(\zeta,k_F)}=\frac{1}{2{\rm Im}\Sigma(\zeta,k_F)}.
\end{eqnarray}
Writing $n^{\prime}(\zeta)=-\beta n(\zeta)(1-n(\zeta))$, we have for $\omega\sim 0$
\begin{eqnarray}
\sigma\rightarrow  \Lambda^{(0)\;2}k_Fh_1^2\int_{-\infty}^{\infty}\frac{\beta d\zeta}{2\pi}
\frac{n(\zeta)(1-n(\zeta))}{{\rm Im}\Sigma(\zeta,k_F)},
\label{resultcond} 
\end{eqnarray}
where we did not include constants. Note that we get the same result for conductivity when we use $\tanh\frac{x}{2}$
in the contour integral eq.(\ref{contour-integral}) since $n^{\prime}(x)=-2\tanh^{\prime}(\frac{x}{2})$.

For the Landau Fermi liquid $\Sigma(\omega) \sim \omega^2$ at small $T$ \cite{Landau},\cite{Faulkner:2010}. We get
\begin{eqnarray}
\sigma \sim T^{-2},  
\end{eqnarray}
that means we recover the standard result for the resistivity of the Fermi liquid,
$\rho\sim T^2$.

In our case, $\Sigma (\omega)\sim \omega^{2\nu_{k_F}}$,
\begin{eqnarray}
\sigma \sim T^{-2\nu_{k_F}},  
\end{eqnarray}  
where $\nu_{k_F}$is the IR conformal dimension. This result agrees with the DC conductivity obtained in \cite{Faulkner:2010}.
For the marginal liquid, $\nu_{k_F}=\frac{1}{2}$, we recover the resistivity $\rho\sim T$, which is known
for strange metals. It is interesting that the scaling behavior of the DC conductivity is the same as the single particle
scattering rate. In the gravity calculations it is explained by the fact that the dissipative part of the 
current-current correlator is controlled by the rate of the bulk fermion falling in the horizon, given by the single particle
scattering rate.

To check our calculation, we get the DC conductivity using the spectral representation
\begin{eqnarray}
G(i\omega_m,\vec{k})=\int\frac{dk_0}{2\pi}\frac{A(k_0,\vec{k})}{k_0-i\omega_m}, 
\end{eqnarray}
where the spectral function $A(k_0,\vec{k})$ is given in eq.(\ref{spectral}). For the product of the Green functions
we use the following formula
\begin{eqnarray}
T\sum_{m}\frac{1}{i\omega_m-\omega_1}\frac{1}{i\omega_m+i\nu_n-\omega_2}=
\frac{n(\omega_1)-n(\omega_2)}{i\nu_n+\omega_1-\omega_2}. 
\end{eqnarray}
Taking $i\nu_n\rightarrow \omega+i0^{+}$, the polarization operator is given by
\begin{eqnarray}
\Pi_{AA}(\omega,0) = \int\frac{d^2k}{(2\pi)^2}\frac{d\omega_1}{2\pi}\frac{d\omega_2}{2\pi}
\frac{n(\omega_1)-n(\omega_2)}{\omega+\omega_1-\omega_2}\Lambda_A^{(0)2}
A(\omega_1,k_{\perp})A(\omega_2,k_{\perp}).
\end{eqnarray}
Performing integration over $\omega_2$, we have
\begin{eqnarray}
{\rm Im} \Pi_{AA}(\omega,0) = \int\frac{d^2k}{(2\pi)^2}\frac{d\omega_1}{2\pi}
(n(\omega_1)-n(\omega_2))\Lambda_A^{(0)2}
A(\omega_1,k_{\perp})A(\omega_1+\omega,k_{\perp}). 
\end{eqnarray}
In the limit $\omega\sim 0$, the momentum integration is
\begin{eqnarray}
\int\frac{d^2k}{(2\pi)^2}A^2(\omega_1,k_{\perp})\rightarrow
k_F\int\frac{dk_{\perp}}{2\pi}A^2(\omega_1,k_{\perp})\rightarrow \frac{k_Fh_1^2}{\Sigma_2(\omega_1,k_F)}, 
\end{eqnarray}
with $\Sigma_2={\rm Im}\Sigma$. Therefore the DC conductivity given by eq.(\ref{conductivity}) is
\begin{eqnarray}
\sigma \rightarrow \Lambda_A^{(0)2} k_F h_1^2 \int \frac{\beta d\omega_1}{2\pi}
\frac{n(\omega_1)(1-n(\omega_1))}{{\rm Im}\Sigma(\omega_1,k_F)}
\end{eqnarray}
which is the same as eq.(\ref{resultcond}) obtained by the contour integration.

\section{Discussion}

In this article we studied the particle-hole pairing in the context of the magnetic catalysis. 
The Reissner-Nordstrom charged black hole can carry Fermi surfaces \cite{Hartman:2010}, in a sense that one has 
a Fermi liquid in the bulk: there are Fermi surfaces and free fermions fill all the levels up to the Fermi surface.
It is natural to expect pairing between fermions (particle and hole in this case) when an attractive interaction is
introduced. At the $CFT$ boundary, we indeed get pairing instability for the Fermi liquids.
However, quite surprisingly, there is no pairing realized for the non-Fermi liquids. 
We show that in variational calculations where both terms in the r.h.s. of the gap equation vanish near the Fermi surface.
In the bulk Ginsburg-Landau calculations, we obtain the critical temperature which vanish
exactly at the $IR\;CFT$ conformal dimension $\nu_{k_F}=\frac{1}{2}$, which is the border between Fermi and non-Fermi liquids.
$T_c$ stays zero for $\nu_{k_F}<\frac{1}{2}$.
The same conclusion has been reached for the superconducting
instability in \cite{Hartman:2010}. Probably the reason is that there are no well defined quasiparticles in case of
non-Fermi liquids, in particular the residue for the fermion pole vanishes around the Fermi surface \cite{Faulkner1:2009}. 
Note, that we have well defined particle degrees of freedom in the bulk, while the situation
is different when it is projected to the boundary. As suggested in \cite{Hartman:2010}, the momentum/frequency
dependent four-Fermi interaction may help to realize pairing for the non-Fermi liquids.

Our calculations are different from the corresponding calculations in the field theory in several aspects.
We have imagined the existance of an ``experimental knob''
which can be used to adjust the UV scaling dimension $\Delta$ of an operator, and therefore
an ability to get different $CFT$'s, e.g., describing Fermi and non-Fermi liquids. 
As discussed in \cite{Faulkner1:2009},
the most useful knob will likely depend on the UV geometry into which this $AdS_2$ is embedded.
For example, in our case an external magnetic field will allow one to tune the $IR$ scaling dimension $\nu$.
We obtain a radial profile for the order parameter $\Delta(r)\sim \psi^{0}(r)^{\dagger}\sigma^1 \psi^{0}(r)$
where $\psi^{0}(r)$ are the zero modes with $\omega=0$ and $k=k_F$.
For the Fermi liquids, $\nu_{k_F}>\frac{1}{2}$, $\Delta(r)$ is supported near the $CFT$ boundary,
and for the non-Fermi liquids, $\nu_{k_F}<\frac{1}{2}$, $\Delta(r)$ is supported near the horizon
where the Fermi velocity is considerably smaller than the speed of light $v_F\ll c$.
The radial profile of $\Delta(r)$ with correct fall off is important to insure convergence
of the radial integrals. 

In the presence of the particle-hole condensate $<\bar{\psi}i\Gamma^{\hat{2}}\Gamma^{\hat{5}}\psi>$,
the Fermi momenta corresponding to the bulk fermions with spin up and down are shifted in the opposite
directions. We did not introduce spins in the boundary theory. 
Therefore association with the bulk spins is understood as upper and lower components of the 4-component fermion field.
For a fixed fermion charge we obtain multiple Fermi surfaces. This is the consequence of the $AdS$
geometry effectively behaving as a box potential. It is interesting to understand the physical picture
behind the multiple Fermi surfaces, and if pairing between different Fermi surfaces brings new physics.

In this paper the particle-hole pairing was considered on a very general ground. Therefore its evidence can
be relevant for the chiral spirals \cite{Kharzeev:2010} in the chiral magnetic effect \cite{Warringa:2008} and
the spin density waves,
and can serve as a guide to construct an antiferromagnetic and Mott insulating states of the cuprate superconductors.
In particular, it can be useful to describe the coexisting AFM and SC order parameters in the iron pnictides \cite{Schmalian:2010}. 
Recent work on non-abelian holographic superconductors \cite{ads-cond-mat6:2010} makes applications
to color superconductivity possible. 

It will be instructive to obtain, in analogy to the superconducting instability,
to obtain the particle-hole condensation from the classical Einstein gravity calculations
with the neutral scalar field. 

We considered thermodynamic and transport properties of the boundary field theory.
We did not specify the boundary theory.
Using  the ``dressed'' by the gravity fermion propagator $G_R$ we obtained the equation of state, which
includes a sum over all the poles of $G_R$. We use prescription
of \cite{Denef1:2009},\cite{Denef2:2009} to treat imaginary poles. Imaginary self-energy is a consequence
of inelastic scattering by the black hole, and will be present in any gravity calculations.
In particular the imaginary part provides finite transport coefficients.
The scaling behavior for the heat capacity is $c\sim T$ for the Fermi liquid
and $c\sim T^{2\nu}$, while it is the same in both cases for the DC conductivity $\sigma\sim T^{-2\nu}$.  
It reflects the fact that the specific heat depends on the dispersion relation, in particular
on the real part of $\omega$, while the DC conductivity is sensitive only to the scaling
of the self-energy.

The presented approach has an advantage of unifying description
of the Fermi and non-Frmi liquids within one framework. It uses the language
of poles and branch cuts instead of quasiparticles, that may be more adequate description
for some strongly correlated systems.

\section*{Acknowledgements}

I am grateful to Dirk Rischke for giving me an opportunity to complete this work.
The author thanks Tom Faulkner, Daniel Fernandez-Fraile, Tom Hartman, Nabil Iqbal, Tomoi Koide,
Hong Liu, Mark Mezei, Volodya Miransky, Andreas Schmitt and Igor Shovkovy for helpful inputs and discussions, 
and Armen Sedrakian for useful suggestions and work on the manuscript.
The work was supported in part by the Alliance program of the Helmholtz Association, contract HA216/EMMI 
``Extremes of Density and Temperature: Cosmic Matter in the Laboratory'' and ITP of Goethe University, Frankfurt.

\appendix

\section{Dirac equation in the $AdS_4$}\label{appendix:aa}

Here we discuss the Dirac equation in the presence of the magnetic field the $AdS_4$, and show how Landau levels appear
for the $(x,y)$ part, dimensions of the boundar field theory. The $(x,y)$ part of the Dirac equation decouples from the radial part
and can be solved exactly due to the translational invariance in perpendicular directions.
Then we find the conformal dimension of the spinor operator in the $IR\;CFT$.    
Depending on the $IR$ conformal dimension the boundary theory describes
Fermi, marginal and non-Fermi liquids.

\subsection{Dirac equation with magnetic field in a charged black hole geometry ${\rm AdS_4}$}\label{appendix:a}

Here we solve analytically the part of the Dirac which depends on magnetic field and space-time coordinates
of the boundary theory.   
The free spinor action in the geometry eq.(\ref{ads4-metric1}) and in the presence of magnetic field eq.(\ref{ads4-metric2}) is
given by
\begin{eqnarray}
S_{0} = i\int d^4x \sqrt{-g} \bar{\psi}\left(\Gamma^{M}{\mathcal D}_{M}-m\right)\psi,
\end{eqnarray}
where $\bar{\psi}=\psi^{\dagger}\Gamma^{\hat{t}}$, and
\begin{equation}
 {\mathcal D}_M=\partial_M +\frac{1}{4}\omega_{abM}\Gamma^{ab}-iqA_M, 
\end{equation}
with $\omega_{abM}$ the spin connection, and $\Gamma^{ab}=\frac{1}{2}[\Gamma^a,\Gamma^b]$; here $M$
and $a,b$ denote the bulk space-time and tangent space indices respectively, and $\mu,\nu$
denote indices along the boundary directions, i.e. $M=(r,\mu)$; indices with hat 
refer to tangent space ones, i.e. converting from bulk to tangent indices 
$\Gamma^{M}=e^{M}_{\hat{a}}\Gamma^{\hat{a}}$ with $e^{\hat{a}}=e_M^{\hat{a}} dx^M$ are the tetrads defined by the metric
eq.(\ref{ads4-metric1}), $ds^2=g_{MN}dx^Mdx^N=\eta_{\hat{a}\hat{b}}e^{\hat{a}}e^{\hat{b}}$ and 
$\eta_{\hat{a}\hat{b}}=\rm {diag}(-1,1,1,1)$ is the flat metric. 

Using the translational invariance,
\begin{equation}
\psi(t,x,y,r)=\int d\omega dk {\rm e}^{-i\omega t+ikx}\;\psi(\omega,k,y,r), 
\end{equation}
with $k\equiv k_x$, the Dirac equation is given by
\begin{eqnarray}
&&\hspace{-1cm}\left(\frac{1}{\sqrt{-g_{tt}}}\Gamma^{\hat{t}}(-i\omega+\frac{1}{2}\omega_{\hat{t}\hat{r}t}
 \Gamma^{\hat{t}\hat{r}}-iqA_t(r))
+\frac{1}{\sqrt{g_{rr}}}\Gamma^{\hat{r}}\partial_r
+\frac{1}{\sqrt{g_{ii}}}\Gamma^{\hat{x}}(ik+\frac{1}{2}\omega_{\hat{x}\hat{r}x}
\Gamma^{\hat{x}\hat{r}}-iqA_x(y))\right.\nonumber\\
&+&\left. \frac{1}{\sqrt{g_{ii}}}\Gamma^{\hat{y}}(\partial_y+\frac{1}{2}\omega_{\hat{y}\hat{r}y}
\Gamma^{\hat{y}\hat{r}})
-m \right)\psi(\omega,k,y,r)=0,
\end{eqnarray}
where $g_{ii}\equiv g_{xx}=g_{yy}$, and $A_t(r)=\mu(1-r_0/r)$, $A_x(y)=-{\mathcal H}y$. 
From the torsion-free condition, $\omega^a_b\wedge e^b=-de^a$, we find the spin connection \cite{Carroll:2003}
for the metric \ref{ads4-metric1},
\begin{eqnarray}
\omega_{\hat{t}\hat{r}}=-\frac{\partial_r(\sqrt{-g_{tt}})}{\sqrt{g_{rr}}}dt,\;\;
\omega_{\hat{i}\hat{r}}=\frac{\partial_r(\sqrt{g_{ii}})}{\sqrt{g_{rr}}}dx^i,  
\end{eqnarray}
where $i=x,y$. Note that 
\begin{equation}
-\Gamma^{\hat{t}}\Gamma^{\hat{t}\hat{r}}=\Gamma^{\hat{x}}\Gamma^{\hat{x}\hat{r}}=
\Gamma^{\hat{y}}\Gamma^{\hat{y}\hat{r}}=\Gamma^{\hat{r}},
\end{equation}
and
\begin{eqnarray}
\frac{1}{4}e^{M}_{\hat{a}}\Gamma^{\hat{a}}\omega_{\hat{b}\hat{c}M}\Gamma^{\hat{b}\hat{c}} &=&
\frac{1}{4}\frac{1}{\sqrt{-g_{tt}}}\frac{\partial_r(\sqrt{-g_{tt}})}{\sqrt{g_{rr}}}\Gamma^{\hat{r}}+
\frac{2}{4}\frac{1}{\sqrt{g_{ii}}}\frac{\partial_r\sqrt{g_{ii}}}{\sqrt{g_{rr}}}\Gamma^{\hat{r}}\nonumber\\
&=& \frac{1}{\sqrt{g_{rr}}}\Gamma^{\hat{r}}\partial_r \ln \left(-\frac{g}{g_{rr}}\right)^{1/4},
\end{eqnarray}
where $g$ is the determinant of the metric. Therefore, we can rescale the spinor field
\begin{equation}
\psi=\left(-\frac{g}{g_{rr}}\right)^{-1/4}\Phi,
\label{rescale}
\end{equation}
and remove the spin connection completely. The new action is given by
\begin{eqnarray}
S_{0} = \int d^4x \sqrt{g_{rr}}i\bar{\Phi}(\Gamma^{M}{\mathcal D}_{M}^{\prime}-m)\Phi,
\label{action0}
\end{eqnarray}
where the covariant derivative does not contain spin connection,
${\mathcal D}_M^{\prime}=\partial_M -iqA_M$.

In new field variables, the Dirac equation is given by
\begin{eqnarray}
&& \hspace{-2cm} \left(\frac{\sqrt{g_{ii}}}{\sqrt{g_{rr}}}\Gamma^{\hat{r}}\partial_r -
\frac{\sqrt{g_{ii}}}{\sqrt{-g_{tt}}}\Gamma^{\hat{t}}\;i(\omega+q\mu(1-\frac{r_0}{r}))
-\sqrt{g_{ii}}m +\Gamma^{\hat{x}}\;i(k+q\mathcal{H}y)\right.\nonumber\\
&+&\left. \Gamma^{\hat{y}}\partial_y\right)\Phi(\omega,k,y,r)=0.
\label{dirac-equation-start}
\end{eqnarray}
We separate $y$ and $r$ dependences,
\begin{eqnarray}
P(r) &=& \frac{\sqrt{g_{ii}}}{\sqrt{g_{rr}}}\Gamma^{\hat{r}}\partial_r
-\frac{\sqrt{g_{ii}}}{\sqrt{-g_{tt}}}\Gamma^{\hat{t}}\;
i(\omega+q\mu(1-\frac{r_0}{r}))-\sqrt{g_{ii}}m,\nonumber\\ 
Q(y) &=& \Gamma^{\hat{x}}\;i(k+q\mathcal{H}y)+\Gamma^{\hat{y}}\partial_y,
\end{eqnarray}
and the Dirac equation is
\begin{equation}
(P(r)+Q(y))\Phi=0.
\end{equation}
Though, $[P(r),Q(y)]\neq 0$, one can find a transformation matrix $U$ such that $[UP(r),UQ(y)]=0$, and then look for common
eigenvectors of $UP(r)$ and $UQ(y)$ since they are commuting hermitian operators, i.e. the Dirac equation reads
\begin{equation}
 UP(r)\Phi_n=-UQ(y)\Phi_n=\lambda_n\Phi_n,
\label{dirac2}
\end{equation}
where $n$ will label the Landau levels. In the main text we use $l$ for the Landau index,
in order not to confuse with the Matsubara frequency index $n$.
Transformation matrix $U$ should satisfy conditions 
\begin{equation}
\{U,\Gamma^{\hat{r}}\}=0,\;\,
\{U,\Gamma^{\hat{t}}\}=0,\;\;
[U,\Gamma^{\hat{x}}]=0,\;\;
[U,\Gamma^{\hat{y}}]=0,
\end{equation}
which do not fix $U$ completely.
It is convenient to use the following basis \cite{Faulkner1:2009},
\begin{eqnarray}
&& \Gamma^{\hat{r}}= \left(\begin{array}{cc}
-\sigma^3 & 0 \\
0 & -\sigma^3
\end{array}
\right),\;\; 
\Gamma^{\hat{t}}= \left(\begin{array}{cc}
i\sigma^1 & 0 \\
0 & i\sigma^1
\end{array}
\right),\;\;
\Gamma^{\hat{x}}= \left(\begin{array}{cc}
-\sigma^2 & 0 \\
0 & \sigma^2
\end{array}
\right),\;\;
\nonumber\\ 
&& \Gamma^{\hat{y}}= \left(\begin{array}{cc}
0 & \sigma^2 \\
\sigma^2 & 0
\end{array}
\right),\;\;
\Gamma^{\hat{5}}= \left(\begin{array}{cc}
0 & i\sigma^2 \\
-i\sigma^2 & 0
\end{array}
\right).
\label{matrices}
\end{eqnarray}
Note, that the usual relation holds
\begin{equation}
 \Gamma^{\hat{5}}=\Gamma^{\hat{0}}\Gamma^{\hat{1}}\Gamma^{\hat{2}}\Gamma^{\hat{3}},
\end{equation}
with $0\rightarrow t$, $1\rightarrow x$, $2\rightarrow y$, $3\rightarrow r$.
In the representation of eq.(\ref{matrices}), we can choose
\begin{equation}
 U=\left(\begin{array}{cc}
-i\sigma^2 & 0 \\
0 & -i\sigma^2
\end{array}
\right).
\label{transform}
\end{equation}
Writing $\Phi=(F_1,F_2)^{T}$, and using eq.(\ref{transform}), 
we get the Dirac equation written in a compact form, eq.(\ref{dirac2}),  
\begin{eqnarray}
&& \hspace{-1cm} \left(-\frac{\sqrt{g_{ii}}}{\sqrt{g_{rr}}}\sigma^1\partial_r+\sqrt{g_{ii}}i\sigma^2 m
-\frac{\sqrt{g_{ii}}}{\sqrt{-g_{tt}}}\sigma^3(\omega+q\mu(1-r_0/r))
-\lambda_n\right) \otimes 1 
\left(\begin{array}{c}
F_1 \\
F_2
\end{array}
\right)=0  \label{dirac-equationr}\\
&&\hspace{-1cm}  1 \otimes \left(\begin{array}{cc}
-(k+q\mathcal{H}y)+\lambda_n & -i\partial_y  \\                       
-i\partial_y & (k+q\mathcal{H}y)+\lambda_n
\end{array}
\right)
\left(\begin{array}{c}
F_1 \\
F_2
\end{array}
\right)=0,
\label{dirac-equationy}
\end{eqnarray}
where in $X\otimes Y$, $X$ acts inside $F_1$ or $F_2$ and $Y$ acts between $F_1$ and $F_2$.
In eq.(\ref{dirac-equation}), the $1$ in the first equation shows that there is no mixing of $F_1$ and $F_2$
by the operator $UP(r)$ and the $1$ in the second equation means that there is no mixing inside 
$F_1$ or $F_2$ by the operator $UQ(y)$. Therefore solution can be represented as
\begin{equation}
\left(\begin{array}{c}
F_1\\
F_2
\end{array}
\right)=
\left(\begin{array}{c}
f^{(1)}_n(r)g^{(1)}_n(y)\\
f^{(2)}_n(r)g^{(1)}_n(y)\\
f^{(1)}_n(r)g^{(2)}_n(y)\\
f^{(2)}_n(r)g^{(2)}_n(y)
\end{array}
\right),
\end{equation}
we do not write $\omega$ and $k$ dependences.
Dirac equations for each component are
\begin{eqnarray}
\hspace{-1.5cm} \left(\frac{\sqrt{g_{ii}}}{\sqrt{g_{rr}}}\partial_r +\sqrt{g_{ii}} m \right)f^{(1)}_n(r) 
+ \left(-\frac{\sqrt{g_{ii}}}{\sqrt{-g_{tt}}}(\omega+q\mu(1-r_0/r))
+\lambda_n\right)f^{(2)}_n(r) &=& 0, \nonumber\\
\hspace{-1.5cm} \left(\frac{\sqrt{g_{ii}}}{\sqrt{g_{rr}}}\partial_r -\sqrt{g_{ii}} m \right)f^{(2)}_n(r) 
+ \left(\frac{\sqrt{g_{ii}}}{\sqrt{-g_{tt}}}(\omega+q\mu(1-r_0/r))
+\lambda_n\right) f^{(1)}_n(r) &=& 0, 
\label{eqr} \\
\hspace{-1.5cm} -i\partial_yg^{(1)}_n(y) +\left((k+q{\mathcal H}y)+\lambda_n\right)g^{(2)}_n &=& 0, \nonumber\\
\hspace{-1.5cm} -i\partial_yg^{(2)}_n +\left(-(k+q{\mathcal H}y)+\lambda_n\right)g^{(1)}_n &=& 0.
\label{eqy}
\end{eqnarray}
In equations \ref{eqy} for the $y$ dependence, we rescale $\tilde{y}=\sqrt{|Q{\mathcal H}|}\,(y+k/q{\mathcal H})$
and $\lambda_n=\sqrt{|q{\mathcal H}|}\,\tilde{\lambda}_n$, and get
\begin{eqnarray}
&& -i\partial_{\tilde{y}}g^{(1)}_n +(\tilde{y}+\tilde{\lambda}_n)g^{(2)}_n =0, \nonumber\\
&& -i\partial_{\tilde{y}}g^{(2)}_n +(-\tilde{y}+\tilde{\lambda}_n)g^{(1)}_n =0.
\end{eqnarray}
The second order ODE
\begin{eqnarray}
&& \partial_{\tilde{y}}^2g^{(1)}-\frac{1}{\tilde{y}+\tilde{\lambda}}\partial_{\tilde{y}}g^{(1)}
+(\tilde{\lambda}^2-\tilde{y}^2)g^{(1)}=0, \nonumber\\
&& \partial_{\tilde{y}}^2g^{(2)}-\frac{1}{-\tilde{y}+\tilde{\lambda}}
\partial_{\tilde{y}}g^{(2)}
+(\tilde{\lambda}^2-\tilde{y}^2)g^{(2)}=0, 
\end{eqnarray}
are solved by substitution $g^{(1)}={\rm e}^{-\tilde{y}^2/2}\tilde{g}^{(1)}$
and $g^{(2)}=\pm i{\rm e}^{-\tilde{y}^2/2}\tilde{g}^{(2)}$. The eigenfunctions
are Hermite polynomials.
We get the same eigenvalues, but slightly different eigenfunctions for different signs of $q{\mathcal H}$.
Putting all together, for $q{\mathcal H}>0$, we have
\begin{eqnarray}
&&  \tilde{\lambda}_{-1}=0\,: \nonumber\\
&& g^{(1)}_{-1}(\tilde{y})={\rm e}^{-\tilde{y}^2/2},\;\;
g^{(2)}_{-1}(\tilde{y})=-i{\rm e}^{-\tilde{y}^2/2}  \\
&& \tilde{\lambda}^{\pm}_n =\pm\sqrt{2(n+1)}\, :  \nonumber\\
&& g^{(1)\pm}_n(\tilde{y})= {\rm e}^{-\tilde{y}^2/2}
\left(H_n(\tilde{y})\pm\frac{1}{\sqrt{2(n+1)}}H_{n+1}(\tilde{y})\right),\nonumber\\
&& g^{(2)\pm}_n(\tilde{y})= i{\rm e}^{-\tilde{y}^2/2}
\left(H_n(\tilde{y})\mp\frac{1}{\sqrt{2(n+1)}}H_{n+1}(\tilde{y})\right),\nonumber
\label{solutionyplus}
\end{eqnarray}
and for $q{\mathcal H}<0$, we have
\begin{eqnarray}
&&  \tilde{\lambda}_{-1}=0\, : \nonumber\\
&& g^{(1)}_{-1}(\tilde{y})={\rm e}^{-\tilde{y}^2/2},\;\;
g^{(2)}_{-1}(\tilde{y})=i{\rm e}^{-\tilde{y}^2/2} \\
&& \tilde{\lambda}^{\pm}_n =\pm\sqrt{2(n+1)}\, :  \nonumber \\
&& g^{(1)\pm}_n(\tilde{y})= {\rm e}^{-\tilde{y}^2/2}
\left(H_n(\tilde{y})\mp\frac{1}{\sqrt{2(n+1)}}H_{n+1}(\tilde{y})\right),\nonumber\\
&& g^{(2)\pm}_n(\tilde{y})= -i{\rm e}^{-\tilde{y}^2/2}
\left(H_n(\tilde{y})\pm\frac{1}{\sqrt{2(n+1)}}H_{n+1}(\tilde{y})\right), \nonumber
\label{solutionyminus}
\end{eqnarray}
The case $q{\mathcal H}<0$ can be obtained from the case $q{\mathcal H}>0$ by
replacing $g^{(1)}[q{\mathcal H}<0] = -ig^{(2)}[q{\mathcal H}>0)]$
and $g^{(2)}[q{\mathcal H}<0] = -ig^{(1)}[q{\mathcal H}>0]$.
Using the eigenvalues, eqs.(\ref{solutionyplus},\ref{solutionyminus}),
in the equation for the radial part, eq.(\ref{eqr}), we get
\begin{eqnarray}
&& \hspace{-1cm}\left(\frac{\sqrt{g_{ii}}}{\sqrt{g_{rr}}}\partial_r + \sqrt{g_{ii}} m \right)f^{(1)\pm}_n(r)\nonumber\\ 
&+& \left(-\frac{\sqrt{g_{ii}}}{\sqrt{-g_{tt}}}(\omega+q\mu(1-r_0/r))
\pm \sqrt{2|q{\mathcal H}|(n+1)}\right) f^{(2)\pm}_n(r) = 0, \nonumber\\
&& \hspace{-1cm}\left(\frac{\sqrt{g_{ii}}}{\sqrt{g_{rr}}}\partial_r -\sqrt{g_{ii}} m \right)f^{(2)\pm}_n(r)\nonumber\\ 
&+& \left(\frac{\sqrt{g_{ii}}}{\sqrt{-g_{tt}}}(\omega+q\mu(1-r_0/r))
\pm \sqrt{2|q{\mathcal H}|(n+1)}
\right) f^{(1)\pm}_n(r) = 0. 
\end{eqnarray}
These equations can be obtained by replacing 
\begin{equation}
k\rightarrow \pm\sqrt{2|q{\mathcal H}|(n+1)},
\label{replace}
\end{equation} 
in the Dirac equation at zero magnetic field \cite{Denef1:2009}. Equation (\ref{replace})
also gives a prescription how to treat the limit of zero magnetic field, i.e. the ${\mathcal H}\rightarrow 0$
limit is taken keeping $2|q{\mathcal H}|(n+1)\equiv k^2$ fixed as ${\mathcal H}\rightarrow 0$.
In a compact form the Dirac equation, eq.(\ref{dirac-equationr}), in a magnetic field reads
\begin{eqnarray}
&& \hspace{-1.5cm}\left(-\frac{1}{\sqrt{g_{rr}}}\sigma^3\partial_r- m
+\frac{1}{\sqrt{-g_{tt}}}\sigma^1(\omega+q\mu(1-r_0/r))\right.\nonumber\\
&\mp&\left. \frac{1}{\sqrt{g_{ii}}}i\sigma^2\sqrt{2|q{\mathcal H}|(n+1)}\right) \otimes 1 
\left(\begin{array}{c}
F_1 \\
F_2
\end{array}
\right)=0, 
\end{eqnarray}
which coincides with eq. (A14) in \cite{Faulkner1:2009} with the replacement eq.(\ref{replace}).
The starting Dirac equation in a magnetic field, eq.(\ref{dirac-equation-start}), is given by
\begin{eqnarray}
&& \hspace{-1.5cm}\left(\frac{1}{\sqrt{g_{rr}}}\Gamma^{\hat{r}}\partial_r-
\frac{1}{\sqrt{-g_{tt}}}\Gamma^{\hat{t}}\;i(\omega+q\mu(1-\frac{r_0}{r}))-m\right.\nonumber\\ 
&\mp& \left.\frac{1}{\sqrt{g_{ii}}} U^{-1}\sqrt{2|q{\mathcal H}|(n+1)}\right)\Phi(r)=0,
\label{dirac-equation-finish}
\end{eqnarray}
we do not write $\omega$ dependence, $\Phi=(F_1,F_2)^{T}$, $n=-1,0,1,\ldots$;
and where $U^{-1}$ is the inverse matrix
to the transformation matrix eq.(\ref{transform})
\begin{equation}
 U^{-1}=\left(\begin{array}{cc}
i\sigma^2 & 0 \\
0 & i\sigma^2
\end{array}
\right).
\label{transform}
\end{equation}
We use equation (\ref{dirac-equation-finish}) in the main text.

\subsection{Dirac equation. Conformal dimension in the low frequency limit}\label{appendix:a'}

As outlined in \cite{Denef1:2009,Denef2:2009}, the fermion determinant in the black hole background is given by a sum
over the quasinormal frequencies $z_{*}=i\omega_n$, that are obtained as complex frequencies which give zero eigenvalues,
$\lambda=0$, of the Dirac equation. The eigenvalues are given by
\begin{equation}
\left(\Gamma^M {\mathcal D}_M-m -\Delta i\Gamma^{\hat{2}}\Gamma^{\hat{5}}
\pm q{\mathcal H}\Gamma^{\hat{t}}\right)\psi=\lambda\psi.
\label{dirac5}
\end{equation}
In Appendix \ref{appendix:a}, we simplified the Dirac equation obtained from the free fermion action
eq.(\ref{free-action}). We rescaled the original fermion field,
eq.(\ref{rescale}), as was done in \cite{Faulkner1:2009}, and removed the spin connection,
with the result eq.(\ref{action0}) and eq.(\ref{dirac-equation-start}).
In new field variables, the Dirac equation is
\begin{eqnarray}
 \left(\Gamma^M {\mathcal D}_M^{\prime}-m -\Delta i\Gamma^{\hat{2}}\Gamma^{\hat{5}}
\pm q{\mathcal H}\Gamma^{\hat{t}}\right)\Phi=0,
\end{eqnarray}
where ${\mathcal D}_M=\partial_M -iqA_M$.
It is written in the geometry eq.(\ref{ads4-metric1}) and in the Landau gauge as
\begin{eqnarray}
&& \hspace{-1.cm} \left(\frac{1}{\sqrt{g_{rr}}}\Gamma^{\hat{r}}\partial_r -
\frac{1}{\sqrt{-g_{tt}}}i\Gamma^{\hat{t}}(\omega+q\mu(1-\frac{r_0}{r}))
-m +\frac{1}{\sqrt{g_{ii}}}i\Gamma^{\hat{x}}(k+q\mathcal{H}y)+\frac{1}{\sqrt{g_{ii}}}\Gamma^{\hat{y}}\partial_y\right.\nonumber\\
&-&\left.i\Delta\Gamma^{\hat{2}}\Gamma^{\hat{5}} 
\pm q{\mathcal H}\Gamma^{\hat{t}}\right)\Phi=0,
\label{}
\end{eqnarray} 
with $k\equiv k_x$, $k_y=0$. At the boundary, $r\rightarrow\infty$,
the frequency $\omega$ is measured from the effective chemical potential
$q\mu$. The dependence on charge $q$ of $\psi$-field enters only through the combination $q\mu$.
Also as given in Appendix A, in the Landau gauge eq.(\ref{factor})
the radial and $y$ dependences decouple, eq.(\ref{dirac-equationr}) and eq.(\ref{dirac-equationy}),
and $y$ dependence reduces to the harmonic oscillator.
We solve the $y$ dependent part of the Dirac equation, eqs.(\ref{solutionyplus}) and (\ref{solutionyminus}),
and use the result in the radial part of the Dirac equation.    
The procedure of obtaining the Dirac equation
at nonzero magnetic field amounts to replacing 
\begin{equation}
k\rightarrow \pm\sqrt{2|q{\mathcal H}|l}
\label{replace1}
\end{equation}
in the Dirac equation at zero magnetic field. As compared to eq.(\ref{replace}), eq.(\ref{replace1})
takes into account the Zeeman splitting term. 
The result for non-interacting fermions is given
by eq.(\ref{dirac-equation-finish}), which agrees with \cite{Denef1:2009,Denef2:2009}. The Dirac equation
is written as
\begin{eqnarray}
\left(\frac{1}{\sqrt{g_{rr}}}\Gamma^{\hat{r}}\partial_r-
\frac{1}{\sqrt{-g_{tt}}}i\Gamma^{\hat{t}}(\omega+q\mu(1-\frac{r_0}{r}))
-m \mp \frac{1}{\sqrt{g_{ii}}}U^{-1}\sqrt{2|q{\mathcal H}|l}
-\Delta i\Gamma^{\hat{2}}\Gamma^{\hat{5}} \right)\Phi=0,\nonumber\\
\label{dir2}
\end{eqnarray} 
where $l=0,1,\ldots$. The $\Gamma$ matrices are defined in eq.(\ref{matrices}) as in \cite{Faulkner1:2009}
and $U^{-1}$ is given by eq.(\ref{transform}).

We split the $4$-component spinors
into two $2$-component spinors (we do not write zero entries) 
$F=(F_1,F_2)^{T}$ where the index $\alpha=1,2$
is the Dirac index of the boundary theory, using projectors 
\begin{equation}
\Pi_{\alpha}=\frac{1}{2}(1-(-1)^{\alpha}\Gamma^{\hat{r}}\Gamma^{\hat{t}}\Gamma^{\hat{1}}),\;\;
\alpha=1,2,\;\; \Pi_1+\Pi_2=1,
\label{projection} 
\end{equation}
which commute with the Dirac operator of eq.(\ref{dir2}), and $F_{\alpha}=\Pi_{\alpha}\Phi$,
$\alpha=1,2$, decouple from each other. Gamma matrices were chosen in such a way that this decoupling
is possible.
The Dirac equation for two components is given by
\begin{eqnarray}
&& \left(-\frac{1}{\sqrt{g_{rr}}}\sigma^3\partial_r- m
+\frac{1}{\sqrt{-g_{tt}}}\sigma^1(\omega+q\mu(1-r_0/r))
\mp\frac{1}{\sqrt{g_{ii}}}i\sigma^2\sqrt{2|q{\mathcal H}|l}
+(-1)^{\alpha}s\Delta \right)F_{\alpha}=0,\nonumber\\
\label{dir}
\end{eqnarray}
where $s=\pm sgn(q{\mathcal H})$.
To obtain the retarded Green function for fermionic operator $O$ in the boundary theory,
we need to find a solution $\Phi$ which satisfies the ingoing boundary conditions at the horizon,
and to expand it near the boundary at $r\rightarrow\infty$ as 
\begin{eqnarray}
F_1 &\approx&  a_1 r^{m_1 R}
\left(\begin{array}{c}
0 \\
1 
\end{array}
\right)
+  b_1 r^{-m_1 R}
\left(\begin{array}{c}
1 \\
0 
\end{array}
\right),
\nonumber\\
F_2 &\approx&  a_2 r^{m_2 R}
\left(\begin{array}{c}
0 \\
1 
\end{array}
\right)
+  b_2 r^{-m_2 R}
\left(\begin{array}{c}
1 \\
0 
\end{array}
\right),
\label{boundary}
\end{eqnarray}
where $a_{\alpha}$ is the source and $b_{\alpha}$ is the v.e.v. in the Green function terminology,
and the two masses are $m_1 \equiv m+ \Delta$, $m_2 \equiv |m-\Delta|$. The conformal dimension $\Delta_{\psi}$
of ${\mathcal O}$ is given in terms of the mass of $\psi$ field by eq.(\ref{conformal-dimension}),
$\Delta_{\psi}=\frac{3}{2}+m_iR$.
The two spinors $(1,0)$ and $(0,1)$ are eigenspinors of $\Gamma^{\hat{r}}$ with opposite eigenvalues,
implying that $a_{\alpha}$ and $b_{\alpha}$ are canonically conjugate 
(in a radial Hamiltonian slicing) \cite{Iqbal:2009}. Depending on the value of the exponent, we impose different
boundary condition for $a_{\alpha}$ and/or $b_{\alpha}$. 
The retarded Green function is given by \cite{Son:2002,Iqbal:2009,Ching:1995}
\begin{equation}
G_R(\omega,k)=
\left(\begin{array}{cc}
G_1(\omega,k) & 0 \\
0 & G_2(\omega,k)
\end{array}
\right), \;\; {\rm with}\;\; 
G_{\alpha}(\omega,k)=\frac{b_{\alpha}}{a_{\alpha}}, \;\; 
\alpha=1,2, 
\label{block-diagonal}
\end{equation}
with $k\rightarrow \sqrt{2|q{\mathcal H}|l}$.
 
For a general Reissner-Nordstrom black hole background there is no mixing between $F_1$ and $F_2$
(interaction term  $S_{int}$ is diagonal),
and in the absence of $\Delta$-term two fields $F_1$ and $F_2$ have coincident Fermi surfaces
(at $\omega+q\mu=0$ at the boundary $r\rightarrow\infty$). With the $\Delta$ term, there is
a relative shift of the Fermi surface in the spectrum of $F_1$ and $F_2$.
In the following, we consider only one component, e.g., $F_1$.

We could have solved eq.(\ref{dir}) with the ingoing boundary conditions at the horizon numerically.
Instead, we use analytical results
of \cite{Faulkner1:2009}, and for that we need to extract information from the low-energy limit of the theory. 
In section \ref{section:4},
we follow \cite{Hartman:2010} and solve the Dirac equation at zero frequency, and then use the matching procedure
to extract the retarded Green function of the boundary theory. For now we use results from \cite{Faulkner1:2009,Vegh:2009}. 
We will be interested in the poles of the retarded Green function.
As was shown in \cite{Faulkner1:2009}, the nonanalytic part of the retarded Green function in the extremal 
Reissner-Nordstrom black hole background
eq.(\ref{ads4-metric1}) comes from the IR region, which has a simpler geometry than ${\rm AdS_4}$.
At small frequencies, $\omega\rightarrow 0$, the metric reduces to
$AdS_4\rightarrow AdS_2\times R_2$ at the horizon 
\begin{equation}
ds^2=\frac{R_2^2}{\zeta^2}(-d\tau^2+d\zeta^2)+\frac{r_{**}^2}{R^2}d\vec{x}^2,\;\;
A_{\tau}=\frac{g_F}{\sqrt{12}}\frac{r_{*}^2}{r_{**}^2}\frac{1}{\zeta},\;\;
A_x=-{\mathcal H}y, 
\end{equation} 
which is called the IR region \cite{Faulkner1:2009}.
In the IR region, the Dirac equation ({\ref{dir}) becomes
\begin{eqnarray}
\left(-\frac{1}{\sqrt{g_{\zeta\zeta}}}\sigma^3\partial_{\zeta}- m
+\frac{1}{\sqrt{-g_{\tau\tau}}}\sigma^1(\omega+\frac{qg_F}{\sqrt{12}}\frac{r_{*}^2}{r_{**}^2}\frac{1}{\zeta})
-\frac{1}{\sqrt{g_{ii}}}i\sigma^2\sqrt{2|q{\mathcal H}|l}
-s\Delta\right)F=0,\nonumber\\
\label{dir2}
\end{eqnarray}
where we omit index by the spinor field.
We can write explicitly
\begin{equation}
\left(\begin{array}{cc}
\frac{\zeta}{R_2}\partial_{\zeta}+m+s\Delta &  -\frac{\zeta}{R_2}(\omega+
\frac{qg_F}{\sqrt{12}}\frac{r_{*}^2}{r_{**}^2}\frac{1}{\zeta})+\frac{R}{r_{**}}\sqrt{2|q{\mathcal H}|l}
\\
\frac{\zeta}{R_2}(\omega+\frac{qg_F}{\sqrt{12}}\frac{r_{*}^2}{r_{**}^2}\frac{1}{\zeta})
+\frac{R}{r_{**}}\sqrt{2|q{\mathcal H}|l} & 
\frac{\zeta}{R_2}\partial_{\zeta} -m -s\Delta 
\end{array}
\right) 
\left(\begin{array}{c}
y \\
z
\end{array}
\right) =0.
\label{}
\end{equation}
At the $AdS_2$ boundary, $\zeta\rightarrow 0$, the Dirac equation to the leading order is
\begin{equation}
\zeta\partial_{\zeta}\Phi=-U\Phi,\;\; 
U=\left(\begin{array}{cc}
R_2(m+s\Delta) &  -\frac{qg_F}{\sqrt{12}}\frac{r_{*}^2}{r_{**}^2}
+\frac{R_2R}{r_{**}}\sqrt{2|q{\mathcal H}|l},
\\
\frac{qg_F}{\sqrt{12}}\frac{r_{*}^2}{r_{**}^2}+\frac{R_2R}{r_{**}}\sqrt{2|q{\mathcal H}|l}
 & -R_2(m +s\Delta). 
\end{array}
\right) 
\end{equation}
Diagonalizing matrix $U$, we obtain the conformal dimension for the operator ${\mathcal O}$
in the ${\rm IR\; CFT}$ dual to $\Phi$,
\begin{eqnarray}
\delta_{\psi} &=& \frac{1}{2}+\nu, \nonumber\\
\nu &=& \sqrt{\left((m+s\Delta)^2+\frac{R^2}{r_{**}^2}2|q{\mathcal H}|l\right)R_2^2
-\frac{q^2g_F^2}{12}\frac{r_{*}^4}{r_{**}^4}},
\label{conformal}
\end{eqnarray}
where the $AdS_2$ curvature radius is $R_2=R/\sqrt{6}$, the Landau level index
$l=0,1,\ldots$ and $s=\pm sign(q{\mathcal H})$.
The conformal dimension of the fermionic operator in the $IR\;CFT$ in the chiral limit $m=0$ and at the Fermi surface
is given by
\begin{equation}
\nu_{k_F}=\sqrt{(\frac{k_F^2R^2}{r_{**}^2}+\Delta^2)R_2^2-\frac{q^2g_F^2}{12}\frac{r_{*}^4}{r_{**}^4}}, 
\label{conformal2}
\end{equation}
with $k_F=q\mu$, $\mu=\sqrt{3}g_F\frac{r_{*}^2}{R^2r_0}$. We use the $IR$ conformal dimension
of the fermionic operator in the main text.

\section{Dirac equation in the $AdS_2$}\label{appendix:bb}

Here we discuss the Dirac equation in the $AdS_2$. We obtain the conformal dimension of the spinor operator in the $CFT_1$,
which gives (with trivial modifications) the conformal dimension of the operator in the original $CFT_{2+1}$
in the IR region. It controlls the behavior of the theory, e.g., Fermi liquid or non-Fermi liquid regimes.
Then, we derive analytically the two point Green function,
which up to a nonimportant constant defines the self-energy in the fermion propagator of the $CFT_{2+1}$.  
 
\subsection{Dirac equation  and conformal dimension}\label{appendix:b}

We consider the following action for a two-component spinor field $\psi$
\begin{equation}
 S= i\int d^2 x\sqrt{-g}\bar{\psi}\left(\Gamma^{\alpha}D_{\alpha}-m+i\Gamma\lambda
-\tilde{\Gamma}\Delta\right)\psi,
\label{action-ads2}
\end{equation}
with $D_{\alpha}=\partial_{\alpha}-iqA_{\alpha}$ and the background metric and gauge field given by
\begin{eqnarray}
 ds^2=\frac{R_2^2}{\zeta^2}(-d\tau^2 +d\zeta^2),\;\; A_{\tau}=\frac{g_Fr_{*}^2}{\sqrt{12}r_{**}^2}\frac{1}{\zeta}.
\end{eqnarray}
In eq.(\ref{action-ads2}), we included a time-reversal violating $\lambda$ term which in our application
will be related to $k_x$ momentum in ${\rm R}^2$, and hence the eigenvalue $\lambda_n$ incorporating effect
of the magnetic field; we also included $\Delta$ term to mimic the gap in ${\rm R}^2$.  
We choose the following Gamma matrices
\begin{equation}
\Gamma^{\hat{\zeta}}=-\sigma^3,\;\;
\Gamma^{\hat{\tau}}=i\sigma^1,\;\;
\Gamma =-\sigma^2,\;\;
\tilde{\Gamma}=1, 
\end{equation}
where the hat indices denote those in tangent frame and $\sigma^i$ are standard sigma matrices.
Writing $\psi(\tau,\zeta)=(-g/g_{\zeta\zeta})^{-1/4}\int d\omega {\rm e}^{-i\omega}\Phi(\omega,\zeta)$
the equations of motion are
\begin{eqnarray}
&& \left(-\frac{1}{\sqrt{g_{\zeta\zeta}}}\sigma^3\partial_{\zeta}- m_{**}
+\frac{1}{\sqrt{-g_{\tau\tau}}}\sigma^1(\omega+qA_{\tau})
-i\sigma^2\lambda\right)\Phi=0,
\label{dirac2-ads2}
\end{eqnarray}
where we introduced $m_{**}=m+\Delta$.
In a matrix form we have
\begin{equation}
\left(\begin{array}{cc}
\frac{\zeta}{R_2}\partial_{\zeta}+m_{**} &\;\;\;  -\frac{\zeta}{R_2}(\omega+\frac{qg_Fr_{*}^2}{\sqrt{12}r_{**}^2}\frac{1}{\zeta})
+\lambda
\\
\frac{\zeta}{R_2}(\omega+\frac{qg_Fr_{*}^2}{\sqrt{12}r_{**}^2}\frac{1}{\zeta}) +\lambda &\;\;\; \frac{\zeta}{R_2}\partial_{\zeta} -m_{**} 
\end{array}
\right) 
\left(\begin{array}{c}
y \\
z
\end{array}
\right) =0.
\label{dirac}
\end{equation}
To find the conformal dimension of the operator ${\mathcal O}$ dual to $\Phi$,
we solve the Dirac equation near the boundary.
Near the ${\rm AdS_2}$ boundary $\zeta\rightarrow 0$, equation (\ref{dirac}) to the leading order is
\begin{equation}
\zeta\partial_{\zeta}\Phi=-U\Phi,\;\; 
U=\left(\begin{array}{cc}
m_{**}R_2 &\;\;\;  -\frac{qg_Fr_{*}^2}{\sqrt{12}r_{**}^2}+ \lambda R_2
\\
\frac{qg_Fr_{*}^2}{\sqrt{12}r_{**}^2} + \lambda R_2 &\;\;\; -m_{**} R_2 
\end{array}
\right). 
\label{dirac-matrix}
\end{equation}
As $\zeta\rightarrow 0$, $\Phi$ can be written to the leading order 
in the form
\begin{equation}
\Phi=Av_{+}\zeta^{-\nu}+Bv_{-}\zeta^{\nu},
\label{general-solution}
\end{equation}
where $v_{\pm}$ are real eigenvectors of $U$ with eigenvalues $\pm\nu$ respectively
\begin{equation}
\nu=\sqrt{\left((m+\Delta)^2+\lambda^2\right)R_2^2 -\frac{q^2g_F^2}{12}\frac{r_{*}^4}{r_{**}^4}}, 
\label{dimension-ads2}
\end{equation}
with $R_2=R/\sqrt{6}$. 
Here $\nu$ determines the scaling dimension of the fermionic operator in the $CFT_1$ dual to the
spinor field in the $AdS_2$. With small modifications this gives the scaling dimension of the $CFT_3$
in the $IR$ region, since 
as $\omega\rightarrow 0$ the metric reduces $AdS_4\rightarrow  AdS_2\times R_2$
near the $AdS_4$ horizon.
The eigenvectors of $U$ can be chosen from eq.(\ref{dirac-matrix}) as
\begin{equation}
v_{\pm}=\left(\begin{array}{c}
(m+\Delta)R_2\pm\nu_{\psi}
\\
\lambda R_2+\frac{qg_Fr_{*}^2}{\sqrt{12}r_{**}^2}  
\end{array}
\right). 
\end{equation}
Imposing the ingoing boundary condition for $\Phi$ at the $AdS_2$ horizon,
the retarded Green function for the boundary operator in the $IR \;\;CFT_1$ dual to $\Phi$
can be written as
\begin{equation}
G^{IR}_{R}(\omega)=\frac{B}{A}\omega^{2\nu}, 
\label{formg}
\end{equation}
meaning that the operator conformal dimension is given by 
\begin{equation}
\delta_{\psi}=\frac{1}{2}+\nu,
\end{equation}
where $\nu$ is the scaling exponent with respect to $\omega$, given by eq.(\ref{dimension-ads2}).
In eq.(\ref{formg}), we used the fact that in the solution $\Phi$, $\omega$ and $\zeta$ 
scale with the same power \cite{Faulkner1:2009}.

\subsection{Two-point functions for charged fermions in ${\bf AdS_2}$}\label{appendix:b'}

We calculate $G^{IR}$ analytically.
The equation of motion (\ref{dirac2-ads2}) can be written as
\begin{equation}
\partial_{\zeta}\Phi=i\sigma^2\left(\omega+\frac{qg_Fr_{*}^2}{\sqrt{12}r_{**}^2}\frac{1}{\zeta}\right)\Phi
-\frac{R_2}{\zeta}\left(\sigma^3 m_{**} + \sigma^1\lambda\right)\Phi,
\end{equation}
where $m_{**}=m+\Delta$.
The equation of motion for two components $\Phi=(y,z)^T$ is given by 
\begin{eqnarray}
(\zeta\partial_{\zeta}+m_{**}R_2)y-(\omega\zeta+\frac{qg_Fr_{*}^2}{\sqrt{12}r_{**}^2}-\lambda R_2)z &=&0,\nonumber\\
(\zeta\partial_{\zeta}-m_{**}R_2)z+(\omega\zeta+\frac{qg_Fr_{*}^2}{\sqrt{12}r_{**}^2}+\lambda R_2)y &=&0,
\label{originalode}
\end{eqnarray}
which contain $\zeta$ dependence in the mixing term proportional to $i\sigma^2$, that makes
it difficult to solve for one component. We therefore transform to another basis to make
all $\zeta$ dependent terms diagonal \cite{TomFaulkner}.
We make the following basis rotation
\begin{eqnarray}
 \left(\begin{array}{c}
\tilde{y}\\
\tilde{z}
\end{array}
\right) = M \left(\begin{array}{c}
 y \\
 z
\end{array}
\right),\;\; {\rm with}\;\;
 M = \left(
\begin{array}{cc}
1 & -i \\
-i &  1
\end{array} \right),\;\;
M^{-1}=\frac{1}{2}\left(
\begin{array}{cc}
 1 & i \\
 i & 1
\end{array} \right).
\label{transformation}
\end{eqnarray}
Transforming the sigma matrices, $M\sigma^{i}M^{-1}$, we have
$i\sigma^2 \rightarrow i\sigma^3$,
$\sigma^3  \rightarrow -\sigma^2$,
$\sigma^1  \rightarrow \sigma^1$.
The equation of motion becomes
\begin{equation}
\partial_{\zeta}\tilde{\Phi}=i\sigma^3\left(\omega+\frac{qg_Fr_{*}^2}{\sqrt{12}r_{**}^2}\frac{1}{\zeta}\right)\tilde{\Phi}
+\frac{R_2}{\zeta}\left(\sigma^2 m_{**} - \sigma^1\lambda\right)\tilde{\Phi},
\end{equation}
and for two components $\tilde{\Phi}=(\tilde{y},\tilde{z})^T$,
\begin{eqnarray}
(\zeta\partial_{\zeta}-i(\omega\zeta+\frac{qg_Fr_{*}^2}{\sqrt{12}r_{**}^2}))\;\tilde{y}
+(im_{**}+\lambda)R_2\;\tilde{z} &=& 0, \nonumber\\
(\zeta\partial_{\zeta}+i(\omega\zeta+\frac{qg_Fr_{*}^2}{\sqrt{12}r_{**}^2}))\;\tilde{z}
-(im_{**}-\lambda)R_2\;\tilde{y} &=& 0.
\label{firstode}
\end{eqnarray}
Expressing $\tilde{y}$ from the second equation
\begin{equation}
\tilde{y}=\frac{\zeta\partial_{\zeta}+i(\omega\zeta+\frac{qg_Fr_{*}^2}{\sqrt{12}r_{**}^2})}{(im_{**}-\lambda)R_2}\;\tilde{z},
\end{equation}
we get the equation for $\tilde{z}$. The equations for both components are given by
\begin{eqnarray}
\left(\zeta\partial_{\zeta}+i(\omega\zeta+\frac{qg_Fr_{*}^2}{\sqrt{12}r_{**}^2})\right)
\left(\zeta\partial_{\zeta}-i(\omega\zeta+\frac{qg_Fr_{*}^2}{\sqrt{12}r_{**}^2})\right)\;\tilde{y}
&=& (m_{**}^2+\lambda^2)R_2^2\tilde{y},
\nonumber\\
\left(\zeta\partial_{\zeta}-i(\omega\zeta+\frac{qg_Fr_{*}^2}{\sqrt{12}r_{**}^2})\right)
\left(\zeta\partial_{\zeta}+i(\omega\zeta+\frac{qg_Fr_{*}^2}{\sqrt{12}r_{**}^2})\right)\;\tilde{z}
&=& (m_{**}^2+\lambda^2)R_2^2\tilde{z}.
\end{eqnarray}
Rewriting these equations, we get
\begin{eqnarray}
&& \zeta^2\partial^2_{\zeta}\tilde{y}+\zeta\partial_{\zeta}\tilde{y}
+\left(-i\zeta\omega+(\omega\zeta+\frac{qg_Fr_{*}^2}{\sqrt{12}r_{**}^2})^2-(m_{**}^2+\lambda^2)R_2^2\right)\tilde{y}=0,\nonumber\\
&& \zeta^2\partial^2_{\zeta}\tilde{z}+\zeta\partial_{\zeta}\tilde{z}
+\left(i\zeta\omega+(\omega\zeta+\frac{qg_Fr_{*}^2}{\sqrt{12}r_{**}^2})^2-(m_{**}^2+\lambda^2)R_2^2\right)\tilde{z}=0.
\label{secondode}
\end{eqnarray}
MATHEMATICA gives the following solutions for equations (\ref{secondode})
\begin{eqnarray}
\tilde{y}(\zeta) &=& {\rm e}^{-i\omega\zeta}\zeta^{\nu}\left(c_1\, U(1+\nu+i\frac{qg_Fr_{*}^2}{\sqrt{12}r_{**}^2},1+2\nu,2i\omega\zeta)
+c_2\, L(-1-\nu-i\frac{qg_Fr_{*}^2}{\sqrt{12}r_{**}^2},2\nu,2i\omega\zeta)\right),\nonumber\\
\tilde{z}(\zeta) &=& {\rm e}^{-i\omega\zeta}\zeta^{\nu}\left(c_3\, U(\nu+i\frac{qg_Fr_{*}^2}{\sqrt{12}r_{**}^2},1+2\nu,2i\omega\zeta)
+c_4\, L(-\nu-i\frac{qg_Fr_{*}^2}{\sqrt{12}r_{**}^2},2\nu,2i\omega\zeta)\right),
\label{solution}
\end{eqnarray}
where $\nu=\sqrt{(m_{**}^2+\lambda^2)R_2^2-\frac{q^2g_F^2r_{*}^4}{12r_{**}^4}}$,
$U(a,b,z)$ is the tricomi confluent hypergeometric function (of the second kind)
and $L(\nu,\lambda,z)\equiv L^{\lambda}_{\nu}(z)$ is the generalized Laguerre function
(for $\nu=n$ the associated Laguerre polynomial).
We substitute solutions (\ref{solution}) into the system of first order ODE, eq.(\ref{firstode}), 
and consider this system at the ${\rm AdS_2}$ boundary, $\zeta\rightarrow 0$, where it is considerably simplified.
%we have 
% \begin{eqnarray}
% &c_1& \left((\nu-2i\omega\zeta-i\frac{qg_Fr_{*}^2}{\sqrt{12}r_{**}^2})U(1+\nu+i\frac{qg_Fr_{*}^2}{\sqrt{12}r_{**}^2},1+2\nu,2i\omega\zeta)
% -2i\omega\zeta(1+\nu+i\frac{qg_Fr_{*}^2}{\sqrt{12}r_{**}^2})U(2+\nu+i\frac{qg_Fr_{*}^2}{\sqrt{12}r_{**}^2},2+2\nu,2i\omega\zeta)\right) \nonumber\\
% &+& c_2\left((\nu-2i\omega\zeta-i\frac{qg_Fr_{*}^2}{\sqrt{12}r_{**}^2})L(-1-\nu-i\frac{qg_Fr_{*}^2}{\sqrt{12}r_{**}^2},2\nu,2i\omega\zeta)
% -2i\omega\zeta L(-2-\nu-i\frac{qg_Fr_{*}^2}{\sqrt{12}r_{**}^2},1+2\nu,2i\omega\zeta)\right)\nonumber\\
% &+&(im_{**}+\lambda)R_2\left(c_3U(\nu+i\frac{qg_Fr_{*}^2}{\sqrt{12}r_{**}^2},1+2\nu,2i\omega\zeta)
% +c_4L(-\nu-i\frac{qg_Fr_{*}^2}{\sqrt{12}r_{**}^2},2\nu,2i\omega\zeta)\right)=0\nonumber\\
% &c_3&\left((\nu+i\frac{qg_Fr_{*}^2}{\sqrt{12}r_{**}^2})U(\nu+i\frac{qg_Fr_{*}^2}{\sqrt{12}r_{**}^2},1+2\nu,2i\omega\zeta)
% -2i\omega\zeta(\nu+i\frac{qg_Fr_{*}^2}{\sqrt{12}r_{**}^2})U(1+\nu+i\frac{qg_Fr_{*}^2}{\sqrt{12}r_{**}^2},2+2\nu,2i\omega\zeta)\right)\nonumber\\
% &+& c_4\left((\nu+i\frac{qg_Fr_{*}^2}{\sqrt{12}r_{**}^2})L(-\nu-i\frac{qg_Fr_{*}^2}{\sqrt{12}r_{**}^2},2\nu,2i\omega\zeta)
% -2i\omega\zeta L(-1-\nu-i\frac{qg_Fr_{*}^2}{\sqrt{12}r_{**}^2},1+2\nu,2i\omega\zeta)\right)\nonumber\\
% &-&(im_{**}-\lambda)R_2\left(c_1U(1+\nu+i\frac{qg_Fr_{*}^2}{\sqrt{12}r_{**}^2},1+2\nu,2i\omega\zeta)
% +c_2L(-1-\nu-i\frac{qg_Fr_{*}^2}{\sqrt{12}r_{**}^2},2\nu,2i\omega\zeta)\right)=0 
% \label{check}
% \end{eqnarray}
%At the ${\rm AdS_2}$ boundary, $\zeta\rightarrow 0$, this system is considerably simplified. 
As a result we get the relations between the constants
\begin{eqnarray}
\frac{c_1}{c_3} &=& (im_{**}+\lambda) R_2\nonumber\\
\frac{c_2}{c_4} &=& \frac{(im_{**}+\lambda) R_2}{\nu+i\frac{qg_Fr_{*}^2}{\sqrt{12}r_{**}^2}}
=  \frac{\nu-i\frac{qg_Fr_{*}^2}{\sqrt{12}r_{**}^2}}{(-im_{**}+\lambda) R_2},
\label{relation-one}
\end{eqnarray}  
to simplify we use $\Gamma(z)\Gamma(1-z)=\pi/\sin(\pi z)$.
Both equations in the system give the same relations eq.(\ref{relation-one}). As a consistency check, we found
the same relations considering the system at the ${\rm AdS_2}$ horizon, $\zeta\rightarrow\infty$.
In order to insure the ingoing wave $\sim {\rm e}^{i\omega \zeta}$ 
at the horizon $\zeta\rightarrow\infty$ for each of the solutions, we have
one more relation between the constants 
\begin{eqnarray}
\frac{c_1}{c_2} &=& -\frac{\Gamma(1+\nu+i\frac{qg_Fr_{*}^2}{\sqrt{12}r_{**}^2})}{\pi}. 
\label{relation-two}
\end{eqnarray}
Relations (\ref{relation-one},\ref{relation-two}) fix all the constants up to overall normalization constant.
Using the relations (\ref{relation-one},\ref{relation-two}), 
the solution of the system (\ref{firstode}) at $\zeta=0$ becomes
\begin{eqnarray}
\tilde{y} &\rightarrow & (2i\omega)^{-2\nu}\frac{\Gamma(2\nu)\Gamma(1-\nu-i\frac{qg_Fr_{*}^2}{\sqrt{12}r_{**}^2})}{\nu+i\frac{qg_Fr_{*}^2}{\sqrt{12}r_{**}^2}}\zeta^{-\nu}
+\frac{\Gamma(-2\nu)\Gamma(1+\nu-i\frac{qg_Fr_{*}^2}{\sqrt{12}r_{**}^2})}{-\nu+i\frac{qg_Fr_{*}^2}{\sqrt{12}r_{**}^2}}\zeta^{\nu},\\
\tilde{z} &\rightarrow & (2i\omega)^{-2\nu}\frac{\Gamma(2\nu)\Gamma(1-\nu-i\frac{qg_Fr_{*}^2}{\sqrt{12}r_{**}^2})}{(im_{**}+\lambda) R_2}\zeta^{-\nu}
+\frac{\Gamma(-2\nu)\Gamma(1+\nu-i\frac{qg_Fr_{*}^2}{\sqrt{12}r_{**}^2})}{(im_{**}+\lambda) R_2}\zeta^{\nu}.
\label{ads2-solutions}
\end{eqnarray}
At  $\zeta\rightarrow 0$, solution to the original system of equations (\ref{originalode}) is given by
\begin{eqnarray}
 \left(\begin{array}{c}
y\\
z
\end{array}
\right) = \frac{1}{2} \left(\begin{array}{c}
 \tilde{y}+i\tilde{z} \\
 \tilde{z}+i\tilde{y}
\end{array}
\right) 
 &=& (2i\omega)^{-2\nu}
 \frac{\Gamma(2\nu)\Gamma(1-\nu-i\frac{qg_Fr_{*}^2}{\sqrt{12}r_{**}^2})}{2(\nu+i\frac{qg_Fr_{*}^2}{\sqrt{12}r_{**}^2})}
 \left(\begin{array}{c}
 1+i\frac{\nu+i\frac{qg_Fr_{*}^2}{\sqrt{12}r_{**}^2}}{(im_{**}+\lambda)R_2}
    \\
  \frac{\nu+i\frac{qg_Fr_{*}^2}{\sqrt{12}r_{**}^2}}{(im_{**}+\lambda)R_2}+i
  \end{array}\right)\zeta^{-\nu} 
 \nonumber\\
  &+& \frac{\Gamma(-2\nu)\Gamma(1+\nu-i\frac{qg_Fr_{*}^2}{\sqrt{12}r_{**}^2})}{2(-\nu+i\frac{qg_Fr_{*}^2}{\sqrt{12}r_{**}^2})}
  \left(\begin{array}{c}
  1+i\frac{-\nu+i\frac{qg_Fr_{*}^2}{\sqrt{12}r_{**}^2}}{(im_{**}+\lambda)R_2}
  \\
  \frac{-\nu+i\frac{qg_Fr_{*}^2}{\sqrt{12}r_{**}^2}}{(im_{**}+\lambda)R_2} +i
  \end{array}
  \right)\zeta^{\nu}.
\label{solution1}
\end{eqnarray}
The system of equations (\ref{originalode}) near the ${\rm AdS_2}$ boundary has a solution
which can be written in a general form as, eq.(\ref{general-solution}), 
\begin{equation}
\left(\begin{array}{c}
y\\
z
\end{array}
\right) = Av_{+}\zeta^{-\nu}+Bv_{-}\zeta^{\nu}
=A\left(\begin{array}{c}
 m_{**}R_2+\nu \\
 \lambda R_2+\frac{qg_Fr_{*}^2}{\sqrt{12}r_{**}^2}
\end{array}
\right)\zeta^{-\nu}
+B\left(\begin{array}{c}
 m_{**}R_2-\nu \\
 \lambda R_2+\frac{qg_Fr_{*}^2}{\sqrt{12}r_{**}^2}
\end{array}
\right)\zeta^{\nu}. 
\end{equation}
then the Green function is $G_R^{IR}(\omega)=B/A$.
We put solution eq.(\ref{solution1}) into this form,
\begin{eqnarray}
\left(\begin{array}{c}
y\\
z
\end{array}
\right) &=& 
(2i\omega)^{-2\nu} \Gamma(2\nu)\Gamma(1-\nu-i\frac{qg_Fr_{*}^2}{\sqrt{12}r_{**}^2})
\frac{1+\frac{\nu+i\frac{qg_Fr_{*}^2}{\sqrt{12}r_{**}^2}}{(m_{**}-i\lambda)R_2}}{2(\nu+i\frac{qg_Fr_{*}^2}{\sqrt{12}r_{**}^2})(m_{**}R_2+\nu)}
\left(\begin{array}{c}
 m_{**}R_2+\nu \\
 \lambda R_2+\frac{qg_Fr_{*}^2}{\sqrt{12}r_{**}^2}
\end{array}
\right)\zeta^{-\nu} \nonumber\\
&+& \Gamma(-2\nu)\Gamma(1+\nu-i\frac{qg_Fr_{*}^2}{\sqrt{12}r_{**}^2})
\frac{1+\frac{-\nu+i\frac{qg_Fr_{*}^2}{\sqrt{12}r_{**}^2}}{(m_{**}-i\lambda)R_2}}{2(-\nu+i\frac{qg_Fr_{*}^2}{\sqrt{12}r_{**}^2})(m_{**}R_2-\nu)}
\left(\begin{array}{c}
 m_{**}R_2-\nu \\
 \lambda R_2+\frac{qg_Fr_{*}^2}{\sqrt{12}r_{**}^2}\\
\end{array}
\right)\zeta^{-\nu},\nonumber\\
\label{result}
\end{eqnarray}
and extract the IR Green function to be 
\begin{eqnarray}
G_R^{IR}(\omega) &=& {\rm e}^{-i\pi\nu}
\frac{\Gamma(-2\nu)\Gamma(1+\nu-i\frac{qg_Fr_{*}^2}{\sqrt{12}r_{**}^2})}{\Gamma(2\nu)\Gamma(1-\nu-i\frac{qg_Fr_{*}^2}{\sqrt{12}r_{**}^2})}
\times \nonumber\\
&& \frac{((m_{**}-i\lambda) R_2+i\frac{qg_Fr_{*}^2}{\sqrt{12}r_{**}^2}-\nu)(i\frac{qg_Fr_{*}^2}{\sqrt{12}r_{**}^2}+\nu)(m_{**}R_2+\nu)}
{((m_{**}-i\lambda)R_2+i\frac{qg_Fr_{*}^2}{\sqrt{12}r_{**}^2}+\nu)(i\frac{qg_Fr_{*}^2}{\sqrt{12}r_{**}^2}-\nu)(mR_2-\nu)}
(2\omega)^{2\nu}.
\end{eqnarray}
Simplifying the following ratio
\begin{equation}
\frac{((m_{**}-i\lambda) R_2+i\frac{qg_Fr_{*}^2}{\sqrt{12}r_{**}^2}-\nu)(i\frac{qg_Fr_{*}^2}{\sqrt{12}r_{**}^2}+\nu)}
{((m_{**}-i\lambda)R_2+i\frac{qg_Fr_{*}^2}{\sqrt{12}r_{**}^2}+\nu)(i\frac{qg_Fr_{*}^2}{\sqrt{12}r_{**}^2}-\nu)}
= \frac{((m_{**}+i\lambda) R_2-i\frac{qg_Fr_{*}^2}{\sqrt{12}r_{**}^2}-\nu)}
{((m_{**}+i\lambda)R_2-i\frac{qg_Fr_{*}^2}{\sqrt{12}r_{**}^2}+\nu)},
\end{equation}
we get the retarded IR Green function given by
\begin{eqnarray}
G_R^{IR}(\omega) = {\rm e}^{-i\pi\nu}
\frac{\Gamma(-2\nu)\Gamma(1+\nu-i\frac{qg_Fr_{*}^2}{\sqrt{12}r_{**}^2})}{\Gamma(2\nu)\Gamma(1-\nu-i\frac{qg_Fr_{*}^2}{\sqrt{12}r_{**}^2})}
\frac{((m_{**}+i\lambda) R_2-i\frac{qg_Fr_{*}^2}{\sqrt{12}r_{**}^2}-\nu)}
{((m_{**}+i\lambda)R_2-i\frac{qg_Fr_{*}^2}{\sqrt{12}r_{**}^2}+\nu)}
(2\omega)^{2\nu},
\label{green-result}
\end{eqnarray}
with $m_{**}=m+\Delta$. In eq.(\ref{green-result}),
we did not include a ratio $\frac{m_{**}R_2+\nu}{m_{**}R_2-\nu}$, since there is an ambiguity
in definition of the Green function up to a real function of $\lambda$ (or $k$ with no magnetic field) and $q,m$.
If the matching is done using our basis then this difference should not matter.
This expression for the IR Green function agrees with the one obtained in \cite{Faulkner1:2009}.

\section{One-loop calculations in a $(2+1)$ dimensional field theory}\label{appendix:c}

We calculate here the free fermion energy and the gap equation.
One-loop fermion effective action in the chiral limit, $m=0$, is given by
\begin{equation}
S_{eff}^{1loop}=-i\ln\det(i\slashchar{D}-\Delta)=
-\frac{i}{2}\ln\det(\slashchar{D}^2+\Delta^2), 
\label{det}
\end{equation}
where $i\slashchar{D}=(i\partial_t+\mu)\gamma^0-v_F\vec{K}\vec{\gamma}$,
and $\vec{K}=i\vec{\nabla}+q\vec{A}$. For simplicity, we added to the free part the interaction
$G_{int}(\bar{\psi}\psi)(\bar{\psi}\psi)\rightarrow (\Delta(\bar{\psi}\psi)+h.c.)-\Delta^2/4G_{int}$,
where the strength of intercation in $(2+1)$-d is $G_{int}\sim\frac{1}{M_F}$.
Here, the order parameter is $\Delta=2G_{int}<\bar{\psi}\psi>$.
In the Landau gauge $\vec{A}=(-{\mathcal H}y,0)$, and after the Fourier transform, we have
\begin{equation}
-\slashchar{D}^2 = (\omega+\mu)^2-v_{F}^2\vec{K}^2 -iq{\mathcal H}v_F^2\gamma^1\gamma^2. 
\end{equation}
To calculate the fermion determinant, eq.(\ref{det}), we use $\ln\det G^{-1}={\rm Tr}\ln G^{-1}$.
The eigenvalues of operator $\vec{K}^2$ are known $(2l+1)|q{\mathcal H}|$ 
(we also calculated them in Appendix \ref{appendix:a});
the eigenvalues of operator $iv_F^2q{\mathcal H}\gamma^1\gamma^2$ are $\pm v_F^2|q{\mathcal H}|$ 
(in standard representation
for $\gamma$ matrices); i.e., the ${\mathcal H}$ dependent part 
is $(2l+1)v_F^2|q{\mathcal H}|\pm v_F^2|q{\mathcal H}|$. 
One can rescale $l\rightarrow l-1$ for one of the signs and combine two terms
with both signs together with the result $v_{F}^22|q{\mathcal H}|l$. 
After rescaling there will be however different prefactors for two signs from taking matrix elements 
under the trace, ${\rm Tr}$
(see \cite{Shovkovy:2d} for details). Since we will consider only the lowest Landau level, we can ignore the difference
in prefactors, and moreover
\begin{equation}
\int\frac{d^2k}{(2\pi)^2} \rightarrow  \frac{V_2|q{\mathcal H}|}{(2\pi)}, 
\end{equation}
that takes into account the degeneracy of Landau levels, since the Dirac equation eigenvalue $\lambda$
and hence the quasiparticle spectrum do not depend on momentum $k$.  
Here $V_2=L_x\times L_y$ is the size of the sample.    
We therefore have
\begin{equation}
S_{eff}^{1loop}=-\frac{V_2|q{\mathcal H}|}{2\pi}\sum_n \ln\frac{(\omega_n+i\mu)^2+E_l^2}{T^2}, 
\end{equation}
where the fermionic Matsubara frequencies at temperature $T$ are 
$\omega_n=(2n+1)\pi T$ (we changed to Matsubara frequency by Wick rotation
$\omega_n=i\omega$), and $E_l=\sqrt{2v_{F}^2|q{\mathcal H}|l+\Delta^2}$. 
Sum over the Landau levels $l$ is implied.
The Dirac equation eigenvalue is $\lambda=(\omega_n+i\mu)^2+E_l^2$, which gives
quasiparticle poles $z_{*}(l)=i\omega_n$ at $\lambda=0$ equal to
$z_{*}(l)=\mu\pm E_l$.
We rewrite the Matsubara sum as a contour integral
\begin{equation}
\sum_n \ln\frac{(\omega_n+i\mu)^2+E_l^2}{T^2}=
\frac{i}{2}\int_C \frac{d z}{2\pi} \ln\frac{-(z-\mu)^2+E_l^2}{T^2}\tanh\frac{z}{2 T},
\end{equation}
due to the fact that the poles of $\tanh$ are situated along the imaginary axis at $z=i(2n+1)\pi T$. 
Differentiating both sides with respect to $E_l$, we take the r.h.s. integral
\begin{equation}
\sum_n\frac{2E_l}{(\omega_n+i\mu)^2+E_l^2}=\frac{1}{2}\sum_{z_{*}}\tanh\frac{|z_{*}(l)|}{2T}
= \frac{1}{2}\left(\tanh\frac{E_l-\mu}{2T}+\tanh\frac{E_l+\mu}{2T}\right).
\label{matsubara} 
\end{equation}
Integrating back over $E_l$, we have 
\begin{eqnarray}
T\sum_n \ln\frac{(\omega_n+i\mu)^2+E_l^2}{T^2}&=&T\sum_{z_{*}(l)}\ln\left(2\cosh\frac{|z_{*}(l)|}{2T}\right) 
= T \sum_{z_{*}(l)}\left(\frac{|z_{*}(l)|}{2T}+
\ln(1+{\rm e}^{-|z_{*}(l)|/T})\right)  
\nonumber\\
&=& \frac{E_l-\mu}{2}+T\ln(1+{\rm e}^{-(E_l-\mu)/T})
+ \frac{E_l+\mu}{2}+T\ln(1+{\rm e}^{-(E_l+\mu)/T}).\nonumber\\
\end{eqnarray}
A useful formula following from eq.(\ref{matsubara}),
\begin{eqnarray}
T\sum_n\frac{1}{(\omega_n+i\mu)^2+E_l^2}&=&\frac{1}{2E_l}\sum_{z_{*}(l)}\frac{1}{2}\tanh\frac{|z_{*}(l)|}{2T}
= \frac{1}{2E_l}\frac{\sinh\frac{E_l}{T}}{\cosh\frac{E_l}{T}+\cosh\frac{\mu}{T}}. 
\end{eqnarray}
Putting all together, an effective action for $\Delta$ is given by
\begin{equation}
S_{eff}= \frac{V_2}{T}\left(\frac{|\Delta|^2}{4G_{int}}-\frac{T|q{\mathcal H}|}{2\pi}
\sum_{z_{*}(l)}\ln\left(2\cosh\frac{z_{*}(l)}{2T}\right)\right),
\label{action2}
\end{equation}
with $z_{*}(l)=\mu\pm E_l$, $E_l=\sqrt{2|q{\mathcal H}|l+\Delta^2}$, sum over the Landau levels $l$ is implied.
The free fermion energy can be obtained by dividing $S_{eff}$ by the space-time volume, i.e.,
$\Omega_F=-S_{eff}/(TV_2)$.
Minimizing effective action, $\delta S_{eff}/\delta\Delta =0$, we get the gap equation
\begin{equation}
 \Delta=\frac{G_{int}|q{\mathcal H}|}{\pi}\frac{\Delta}{E_l} \frac{\sinh\frac{E_l}{T}}{\cosh\frac{E_l}{T}+\cosh\frac{\mu}{T}}, 
\end{equation}
with $E_l=\sqrt{2|q{\mathcal H}|l+\Delta^2}$, sum over $l$ is implied. 
For the lowest Landau level, $l=0$, the gap equation reads
\begin{equation}
 \Delta=\frac{G_{int}|q{\mathcal H}|}{\pi}\frac{\sinh\frac{\Delta}{T}}{\cosh\frac{\Delta}{T}+\cosh\frac{\mu}{T}}.
\label{gap-equation2} 
\end{equation}
At $T=0$, the solution is given by
\begin{equation}
\Delta= \frac{1}{\pi}G_{int}|q{\mathcal H}|,
\label{solution-gap}
\end{equation}
provided $\Delta>\mu$, and where $G_{int}=\frac{1}{M_F}$.
At $T\neq 0$, from eq.(\ref{gap-equation2}),
there is the second solution $\Delta=0$, and the phase transition
between $\Delta\neq 0$ and $\Delta=0$. The character of the phase transition,
first or second order depends on the values of parameters \cite{Shovkovy:2d}.

We calculate the critical temperature of the phase transition.
We fix the charge density, $n$, and express the chemical potential through $n$.
From the effective action eq.(\ref{action2}), the charge density and the gap equation for the lowest
Landau level are given by 
\begin{eqnarray}
n &=& \frac{|q{\mathcal H}|}{2\pi}\frac{\sinh(\frac{\mu}{T})}{\cosh(\frac{\Delta}{T})+\cosh(\frac{\mu}{T})},
\nonumber\\
\Delta &=& \frac{G_{int}|q{\mathcal H}|}{\pi}\frac{\sinh(\frac{\Delta}{T})}{\cosh(\frac{\Delta}{T})+\cosh(\frac{\mu}{T})}. 
\end{eqnarray}
We introduce the filling factor
\begin{equation}
\eta_{\mathcal H}=\frac{2\pi n}{|q{\mathcal H}|}\equiv \frac{\mathcal H_c}{\mathcal H}, 
\end{equation}
then from the expression for the charge density, we have
\begin{equation}
\cosh(\frac{\mu}{T}) =\frac{\eta_{\mathcal H}^2\cosh(\frac{\Delta}{T})+\sqrt{1+\eta_{\mathcal H}^2\sinh^2(\frac{\Delta}{T})}}
{1-\eta_{\mathcal H}^2}. 
\end{equation}
Therefore the gap equation becomes
\begin{equation}
\Delta = \frac{G_{int}|q{\mathcal H}|}{\pi} \frac{(1-\eta_{\mathcal H}^2)\sinh(\frac{\Delta}{T})}
{\cosh(\frac{\Delta}{T})+\sqrt{1+\eta_{\mathcal H}^2\sinh^2(\frac{\Delta}{T})}}.
\end{equation}
At T=0, the solution is given by
\begin{equation}
\Delta=\frac{G_{int}|q{\mathcal H}|}{\pi}(1-\eta_{\mathcal H}). 
\end{equation}
There is no nonzero gap for the filling factor $\eta_{\mathcal H}>1$. The condition $\eta_{\mathcal H}<1$ to
have a nonzero gap translates for the charge density to be smaller than critical one, $n<n_c$,
with $n_c=n(\eta_{\mathcal H}=1)$, or
for the magnetic field to be larger than the critical one, ${\mathcal H}>{\mathcal H}_c$.
For $\eta_{\mathcal H}>1$, i.e. $n>n_c$ or ${\mathcal H}<{\mathcal H}_c$ the symmetry is restrored, $\Delta=0$.
Around the critical temperature, when the gap is vanishing, the gap equation gives the following critical temperature 
\begin{equation}
T_c= \frac{G_{int}|q{\mathcal H}|}{2\pi}(1-\eta_{\mathcal H}^2), 
\end{equation}
where $T_c=0$ for $\eta_{\mathcal H}>1$, i.e. for ${\mathcal H}<{\mathcal H}_c$. 
For ${\mathcal H}>{\mathcal H}_c$, $T_c$ grows linearly with magnetic field, $T_c\sim |q{\mathcal H}|$,
in the vicinity of the phase transition. Away from the phase transition one should solve 
the following gap equation for the lowest Landau level numerically
\begin{equation}
\Delta=\frac{2T_c\sinh(\frac{\Delta}{T})}{\cosh(\frac{\Delta}{T})+\sqrt{1+\eta_{\mathcal H}^2\sinh^2(\frac{\Delta}{T})}}. 
\end{equation}
We use this procedure to derive the gap equation and to calculate $T_c$ in the ${\rm AdS_4}$.

\section{Critical temperature from the $ADS_4$ variational calculations}\label{appendix:d}

We calculte the critical temperature $T_c$ for the case of Ladau Fermi liquid, $\nu_{k_F}>\frac{1}{2}$.
Let us introduce an analog of the charge density in the $AdS_4$ by differentiating an effective action eq.(\ref{action3})
with respect to the Fermi momentum $k_F$, $n(r)=\frac{\delta S_{eff}}{\delta (v_Fk_F)}$. Together with the gap equation,
$\frac{\delta S_{eff}}{\delta\Delta(r)}=0$, we have 
\begin{eqnarray}
n &=& \frac{|q{\mathcal H}|}{2\pi R}
\frac{1}{\pi}
\sum_{z_{*}[\Delta(r)]}\left(\frac{\delta\omega_{*}[\Delta(r)]}{\delta(v_Fk_F)}
{\rm Im}  \Psi(\frac{iz_{*}[\Delta(r)]}{2\pi T}+\frac{1}{2})\right.\nonumber\\
&-&\left.\frac{\delta\Gamma[\Delta(r)]}{\delta(v_Fk_F)}
{\rm Re} \Psi(\frac{iz_{*}[\Delta(r)]}{2\pi T}+\frac{1}{2})\right), 
\label{charge-density2}
\\
\Delta(r) &=& \frac{G_{int}|q{\mathcal H}|}{\pi}
\frac{1}{\pi}\sum_{z_{*}(\Delta(r))}\left(\frac{\delta\omega_{*}[\Delta(r)]}{\delta\Delta(r)}
{\rm Im}  \Psi(\frac{iz_{*}[\Delta(r)]}{2\pi T}+\frac{1}{2})\right.\nonumber\\
&-&\left.\frac{\delta\Gamma[\Delta(r)]}{\delta\Delta(r)}
{\rm Re} \Psi(\frac{iz_{*}[\Delta(r)]}{2\pi T}+\frac{1}{2})\right). 
\label{gap-equation2}
\end{eqnarray}
Here sum goes over the two poles.
For the lowest Landau level, $l=0$,
\begin{eqnarray}
n &=& \frac{|q{\mathcal H}|}{2\pi R}
\frac{1}{\pi}{\rm Im}\left(
-\Psi(\frac{iv_F(\delta k_F[\Delta(r)]-k_F)}{2\pi T}+\frac{1}{2})\right.\nonumber\\
&+&\left. \Psi(\frac{iv_F(\delta k_F[\Delta(r)]+k_F)}{2\pi T}+\frac{1}{2})\right),
\\
\Delta(r) &=& \frac{G_{int}|q{\mathcal H}|}{\pi}
\frac{1}{\pi}{\rm Im}\left(
\Psi(\frac{iv_F(\delta k_F[\Delta(r)]-k_F)}{2\pi T}+\frac{1}{2})\right.\nonumber\\
&+&\left.\Psi(\frac{iv_F(\delta k_F[\Delta(r)]+k_F)}{2\pi T}+\frac{1}{2})\right)
\frac{\psi^{0}(r)^{\dagger}\sigma^{1}\psi^{0}(r)}{R^4},  
\label{}
\end{eqnarray}
where the shift of the Fermi momentum is given by
\begin{equation}
\delta k_F[\Delta(r)]=\frac{1}{v_F R^4}\int dr \sqrt{-g} \psi^{0}(r)^{\dagger}\sigma^1\psi^{0}(r)\Delta(r). 
\end{equation}
For $T\sim T_c$, we expand in $\Delta \ll T$, 
\begin{eqnarray}
n &=& \frac{|q{\mathcal H}|}{2\pi R}
\frac{1}{\pi}{\rm Im}\left(
-\Psi(\frac{-iv_Fk_F}{2\pi T}+\frac{1}{2})
+\Psi(\frac{iv_Fk_F}{2\pi T}+\frac{1}{2})\right),
\label{density2} 
\\
\Delta(r) &=& \frac{G_{int}|q{\mathcal H}|}{\pi}
\frac{1}{\pi}{\rm Im}\frac{i\int dr\sqrt{-g} \psi^{0}(r)^{\dagger}\sigma^1\psi^{0}(r)\Delta(r)}{2\pi T R^4}
\times\nonumber\\
&& \left(\Psi'(\frac{-iv_Fk_F}{2\pi T}+\frac{1}{2})
+\Psi'(\frac{iv_Fk_F}{2\pi T}+\frac{1}{2})\right)
\frac{\psi^{0}(r)^{\dagger}\sigma^{1}\psi^{0}(r)}{R^4}, 
\label{tc2}
\end{eqnarray}
where $\Psi'(x)$ is the derivative of the digamma function $\Psi'(x)=\frac{d^2\ln\Gamma(x)}{dx^2}$;
the subleading term $\sim\Delta$ in $n$ and the leading term $\sim 1$ in $\Delta$
vanish due to the imaginary part.
We use that the solution of the gap equations at zero temperature is given by eq.(\ref{solution-gap2}).
Therefore the radial profile is given by
\begin{equation}
\Delta(r) \sim \psi^{0}(r)^{\dagger}\sigma^1\psi^{0}(r). 
\end{equation}
Substituting it into eq.(\ref{tc2}), we have
\begin{equation}
1 =  \frac{G_{int}|q{\mathcal H}|}{\pi}
\frac{1}{\pi}{\rm Im}\frac{i}{2\pi T} 
\left(\Psi'(\frac{-iv_Fk_F}{2\pi T}+\frac{1}{2})
+\Psi'(\frac{iv_Fk_F}{2\pi T}+\frac{1}{2})\right)
\frac{\int dr\sqrt{-g} (\psi^{0}(r)^{\dagger}\sigma^1\psi^{0}(r))^2}{R^8}.
\end{equation}
Simplifying the digamma functions and their derivatives,    
we obtain
\begin{eqnarray}
n &=& \frac{|q{\mathcal H}|}{2\pi R}\tanh\frac{v_Fk_F}{2T},
\label{density}
\\
1 &=& \frac{G_{int}|q{\mathcal H}|}{\pi}\frac{1}{2T}\frac{1}{\cosh^2\frac{v_Fk_F}{2T}}
\frac{\int dr \sqrt{-g} (\psi^{0}(r)^{\dagger}\sigma^{1}\psi^{0}(r))^2}{R^8}.
\label{gap-equation5}
\end{eqnarray}
We introduce the filling factor
\begin{equation}
\eta_{\mathcal H}(r) =\frac{2\pi R n}{|q{\mathcal H}|} \equiv \frac{\mathcal H_c}{\mathcal H}.
\end{equation}
From the equation (\ref{density}) for the charge density, we have
\begin{equation}
\cosh^2(\frac{v_Fk_F}{2T})=\frac{1}{1-\eta_{\mathcal H}^2}. 
\end{equation}
Using it in the gap equation (\ref{gap-equation5}), we get the critical temperature 
for the lowest Landau level
\begin{equation}
T_c =\frac{G_{int}|q{\mathcal H}|}{2\pi R^8}(1-\eta_{\mathcal H}^2)
\int dr \sqrt{-g} (\psi^{0}(r)^{\dagger}\sigma^{1}\psi^{0}(r))^2. 
\end{equation}
For the filling factor $\eta_{\mathcal H}>1$, the critical temperature vanishes, $T_c=0$,
and for $\eta_{\mathcal H}<1$, which means either ${\mathcal H}>{\mathcal H_c}$ or $n<n_c$,
the critical temperature grows with the magnetic field 
in the vicinity of the phase transition. The integral over the profile agrees with
the critial temperature given in eq.(\ref{critical-temperature}). 
In eq.(\ref{critical-temperature}), $v_Fh_1$ introduces the dependence
$v_Fh_1\sim 1/\int dr \sqrt{g/g_{tt}} \psi^{0}(r)^{\dagger}\psi^{0}$, which probably
follows from a more careful definition for the density $n$ in the above calculations.

\bibliographystyle{JHEP}

\begin{thebibliography}{99}
\bibitem{Tomoi}
Unpublished notes on branch cut in the integrals 
$\ln z=\int_1^z \frac{d\zeta^{\prime}}{\zeta^{\prime}}$ where $z$ is imaginary.
\end{thebibliography}

\end{document}